%% file: chr_survey.tex
\documentclass{tlp}

\submitted{27 March 2008}
\revised{9 March 2009}
\accepted{24 June 2009}

\usepackage{stmaryrd}
\usepackage{amsmath}
\usepackage{amssymb}
\usepackage{fancyvrb}
\usepackage[usenames]{color}
\usepackage{makeidx}
\makeindex

\input{macros}

\newcommand\thetitle{As Time Goes By: Constraint Handling Rules}
\newcommand\thesubject{A Survey of CHR Research from 1998 to 2007}
\newcommand\thekeywords{Constraint Handling Rules, CHR, survey}

\usepackage{ifpdf}
\ifpdf
	\usepackage[pdftex]{graphicx}
	\usepackage{pdfpages}
	\usepackage[
	    pdftex,
	    pdfauthor={Jon Sneyers and Peter Van Weert and Tom Schrijvers and Leslie De Koninck},
	    pdftitle={\thetitle},
	    pdfsubject={\thesubject},
	    pdfkeywords={\thekeywords},
	    breaklinks,
	    pdfstartview=FitH,
	    colorlinks,
	    linkcolor=black,
	    filecolor=black,
	    citecolor=black,
	    urlcolor=black
	]{hyperref}
	\usepackage[all]{hypcap}
\else	
	\usepackage[dvips]{graphicx}
	\usepackage{url}
	\newcommand\href[2]{{\tt #2}}
\fi

\title[\thesubject]{\thetitle\\{\large \thesubject}}
\author[J. Sneyers et al.]
{JON SNEYERS, PETER VAN WEERT, \authorbreak
TOM SCHRIJVERS, and LESLIE DE KONINCK\\
Dept.\ of Computer Science, K.U.Leuven, Belgium\\
\email{\textit{FirstName}.\textit{LastName}@cs.kuleuven.be}
}

\begin{document}
\maketitle
\pagestyle{plain}

\begin{abstract}
Constraint Handling Rules (CHR) is a high-level programming language based on
multi-headed multiset rewrite rules. Originally designed for writing
user-defined constraint solvers, it is now recognized as
an elegant general purpose language.


CHR-related research has surged during the decade following the previous survey 
by \citeN{fru_chr_overview_jlp98}. 
Covering more than 180 publications, this new survey provides an overview of 
recent results in a wide range of research areas, 
from semantics and analysis to systems, extensions and applications.
\end{abstract} 

\begin{keywords}
\thekeywords
\end{keywords}

\setcounter{tocdepth}{2}
\tableofcontents

\input{01-introduction}

\input{02-semantics}

\input{03-analysis}
\input{04-implementation}
\input{05-extensions}
\input{06-other_formalisms}

\input{07-applications}
\input{08-conclusions}

\section*{Acknowledgments}
The authors would like to thank 
Alan Baljeu (Cornerstone), 
Bart Demoen, 
Mike Elston (Scientific Software \& Systems),
Thom Fr\"uhwirth,
Ralf Gerlich (BSSE),
Gerda Janssens, 
Paolo Pilozzi, 
Dean Voets, 
Jonathan Weston-Dawkes (MITRE),
and Pieter Wuille
for their invaluable contributions to this paper.

Jon Sneyers and Leslie De Koninck are
funded by Ph.D. grants of the Institute for the 
Promotion of Innovation through Science and Technology 
in Flanders (IWT-Vlaanderen).
Peter Van Weert is a Research Assistant of the fund for Scientific Research - Flanders (Belgium) (F.W.O. - Vlaanderen).
Tom Schrijvers is a Post-Doctoral Researcher of the fund for Scientific Research - Flanders.

\bibliographystyle{acmtrans2} 
\bibliography{biblio}

\end{document}

%% file: macros.tex
\newcommand{\Prog}{\mathcal{P}}
\newcommand{\simp}{\ \ensuremath{\Longleftrightarrow}\ }
\newcommand{\prop}{\ \ensuremath{\Longrightarrow}\ }
\newcommand{\bs}{\ \ensuremath{\backslash}\ }
\newcommand{\guard}{\ \ensuremath{|}\ }

\newcommand{\HL}{\ensuremath{\mathcal{H}}}

\newcommand{\CT}{\mathcal{D}}
\newcommand{\BT}{\ensuremath{\CT_{\HL}}}
\newcommand{\musthold}{\ensuremath{\BT \models}}
\newcommand{\true}{\ensuremath{\mbox{\it true}}}
\newcommand{\false}{\ensuremath{\mbox{\it fail}}}
\newcommand{\peq}{{\tt ==}}
\newcommand{\pneq}{{\tt \textbackslash==}}
\newcommand{\exist}{\ensuremath{\bar \exists}}
\newcommand{\omegat}{\ensuremath{\omega_t}}
\newcommand{\omegar}{\ensuremath{\omega_r}}
\newcommand{\omegap}{\ensuremath{\omega_p}}
\newcommand{\omegarp}{\ensuremath{\omega_{rp}}}
\newcommand{\omegac}{\ensuremath{\omega_c}}
\newcommand{\omegao}{\ensuremath{\omega_o}}

\newcommand{\omegarnot}{\omeganot{r}}
\newcommand{\omeganot}[1]{\ensuremath{\omega_{#1}^{\!\!\lnot}}}
\newcommand{\omegarv}{\ensuremath{\omega_r^\vee}}

\newcommand{\omegatree}{\ensuremath{\omega_{\Yup}}}
\newcommand{\omegaset}{\ensuremath{\omega_{\mbox{\it set}}}}

\newcommand{\CHRv}{CHR$^{\vee}$}
\newcommand{\CHRnot}{\ensuremath{\mathrm{CHR}^{\mbox{\large$\!\!\lnot$}}\ }}

\newcommand{\chrrp}{CHR$^{\textrm{rp}}$}

\newcommand{\chr}{\mathit{chr}}

\newcommand{\id}{\mathit{id}}

\newcommand{\States}{\ensuremath{\Sigma^{\mbox{\sc chr}}}}

\newcommand{\st}[5]{\ensuremath{\langle #1, #2, #3, #4 \rangle_{#5}}}

\newcommand{\ILetter}[1]{\ensuremath{\mathbb{#1}}}
\newcommand{\IG}{\ILetter{G}}

\newcommand{\IS}{\ILetter{S}}
\newcommand{\IB}{\ILetter{B}}
\newcommand{\IT}{\ILetter{T}}

\newcommand{\num}[2]{#1\##2}

\newcommand{\tr}{\ensuremath{\rightarrowtail}}

\newcommand{\N}{\ensuremath{\mathbb{N}}}
\newcommand{\concat}{\mathrel+\joinrel\mathrel+}

\newcommand{\comment}[1]{}

\newcommand{\jon}[1]{\todo{#1}{jon}{red}}
\newcommand{\tom}[1]{\todo{#1}{tom}{cyan}}

\newcommand{\leslie}[1]{\todo{#1}{leslie}{blue}}
\newcommand{\ldk}[1]{\leslie{#1}}

\newcommand\todo[3]{\textbf{\color{#3}#1}\marginpar{\sc {\color{#3}#2}}}

\newcommand{\eclipse}{ECL$^i$PS$^e$}

\newcommand{\klein}{\scriptsize}

\include{newtheorems}

%% file: newtheorems.tex
\newtheorem{theorem}{Theorem}[section]

\newtheorem{definition}[theorem]{Definition}

%% file: 01-introduction.tex
\section{Introduction}\label{section:intro}

Constraint Handling Rules (CHR) is a high-level programming language 
based on multi-headed, committed-choice, guarded multiset rewrite rules.
Originally designed in 1991 by Fr\"uhwirth
\citeNN{fru_constraint_simplification_rules_techrep92,fru_chr_cp95,fru_chr_overview_jlp98,fru_chr_2008}
for the special purpose of adding user-defined constraint solvers to a host-language, 
CHR has matured over the last decade to a powerful and elegant general-purpose language
with a wide spectrum of application domains.
Its logical semantics and monotonicity properties naturally lead to anytime, online, and concurrent programs.

The previous survey on CHR \cite{fru_chr_overview_jlp98} was written in 1998.
The aim of this paper is to complement that survey
by giving an overview of the last decade of CHR-related research.
We advise readers that are not yet familiar with CHR to read 
\citeN{fru_chr_overview_jlp98}
first as we have kept the amount of overlap minimal.


\paragraph{Overview.}
We start with a short historical overview of the past 10 years of CHR research,
followed by an introduction to the language itself.
Section~\ref{sec:semantics} describes the logical and operational semantics;
Section~\ref{sec:analysis} covers 
program analysis topics such as confluence, termination, and complexity.
Next, in Section~\ref{sec:implementation}, we discuss the different CHR systems
and compilation techniques. 
Extensions and variants of CHR are dealt with in Section~\ref{sec:extensions},
while Section~\ref{sec:other_formalisms} discusses the relation between
CHR and other formalisms.
In Section~\ref{sec:applications} we give an overview of the many applications of CHR.
Finally, Section~\ref{sec:conclusions} concludes this survey.

\subsection{Historical Overview}


Early CHR research is performed 
at the Ludwig Maximilians Universit\"at (LMU) and the European Computer-Industry
Research Centre (ECRC), both in Munich, by 
Fr\"uhwirth (who later moves to Ulm) and his students 
Abdennadher (who later moves to Cairo), 
Meuss, and 
Wolf (in Berlin).

At the end of the nineties, CHR research focusses on
theoretical properties of CHR programs like 
confluence, 
completion, 
operational equivalence, 
termination, 
and complexity 
(Section~\ref{sec:analysis}).
In the same period, the seminal Holzbaur-Fr\"uhwirth CHR system
in SICStus Prolog is developed
(\citeANP{holz_fru_CHR_manual_techrep98}
\citeyearNP{holz_fru_CHR_manual_techrep98,holz_fru_compiling_chr_attr_vars_ppdp99,holz_fru_prolog_chr_compiler_aai00})
and the first CHR systems in Java are created
(Section~\ref{sec:implementation:systems:java}).

\begin{figure}[tb]
        \centering
	\includegraphics[width=\textwidth]{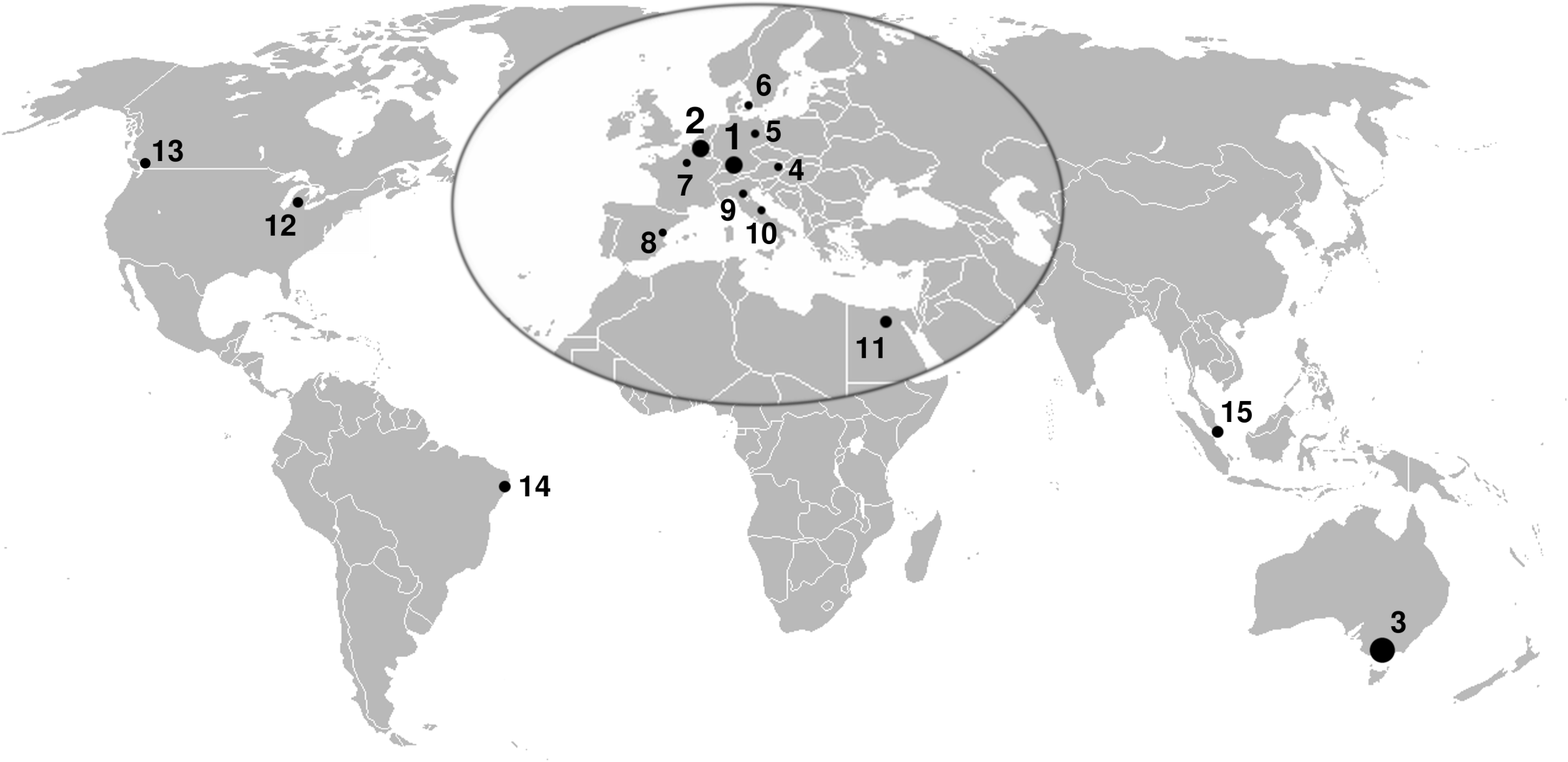}
        \caption{CHR research groups all over the world: 
	1.~Ulm, Germany 	{\klein (Fr\"uhwirth, Meister, Betz, Djelloul, Raiser, \ldots)};
	2.~Leuven, Belgium 	{\klein (Schrijvers, Demoen, Sneyers, De~Koninck, Van~Weert, \ldots)};
	3.~Melbourne, Australia {\klein (Duck, Stuckey, Garc\'ia~de~la~Banda, Wazny, Brand, \ldots)};
	4.~Vienna, Austria 	{\klein (Holzbaur)};
	5.~Berlin, Germany 	{\klein (Wolf)};
	6.~Roskilde, Denmark 	{\klein (Christiansen)};
	7.~Paris, France 	{\klein (Coquery, Fages)};
	8.~Castell\'on, Spain 	{\klein (Escrig, \ldots)};
	9.~Bologna/Ferrara, Italy 	{\klein (Mello, Lamma, Tacchella, \ldots)};
	10.~Chieti, Italy 	{\klein (Meo, \ldots)};
	11.~Cairo, Egypt 	{\klein (Abdennadher, \ldots)};
	12.~Michigan, USA 	{\klein (Sarna-Starosta)};
	13.~Vancouver, Canada 	{\klein (Dahl)};
	14.~Recife, Brazil 	{\klein (Robin, Vitorino, \ldots)};
	15.~Singapore 		{\klein (Sulzmann, Lam)};
	}\label{fig:world}
\end{figure}

Until about 2001, most of the CHR research is still done in Germany and Vienna; 
other groups are discovering CHR,
at first mostly as an implementation language for applications.
For instance, Sulzmann et al.\ use CHR for type systems (Section~\ref{sec:applications:type}) 
and Christiansen and Dahl develop CHR grammars for language processing (Section~\ref{sec:applications:parsing_and_nlp}).
Meanwhile,
Brand, Monfroy, Abdennadher and Rigotti study the automatic generation of CHR programs
(Section~\ref{sec:extensions:codegen}).
Starting around 2002 there is a strong growth of international 
research interest in CHR, leading to a series of workshops on CHR
\cite{pchr04,pchr05,pchr06,pchr07}.
Figure~\ref{fig:world} gives an (incomplete) overview of the most 
active CHR research groups all over the world.


In 2003 and 2004, the groups in Melbourne and Leuven
start working on (static) analysis and optimizing compilation of CHR,
culminating in the Ph.D. theses of 
\citeN{duck_phdthesis05}
and 
\citeN{schr_phdthesis05}.
This work leads to the formulation of the refined operational semantics
(Section~\ref{sec:semantics:operational:omegar})
and the creation of new, highly optimizing CHR systems (Section~\ref{sec:implementation:compilation}).

In Leuven and Brazil, research starts around 2005
on search and Java implementations of CHR
(Sections~\ref{sec:implementation:systems:java}~and~\ref{sec:extensions:disj}),
while the Ulm group
investigates
alternative logical semantics for CHR 
(Section~\ref{sec:semantics:logical}).
The study of an implementation of the union-find algorithm
in CHR 
leads to a focus on
compiler optimizations 
(Section~\ref{sec:implementation:optimization})
and the use of CHR for general-purpose programming 
(Section~\ref{sec:applications:uf}).

Recent trends in CHR-related research include
the relation between CHR and other formalisms
(Section~\ref{sec:other_formalisms}),
extensions and variants of CHR
(Section~\ref{sec:extensions}),
and a renewed interest in theoretical properties like
confluence, 
termination, 
and complexity 
(Section~\ref{sec:analysis}).
In the context of the NICTA project ``G12'', the Melbourne group is currently developing Cadmium,
an ACD term rewriting language which extends CHR (Section~\ref{ACD}).
The Ulm group is currently researching global constraints in the context
of the DFG project ``GLOBCON''; this work is related to
automatic rule generation (Section~\ref{sec:extensions:codegen})
and program transformation (Section~\ref{sec:implementation:specialization_transformation}).

\subsection{Constraint Handling Rules}

To make this survey somewhat self-contained, we briefly introduce the syntax and informal semantics
of Constraint Handling Rules.
For a gentler introduction to CHR, we refer the reader to
\citeN{fru_chr_overview_jlp98}, \citeN{fru_abd_essentials_of_cp_book03},
\citeN{schr_phdthesis05}, \citeN{duck_phdthesis05}, or \citeN{fru_chr_2008}.

CHR is embedded in a host language $\HL$ that provides data types and 
a number of predefined constraints. 
These constraints are called host language constraints or {\em built-in} constraints. 
The traditional host language of CHR is Prolog. 
Its only host language constraint is equality of Herbrand terms;
its data types are Prolog variables and terms.
We denote the host language in which CHR is embedded between round brackets:
i.e.\ CHR($\HL$) denotes CHR embedded in host language $\HL$.
Most systems are CHR(Prolog) systems, 
but there are also several implementations of CHR(Java) and CHR(Haskell), 
and recently a CHR(C) system was developed.
A thorough discussion of existing CHR implementations is given 
in Section~\ref{sec:implementation}.
We require the host language to provide at least the basic constraints $\true$ and $\false$,
and 
syntactic equality (``\peq'') and inequality (``\pneq'') checks.

\subsubsection{Syntax}

CHR constraint symbols are drawn from the set of predicate symbols, 
denoted by a functor$/$arity pair. 
CHR constraints, also called constraint atoms or constraints for short, 
are atoms constructed from these symbols and the data types provided by the host language.
A CHR program $\Prog$ consists of a sequence of CHR rules.
There are three kinds of rules: (where $l,m,n,o \geq 1$)
\begin{itemize}
\item Simplification rules: \hfill
$ h_1,\ldots,h_n \simp g_1,\ldots,g_m \guard b_1,\ldots,b_o$.
\item Propagation rules: \hfill
$ h_1,\ldots,h_n \prop\, g_1,\ldots,g_m \guard b_1,\ldots,b_o$.
\item Simpagation rules: \hfill
$ h_1,\ldots,h_l \bs h_{l+1},\ldots,h_n \simp g_1,\ldots,g_m \guard b_1,\ldots,b_o$.
\end{itemize}
 The sequence, or conjunction, $h_1,\ldots,h_n$ are CHR
constraints; together they are called the {\em head}
or {\em head constraints} of the
rule. A rule with $n$ head constraints is called an {\em $n$-headed} rule and when $n > 1$,
it is a {\em multi-headed} rule.
All the head constraints
of a simplification rule and the head constraints $h_{l+1},\ldots,h_n$ of a
simpagation rule are called {\em removed} head constraints. 
The other head constraints --- all heads of a propagation
rule and $h_1,\ldots,h_l$ of a simpagation rule --- are called {\em kept}
head constraints.
%
%
The conjunction $b_1,\ldots,b_o$ consists of CHR constraints and host
language constraints; it is called the {\em body} of the rule. 
The part of the rule between the arrow and the body is called the
{\em guard}. It is a conjunction of host language constraints. 
The guard ``$g_1,\ldots,g_m \guard$'' is optional; 
if omitted, it is considered to be ``$\true \guard$''.
A rule is optionally preceded by $name~@$ where $name$ is a term. 
No two rules may have the same name, and rules without an explicit
name get a unique name implicitly.

For simplicity, both simplification and propagation rules are often
treated as special cases of simpagation rules.
The following notation is used:
\[
    H_k \bs H_r \simp G \guard B
\]
If $H_k$ is empty, then
the rule is a simplification rule. If $H_r$ is empty, then the rule is
a propagation rule. At least one of $H_r$ and $H_k$ must be non-empty.

\subsubsection{Informal Semantics}
A derivation starts from an initial query: a multiset of constraint atoms, given by the user.
This multiset of constraints is called the {\em constraint store}.
The derivation proceeds by applying the rules of the program, which
modify the constraint store. 
When no more rules can be applied, the derivation ends; the final constraint store 
is called the solution or solved form.

Rules modify the constraint store in the following way.
A simplification rule can be considered as a rewrite rule
which replaces the left-hand side (the head constraints) with the right-hand side (the body constraints), 
on the condition that the guard holds. 
The double arrow indicates that the head is logically equivalent to the body, 
which justifies the replacement. 
The intention is that the body is a simpler, or more canonical form of the head.

In propagation rules, the body is a consequence of the head: 
given the head, the body may be added (if the guard holds). 
Logically, the body is implied by the head so it is redundant.
However, adding redundant constraints may allow simplifications later on.
Simpagation rules are a hybrid between simplification rules and propagation rules:
the constraints before the backslash are kept, while the constraints after 
the backslash are removed.

\subsubsection{Examples}
\begin{figure}[t]
\centering
\fbox{
\begin{minipage}{0.9\textwidth}
\begin{tabbing}
{\it reflexivity} \qquad \= @ {\bf leq}(X,X)  \simp \true.\\
{\it antisymmetry}      \> @ {\bf leq}(X,Y),\ \quad \={\bf leq}(Y,X) \= \simp X = Y.\\
{\it idempotence}       \> @ {\bf leq}(X,Y)  \bs  \>{\bf leq}(X,Y) \> \simp \true.\\
{\it transitivity}      \> @ {\bf leq}(X,Y),      \>{\bf leq}(Y,Z) \> \prop  {\bf leq}(X,Z).
\end{tabbing}
\end{minipage}
}
\caption{The CHR(Prolog) program {\sc leq}, a solver for the less-than-or-equal constraint.}
\label{fig:leq}
\end{figure}
The program {\sc leq} (Fig.~\ref{fig:leq}) 
is a classic example CHR program to solve less-than-or-equal constraints.
The first rule, {\it reflexivity}, replaces the trivial constraint {\bf leq}(X,X) by \true.
Operationally, this entails removing this constraint from the constraint store
(the multiset of all known CHR constraints).
The second rule, {\it antisymmetry}, 
states that {\bf leq}(X,Y) and {\bf leq}(Y,X) are logically equivalent to X = Y.
Operationally this means that constraints matching the left-hand side may be removed from the store, 
after which the Prolog built-in equality constraint solver is used to unify X and Y. 
The third rule, {\it idempotence}, 
removes redundant copies of the same {\bf leq}/2 constraint.
It is necessary to do this explicitly since CHR has multiset semantics.
The last rule, {\it transitivity}, 
is a propagation rule that computes the transitive closure of the {\bf leq}/2 relation. 
An example derivation could be as follows:
\begin{tabbing}
\qquad \qquad \qquad \qquad \=	\qquad \= {\bf leq}(A,B), \underline{{\bf leq}(B,C), {\bf leq}(C,A)}	\\
{\it (transitivity)} \> \tr    \>	\underline{{\bf leq}(A,B)}, {\bf leq}(B,C), {\bf leq}(C,A), \underline{{\bf leq}(B,A)}		     \\
{\it (antisymmetry)}   \> \tr    \>	{\bf leq}(B,C), {\bf leq}(C,A), A = B		     \\
{\it (Prolog)} \> \tr    \>	\underline{{\bf leq}(A,C), {\bf leq}(C,A)}, A = B    \\
{\it (antisymmetry)} \> \tr    \>	A = C, A = B		     \\
\end{tabbing}

\begin{figure}[t]
\centering
\fbox{
\begin{minipage}{0.9\textwidth}
\begin{tabbing}
{\it generate} \qquad \qquad \= @ {\bf upto}(N) \= \simp N \Verb|>| 1 \guard {\bf prime}(N), {\bf upto}(N-1).\\
{\it done}       \> @ {\bf upto}(1) \> \simp \true.\\
{\it remove\_nonprime}     \> @ {\bf prime}(A) \bs {\bf prime}(B) \simp B \Verb|mod| A \Verb|=| 0 \guard \true.
\end{tabbing}
\end{minipage}
}
\caption{The CHR program {\sc primes}, a prime number sieve.}
\label{fig:primes}
\end{figure}
Figure~\ref{fig:primes} lists another simple CHR(Prolog) program called {\sc primes},
a CHR variant of the Sieve of Eratosthenes.
Dating back to at least 1992 \cite{fru_constraint_simplification_rules_techrep92}, this is one of the very first 
examples where CHR is used as a general-purpose programming language.
Given a query of the form ``{\bf upto}($n$)'', where $n$ is a positive integer,
it computes all prime numbers up to $n$.
The first rule ({\it generate}) does the following:
if $n>1$, it `simplifies' {\bf upto}($n$) to {\bf upto}($n-1$) and adds a {\bf prime}($n$) constraint.
The second rule handles the case for $n=1$, removing the {\bf upto}(1) constraint.
Note that removing a constraint is done by simplifying it to the built-in constraint \true.
The third and most interesting rule ({\it remove\_nonprime}) is a simpagation rule.
If there are two {\bf prime}/1 constraints {\bf prime}(A) and {\bf prime}(B),
such that B is a multiple of A, the latter constraint is removed.
The effect of the {\it remove\_nonprime} rule is to remove non-primes.
As a result, if the rules are applied exhaustively, the remaining constraints
correspond exactly to the prime numbers up to $n$.

%% file: 02-semantics.tex
\section{Semantics}
\label{sec:semantics}

In this section, we give an overview of both the logical (declarative) semantics
and the operational semantics of CHR. 
The logical semantics (Section~\ref{sec:semantics:logical})
constitute the formal foundations for the CHR programming language,
whilst the operational semantics (Section~\ref{sec:semantics:operational}) 
determine the behavior of actual implementations.

\subsection{Logical Semantics}
\label{sec:semantics:logical}

\subsubsection{Classical Logic Semantics}
\label{sec:semantics:logical:classic}
Let $\bar x$ denote the variables occurring only in the body of the rule.
We use $\bar \forall (F)$ to denote universal quantification over all free
variables in $F$.
A simplification rule $H \simp G \guard B$ corresponds to a logical equivalence,
under the condition that the guard is satisfied:
$ \bar \forall (G \rightarrow (H \leftrightarrow \exists \bar x B))$.
A propagation rule $H \prop G \guard B$ \linebreak[1]
corresponds to a logical implication
if the guard is satisfied:
$ \bar \forall (G \rightarrow (H \rightarrow \exists \bar x B))$.
A simpagation rule $H_k \bs H_r \simp G \guard B$ corresponds to a conditional
equivalence:
$ \bar \forall (G \rightarrow (H_k \rightarrow (H_r \leftrightarrow \exists \bar x B)))$.
The (classical) logical semantics \cite{fru_chr_overview_jlp98} 
of a CHR program --- also called its logical reading, 
declarative semantics, or declarative interpretation --- is given by 
the built-in constraint theory $\CT_\HL$
(which defines the built-ins of the host language $\HL$) 
in conjunction with the logical formulas for each rule.
As an example, consider the program {\sc leq} of Fig.~\ref{fig:leq}.
The logical formulas corresponding to its rules are the following:
$$
\left\{
\begin{array}{lr}
\forall x,y : x = y \rightarrow (\mbox{\bf leq}(x,y) \leftrightarrow \true)
& \mbox{\it (reflexivity)}\\
\forall x,y,x',y' : x = x' \land y = y' \rightarrow 
(\mbox{\bf leq}(x,y) \land \mbox{\bf leq}(y',x') \leftrightarrow x = y)
& \mbox{\it (antisymmetry)}\\
\forall x,y,x',y' : x = x' \land y = y' \rightarrow 
(\mbox{\bf leq}(x,y) \rightarrow (\mbox{\bf leq}(x',y') \leftrightarrow \true))
& \mbox{\it (idempotence)}\\
\forall x,y,y',z : y = y' \rightarrow 
(\mbox{\bf leq}(x,y) \land \mbox{\bf leq}(y',z) \rightarrow \mbox{\bf leq}(x,z))
& \mbox{\it (transitivity)}
\end{array}
\right.
$$

\noindent{}or equivalently:
$$
\left\{
\begin{array}{lr}
\forall x : \mbox{\bf leq}(x,x) 
& \mbox{\it (reflexivity)}\\
\forall x,y :
\mbox{\bf leq}(x,y) \land \mbox{\bf leq}(y,x) \leftrightarrow x = y
& \mbox{\it (antisymmetry)}\\
\true
& \mbox{\it (idempotence)}\\
\forall x,y,z : 
\mbox{\bf leq}(x,y) \land \mbox{\bf leq}(y,z) \rightarrow \mbox{\bf leq}(x,z)
& \mbox{\it (transitivity)}
\end{array}
\right.
$$

Note the strong correspondence between the syntax of the CHR rules, 
their logical reading, and the natural definition of partial order.

The classical logical reading, however, does not reflect 
CHR's multiset semantics (the \textit{idempotence} rule is logically equivalent to $true$).
%
%
Also, the classical logic reading does not always make sense.
For example, consider the classical logic reading of the {\sc primes} program of Fig.~\ref{fig:primes}:
$$
\left\{
\begin{array}{l}
\forall n : n > 1 \rightarrow \mbox{\bf upto}(n) \leftrightarrow 
        \exists n' \mbox{\bf prime}(n) \land n' = n-1 \land \mbox{\bf upto}(n')
\ \ \ \ \ \hfill  \mbox{\it (generate)}\\
\mbox{\bf upto}(1) \leftrightarrow \true
\hfill  \mbox{\it (done)}\\
\forall a,b : 
a|b \rightarrow \mbox{\bf prime}(a) \rightarrow  (\mbox{\bf prime}(b) \leftrightarrow \true)
\hfill  \mbox{\it (remove\_nonprime)}
\end{array}
\right.
$$
which is equivalent to:
$$
\left\{
\begin{array}{lr}
\forall n > 1 : \mbox{\bf upto}(n) \leftrightarrow \mbox{\bf prime}(n) \land \mbox{\bf upto}(n-1)
& \mbox{\it (generate)}\\
\mbox{\bf upto}(1)
& \mbox{\it (done)}\\
\forall a,b : 
\mbox{\bf prime}(a) \land a|b  \rightarrow \mbox{\bf prime}(b)
& \mbox{\it (remove\_nonprime)}
\end{array}
\right.
$$

\noindent The last formula nonsensically states that a number is prime if it has a prime factor.

\subsubsection{Linear Logic Semantics}
\label{sec:semantics:logical:linear}

\comment{
While the original declarative semantics is meaningful for a large class
of constraint solvers like {\sc leq}, it does not give a satisfactory meaning
when applied to general-purpose programs like {\sc primes}. 
In particular, when rules represent actions or updates, the classical logic 
reading is usually inconsistent with the intended meaning. 
\citeANP{betz_fru_linear_logic_semantics_cp05}
\citeNN{betz_fru_linear_logic_semantics_cp05,betz_fru_linear_logic_chr_disj_chr07}
proposed an alternative declarative semantics based on intuitionistic linear logic.
This linear logic semantics is much more useful than the classical semantics
for CHR programs in which the constraints represent a multi-set of resources
or in which the rules represent one-directional operations.
}

For general-purpose CHR programs such as {\sc primes},
or programs that rely on CHR's multiset semantics,
the classical logic reading is often inconsistent with the intended meaning
(see previous section).
To overcome these limitations,
\citeN{bouissou_chr_for_silcc_2004} and Betz and Fr\"uhwirth~\shortcite{betz_fru_linear_logic_semantics_cp05,betz_fru_linear_logic_chr_disj_chr07}
independently proposed an alternative declarative semantics based on (intuitionistic) linear logic.
The latter, most comprehensive study provides strong soundness and completeness results,
as well as a semantics for the \CHRv{} extension of CHR (see Section~\ref{sec:extensions:disj}).
For CHR programs whose constraints represent a multiset of resources,
or whose rules represent unidirectional actions or updates,
a linear logic semantics proves much more appropriate.
A simple example is the following coin-throwing simulator
(which depends on the nondeterminism in the operational semantics):

\begin{center}
{\bf throw}(Coin) \simp Coin = {\tt head}.\\
{\bf throw}(Coin) \simp Coin = {\tt tail}.
\end{center}

The classical logic reading of this program entails {\tt head} = {\tt tail}.
The linear logic reading of the coin-throwing program 
boils down to the following formula:
$$
!(\mbox{\bf throw}(\mathit{Coin}) \multimap (\mathit{Coin} = \mathtt{head}) \& (\mathit{Coin} = \mathtt{tail}))
$$
In natural language, this formula means
``you can always replace $\mbox{\bf throw}(\mathit{Coin})$ 
with either $(\mathit{Coin} = \mathtt{head})$ or $(\mathit{Coin} = \mathtt{tail})$,
but not both''. This corresponds to the committed-choice and unidirectional rule application of CHR.

\subsubsection{Transaction Logic Semantics}
\label{sec:semantics:logical:transaction}

The linear logic semantics is already 
closer to the operational semantics than the CHR classical logical semantics. 
However, it still does not allow precise reasoning about CHR derivations: 
while derivations correspond to proofs of logic equivalence of the initial and the final state, 
it only allows reasoning on the result of an execution, not on the execution itself.
The transaction logic semantics \cite{meister_djelloul_robin_transaction_logic_semantics_lpnmr07}
bridges the remaining gap between the logical and operational semantics of CHR
by providing a framework for both inside one formal system.

\subsection{Operational Semantics}
\label{sec:semantics:operational}

The behavior of CHR implementations is determined by their operational semantics.
As the original theoretical semantics of CHR (Section~\ref{sec:semantics:operational:omegat}) 
proved too nondeterministic for practical programming,
more deterministic instances have been specified that offer more execution control
(Sections~\ref{sec:semantics:operational:omegar} and \ref{sec:semantics:operational:omegap}).

\subsubsection{Theoretical Operational Semantics $\omegat$}
\label{sec:semantics:operational:omegat}

The operational semantics $\omegat$ of CHR \cite{fru_chr_overview_jlp98}, 
sometimes also called {\em theoretical} or {\em high-level} operational semantics,
is highly nondeterministic.
It is formulated as a state transition system. 

\pagebreak      
\begin{definition}
An \emph{identified} CHR constraint $\num{c}{i}$ is a CHR constraint 
$c$ associated with some unique integer $i$, the {\em constraint identifier}.
This number serves to differentiate between copies of the same constraint.
We introduce the functions $\chr(\num{c}{i}) = c$ and $\id(\num{c}{i}) = i$,
and extend them to sequences and sets of identified CHR constraints in
the obvious manner, e.g., $\id(S) = \{ i |\num{c}{i} \in S\}$.
\end{definition}

\begin{definition}
An \emph{execution state} $\sigma$ is a tuple
$ \langle \IG, \IS, \IB, \IT \rangle_n$.
The \emph{goal} $\IG$ is a multiset of constraints to be rewritten to solved form.
The CHR constraint \emph{store} $\IS$ is a set of \emph{identified} 
CHR constraints that can be matched with rules in the program $\Prog$.
Note that $\chr(\IS)$ is a multiset although $\IS$ is a set.
The \emph{built-in constraint store} $\IB$ is the conjunction of all built-in 
constraints that have been posted to the underlying solver.
These constraints are assumed to be solved (implicitly) by the host language $\HL$.
The \emph{propagation history} $\IT$ is a set of tuples, 
each recording the identities of the CHR constraints that fired a rule,
and the name of the rule itself.
The propagation history is used to prevent trivial non-termination for propagation rules:
a propagation rule is allowed to fire on a set of constraints only if
the constraints have not been used to fire the same rule before.%
\footnote{%
	Early work on CHR, as well as some more recent publications
	(e.g.,\ \citeNP{bouissou_chr_for_silcc_2004,duck_stuck_sulz_observable_confluence_iclp07,haemm_fages_abstract_critical_pairs_rta07}),
	use a \emph{token store} instead of a propagation history
	(this explains the convention of denoting the propagation history with $\IT$).
	A token store contains a token for every potential (future) propagation rule application,
	which is removed when the rule is actually applied.
	The propagation history formulation is dual, but closer to most implementations.
	Confusingly, the term \emph{token store} has also been used 
        for what is commonly referred to as the ``propagation history''
	(e.g.,\ \citeNP{chin_sulzmann_wang_haskell_chr_03,tacchella_gabbrielli_meo_unfolding_ppdp07}).
}
Finally, the counter $n \in \N$ represents the next integer that can be 
used to number a CHR constraint.
We use $\sigma, \sigma_0, \sigma_1, \ldots$ to denote execution states and
$\States$ to denote the set of all execution states.
\end{definition}

For a given CHR program $\Prog$, the transitions are defined by
the binary relation 
$\tr_\Prog \subset \States \times \States$
shown in Figure~\ref{fig:omegat}.
Execution proceeds by exhaustively applying the transition rules, starting
from an initial state.

The \textbf{Solve} transition solves a built-in constraint from the goal, the
\textbf{Introduce} transition inserts a new CHR constraint from the goal into
the CHR constraint store, and the \textbf{Apply} transition fires a rule 
instance. A rule instance instantiates a rule with CHR constraints
matching the heads, using a matching substitution
(a one-way variable substitution). 


Relatively strong soundness and completeness results \cite{fru_chr_overview_jlp98}
link the logical semantics and the operational semantics.
\citeN{propagation_completeness_maher_iclp02} discusses the notion
of {\em propagation completeness}
and proves an impossibility result for CHR.

\begin{figure}
\fbox{
\begin{minipage}{0.97\textwidth}
\begin{description} 
\item[\textbf{1. Solve.}]
$
        \st{\{c\} \uplus \IG}{\IS}{\IB}{\IT}{n}
        \tr_\Prog
        \st{\IG}{\IS}{c \land \IB}{\IT}{n}
$\\
where $c$ is a built-in constraint and
$\musthold \exist_\emptyset \IB$.


\item[\textbf{2. Introduce.}]
$
        \st{\{c\} \uplus \IG}{\IS}{\IB}{\IT}{n}
        \tr_\Prog
        \st{\IG}{\{\num{c}{n}\} \cup \IS}{\IB}{\IT}{n+1}
$\\
where $c$ is a CHR constraint and
$\musthold \exist_\emptyset \IB$.

\item[\textbf{3. Apply.}]
$
        \st{\IG}{H_1 \uplus H_2 \uplus \IS}{\IB}{\IT}{n}
        \tr_\Prog
        \st{B \uplus \IG}{H_1 \uplus \IS}{\theta \land \IB}{\IT \cup \{h\}}{n}
$\\
where $\Prog$ contains a (renamed apart) rule of the form 
$ r\ @\ H'_1\ \backslash\ H'_2 \iff G\ |\ B$,\\
$\theta$ is a matching substitution such that
$chr(H_1) = \theta(H'_1)$ and $chr(H_2) = \theta(H'_2)$,\\
$ h = (r,id(H_1),id(H_2)) \not\in \IT $,
and $\musthold ( \exist_\emptyset \IB ) \land (\IB \rightarrow \exist_\IB (\theta \wedge G))$.
\end{description}
\end{minipage}
}
\caption{
	The transition rules of the theoretical operational semantics $\omegat$, defining $\tr_\Prog$.
	We use $\uplus$ for multiset union.
	For constraint conjunctions $B_1$ and $B_2$, 
	$\exist_{B_2}(B_1)$ denotes $\exists X_1, \ldots, X_n: B_1$, with $\{X_1,\ldots,X_n\} = vars(B_1) \backslash vars(B_2)$.
}
\label{fig:omegat}
\end{figure}

Variants of $\omegat$ have been introduced to formalize extensions and
variants of CHR.  For example, 
the operational semantics of \CHRv \cite{abd_chr_disjunction_rcorp00},
probabilistic CHR \cite{fru_dipierro_wiklicky_probabilistic_chr_wflp02}, 
and CHR with aggregates \cite{sney_vanweert_demoen_aggregates_chr07}
are all based on $\omegat$.
We discuss these and other extensions in Section~\ref{sec:extensions}.

We should also mention the work on an and-compositional semantics for CHR
\cite{delz_gab_meo_comp_sem_chr_ppdp05,gabbr_meo_compos_semantics_tocl08}, 
which allows one to retrieve the semantics of a conjunctive query given the semantics of the conjuncts.
This property is a first step towards incremental and modular analysis and verification tools.

\subsubsection{Refined Operational Semantics $\omegar$}
\label{sec:semantics:operational:omegar}

The refined operational semantics $\omegar$ \cite{duck_stuck_garc_holz_refined_op_sem_iclp04}
instantiates the $\omegat$ operational semantics by removing much of the nondeterminism.
It formally captures the behavior of many CHR implementations (see also Section~\ref{sec:implementation}).
CHR programs often rely on the execution control offered by the $\omegar$ semantics 
for correctness, or to achieve a good time complexity.

The refined operational semantics uses a stack of constraints: 
when a new constraint arrives in the constraint store it is pushed on the stack. 
The constraint on top of the stack is called the \emph{active} constraint.
The active constraint attempts to match rule heads,
together with suitable constraints from the constraint store (\emph{partner} constraints),
All occurrences of the active constraint are tried
in the order in which they occur in the program. 
When all occurrences have been tried, the constraint is popped from the stack. 
When a rule fires, its body is executed immediately from left to right, 
thereby potentially suspending the active constraint because of newly arriving constraints. 
When a constraint becomes topmost again, it resumes its search for matching clauses.

Alternative formalizations of the $\omegar$ semantics have been made for easier
reasoning about certain optimizations or analyses.
Examples are the call-based refined operational semantics 
$\omegac$ \cite{schr_stuck_duck_ai_chr_ppdp05} and
the 
semantics for occurrence representations
$\omegao$ \cite{sney_schr_demoen_guard_and_continuation_opt_iclp05}.
Variants of $\omegar$ have also been introduced to
formalize implementations of proposed extensions to CHR or variants of CHR.
To mention just a few of them:
the 
$\omegarnot$ semantics for \CHRnot \cite{vanweert_sney_schr_demoen_negation_chr06},
the 
$\omegarv$ semantics for \CHRv\
which is equivalent to the tree-based 
$\omegatree$ semantics \cite{dekoninck_schr_demoen_search_chr06},
the set-based 
$\omegaset$ semantics of CHRd \cite{sarnastarosta_ramakrishnan_chrd_padl07},
and the concurrent refined semantics (\citeNP{lam_sulz_concurrent_chr_damp07}).
These extensions and variants are discussed in more detail in Section~\ref{sec:extensions}.

\subsubsection{Priority Semantics $\omegap$} \index{chrrp@{\chrrp}}
\label{sec:semantics:operational:omegap}

While the refined operational semantics reduces most of the nondeterminism
of the $\omegat$ semantics, it arguably does not offer the CHR programmer
an intuitive and predictable way to influence control flow.  
The $\omegar$ semantics in a sense forces the programmer to understand 
and take into account how CHR implementations work, 
to achieve the desired execution control. 
\citeN{dekoninck_schr_demoen_chrrp_ppdp07} introduced the extension \chrrp{},
with a corresponding operational semantics called $\omegap$.
The programmer assigns a priority to every rule. The $\omegap$
semantics is an instantiation of $\omegat$ which ensures that of all
applicable rules, the one with the highest priority is applied first.
This feature gives the programmer a much more precise and high-level
control over program execution compared to the $\omegar$ semantics.

%% file: 03-analysis.tex
\section{Program Analysis}
\label{sec:analysis}

In this section we discuss important properties of CHR programs:
confluence (Section~\ref{sec:analysis:confluence}),
termination (Section~\ref{sec:analysis:termination}),
and complexity (Section~\ref{sec:analysis:complexity}),
as well as (semi-) automatic analysis of these properties.
Program analyses that are mostly used for 
optimizing compilation are discussed in Section~\ref{sec:implementation:optimization}.

\subsection{Confluence}
\label{sec:analysis:confluence}

If for a given CHR program, for all initial states,
any $\omegat$ derivation from that state results in the same final state,
the program is called \emph{confluent}. 
Confluence has been investigated thoroughly in the context of CHR 
\cite{abd_fru_meuss_confluence_semantics_csr_constr99}.
Two important results are discussed already in \cite{fru_chr_overview_jlp98}:
the existence of a decidable, sufficient and necessary test for confluence 
of terminating programs, and the result that confluence implies correctness
(consistency of the logical reading).
Confluence under the refined $\omegar$ semantics
is investigated in Chapter 6 of \cite{duck_phdthesis05}, 
which also discusses a refined confluence test.

Recently, the topic of confluence received renewed attention
because certain problems and limitations of the 
confluence test have surfaced.
Firstly, many programs that are in practice confluent fail this
confluence test because 
non-confluence originates from
unreachable states.
The more powerful 
notion of \emph{observable confluence} 
\cite{duck_stuck_sulz_observable_confluence_iclp07}
takes reachability into account.
Secondly, the 
standard notion of confluence
is only applicable to terminating programs.
\citeN{raiser_tacchella_confluence_non_terminating_chr07} 
extended the
notion of confluence 
to 
non-terminating programs.

\citeN{haemm_fages_abstract_critical_pairs_rta07} develop a notion of abstract
critical pairs for rewriting systems in general. They illustrate this notion
for CHR's theoretical semantics. A particularly interesting result is that some
traditional critical pairs can be disregarded because they are redundant.

\paragraph{Related Analyses.}
\citeN{abd_fru_completion_cp98} showed how to do \emph{completion} 
of CHR programs. Completion is a technique to
transform a non-confluent program into a confluent one by adding rules.
It allows extension, modification and specialization of existing programs.

A very useful notion is that of \emph{operational equivalence} of two CHR programs.
Two programs are operationally equivalent if for each query, 
the answer is the same (modulo variable renaming) according to each program.
A straightforward extension of confluence, called \emph{compatibility} of two programs, 
is shown to be too weak to capture the operational equivalence of CHR programs 
\cite{abd_fru_equivalence_cp99,abd_habilitation_2001}.
Instead, \citeN{abd_fru_equivalence_cp99} give
a decidable, sufficient, and necessary syntactic condition for operational equivalence
of well-behaved 
(confluent and terminating)
CHR \emph{programs}.
A sufficient syntactic condition is also given 
for the equivalence of two CHR \emph{constraints},
defined in two different well-behaved CHR programs.
The latter condition is also shown necessary for an interesting class of CHR programs.

\citeN{abd_fru_integration_lopstr03} investigated
the merging of two well-behaved CHR solvers.
If the two programs are not compatible,
well-behavedness can be regained by completion.
Furthermore, \citeN{abd_fru_integration_lopstr03} 
identify a class of solvers whose union is always confluent,
and argue why finding a class whose union preserves termination is hard.  
Finally, \citeN{abd_fru_integration_lopstr03} present a method to remove 
redundant rules from CHR programs, based on the notion of operational equivalence.

\subsection{Termination}
\label{sec:analysis:termination}
The first work 
on termination analysis of CHR programs
was presented by
\citeN{fru_termination_compulog00}. Fr\"uhwirth demonstrated that 
termination proof techniques from logic programming and term rewrite systems 
can be adapted to the CHR context. 
%
%
%
  Termination of CHR programs is proved 
  by defining a ranking function from computation states to a well-founded domain 
  such that the rank of consecutive computation states decreases.
  A condition on simplification rules guarantees such rank decreases for all consecutive states. 
  This approach, however, cannot prove termination of CHR programs with propagation rules,
  because it is impossible to show
  decreases between 
  consecutive states 
  as these rules do not remove constraints from the store.


Recently, two new results on termination analysis of CHR were presented. 
\citeN{pilozzi_schr_deschreye_termination_wst07} describe
a termination preserving transformation of CHR programs to Prolog programs. 
By reusing termination tools from logic programming and, indirectly, from term rewriting, 
proofs of termination of the transformed CHR programs are generated automatically, 
yielding the first fully automatic termination prover for CHR. 
The transformation, however, does not consider propagation histories. 
As such, it is applicable only to CHR programs without propagation rules. 
A transformation of single-headed propagation rules to equivalent simplification rules 
overcomes this problem partially.

The second contribution is presented by \citeN{voets_pilozzi_deschreye_termination_chr07}.
Compare to previous approaches, theirsapproach is applicable to a much larger class of CHR programs. 
By formulating a new termination condition that 
verifies conditions imposed on the dynamic process of 
adding constraints to the store, they derive conditions for both 
simplification and propagation rules.

\subsection{Complexity}
\label{sec:analysis:complexity}

For various CHR programs --- general purpose programs as well as
constraint solvers --- an accurate, though rather ad hoc, complexity 
analysis has been made. 
We list the most notable examples in Section~\ref{sec:analysis:complexity:adhoc}.
While ad hoc methods give the most accurate results in practice, 
they cannot easily be generalized.
Therefore, more structured approaches to complexity analysis 
have been proposed by means of meta-complexity theorems.
An overview is given in Section~\ref{sec:analysis:complexity:meta}.

\subsubsection{Ad Hoc Analysis} \label{sec:analysis:complexity:adhoc}
A CHR implementation of the classical union-find algorithm was proven optimal by
\citeN{schr_fru_opt_union_find_tplp06}. 
\citeN{sney_schr_demoen_dijkstra_chr_wlp06} showed the optimal complexity of an
implementation of Dijkstra's shortest path algorithm that uses Fibonacci heaps. 
\citeN{fru_lexicographic_chr05} formulated the complexity of a
general-purpose lexicographical order constraint solver 
in terms of the number of ask and tell built-in constraints encountered during
execution. Finally,
\citeN{meister_djelloul_fru_compl_tree_equations_csclp06} derived the complexity of a
solver for existentially quantified 
equations over finite and infinite trees,
using bounds on the derivation length.

\subsubsection{Meta-Complexity Results} \label{sec:analysis:complexity:meta}
Fr\"uhwirth \citeNN{fru_number_entcs01,fru_complexity_kr02,fru_complexity2_entcs02}
investigated the time complexity of simplification rules for naive implementations of CHR.
In this approach, a suitable termination order (called a \emph{tight ranking})
is used as an upper bound on the derivation length.
Combined with a worst-case estimate of the number and cost of rule application attempts,
this results in a complexity meta-theorem which gives a rough upper bound
of the time complexity. 
Recent work on optimizing compilation of CHR (cf.\ Section~\ref{sec:implementation:optimization})
allows meta-theorems that give much tighter complexity bounds. 
We now discuss two distinct approaches.

\citeN{ganzinger_mcallester_la_iclp02} propose a formalism called {\em Logical Algorithms} (LA) 
and prove a meta-complexity result. 
\citeN{dekoninck_schr_demoen_la-chr_iclp07} establish a close correspondence between CHR and LA 
(see also Section~\ref{ssec:LA}),
\comment{ 
They present a translation 
from a large class of \chrrp{} programs into equivalent LA programs. 
In this way,
the meta-complexity result of LA can be applied to these \chrrp{} programs. 
}
allowing the LA meta-complexity result to be applied (indirectly) to a large class of CHR programs.
%
\citeN{dekoninck_schr_demoen_la-chr_iclp07} actually address the meta-complexity 
of \chrrp\ programs, an extension of CHR discussed in Section~\ref{sec:extensions:chrrp}.
All CHR programs are also \chrrp\ programs.
\comment{
The approach 
is often more accurate than Fr\"uhwirth's result, 
as it takes into account the properties of current highly optimized implementations.}
The Logical Algorithms approach was previously used, in a more ad hoc way, by 
\citeN{christ_chr_grammars_tplp05} to derive the complexity of CHR grammars
(see Section~\ref{sec:applications:parsing_and_nlp}). 
\comment{He noted that the CHR implementation he used 
--- SICStus CHR by \citeN{holz_fru_CHR_manual_techrep98} --- does not
satisfy the complexity requirements needed for the meta-complexity theorem to hold.
This problem is remedied using the LA implementation in CHR
proposed by \citeN{dekoninck_schr_demoen_la-chr_iclp07},
which executes LA programs with the required complexity.}

\citeN{sney_schr_demoen_chr_complexity_08} 
explicitly decouple the two steps in the approach of Fr\"uhwirth \citeNN{fru_complexity_kr02,fru_complexity2_entcs02}
by introducing 
abstract CHR machines.
In the first step, the number of rule applications is estimated; this corresponds
to the number
of CHR machine steps.
If a suitable termination order can be found, it can be used to show an upper bound.
However for programs that are non-terminating in general, like a RAM machine simulator,
or for which no suitable ranking can be found, other techniques have to be used 
to prove complexity properties. 
In the second step, the complexity of rule application is computed
for a given CHR program; this corresponds to simulating a CHR machine on a RAM machine.
The first step depends only on the operational semantics of CHR,
whereas the second step depends strongly on the performance of the code
generated by the CHR compiler.
\comment{
Recent work on optimizing compilation of CHR (cf.\ Section~\ref{sec:implementation:optimization})
allows much tighter bounds compared to those of 
Fr\"uhwirth \citeNN{fru_complexity_kr02,fru_complexity2_entcs02}.
}

\subsubsection{Complexity-wise Completeness of CHR} 


\citeN{sney_schr_demoen_chr_complexity_08} also consider
the space complexity of CHR programs.
Some compiler optimizations like memory reuse \cite{sney_schr_demoen_memory_reuse_iclp06}
are crucial to achieve tight space complexity bounds
(cf. Section~\ref{sec:implementation:optimization}).
The most interesting result of \citeN{sney_schr_demoen_chr_complexity_08}
is 
the following ``complexity-wise completeness'' result for CHR,
which implies that ``everything can be done efficiently in CHR'':
%
For every algorithm (RAM machine program) 
which uses at least as much time as space,
a CHR program exists which can be executed in the K.U.Leuven CHR system
with time and space complexity within a constant from the original complexities.
Complexity-wise completeness implies Turing completeness but is a much stronger property.

%% file: 04-implementation.tex
\section{Systems and Implementation}\label{sec:implementation}

CHR is first and foremost a programming language. Hence, a large part of CHR
research has been devoted to the development of CHR systems and efficient
execution of CHR programs. The two most comprehensive works on this
subject are the Ph.D. theses of \citeN{duck_phdthesis05} and \citeN{schr_phdthesis05}. 
In this section, we provide an overview of their work
as well as the many other contributions to the field. 

\subsection{Systems}
\label{sec:implementation:systems}

\begin{figure}[t]
        \centering
	\includegraphics[width=0.95\textwidth]{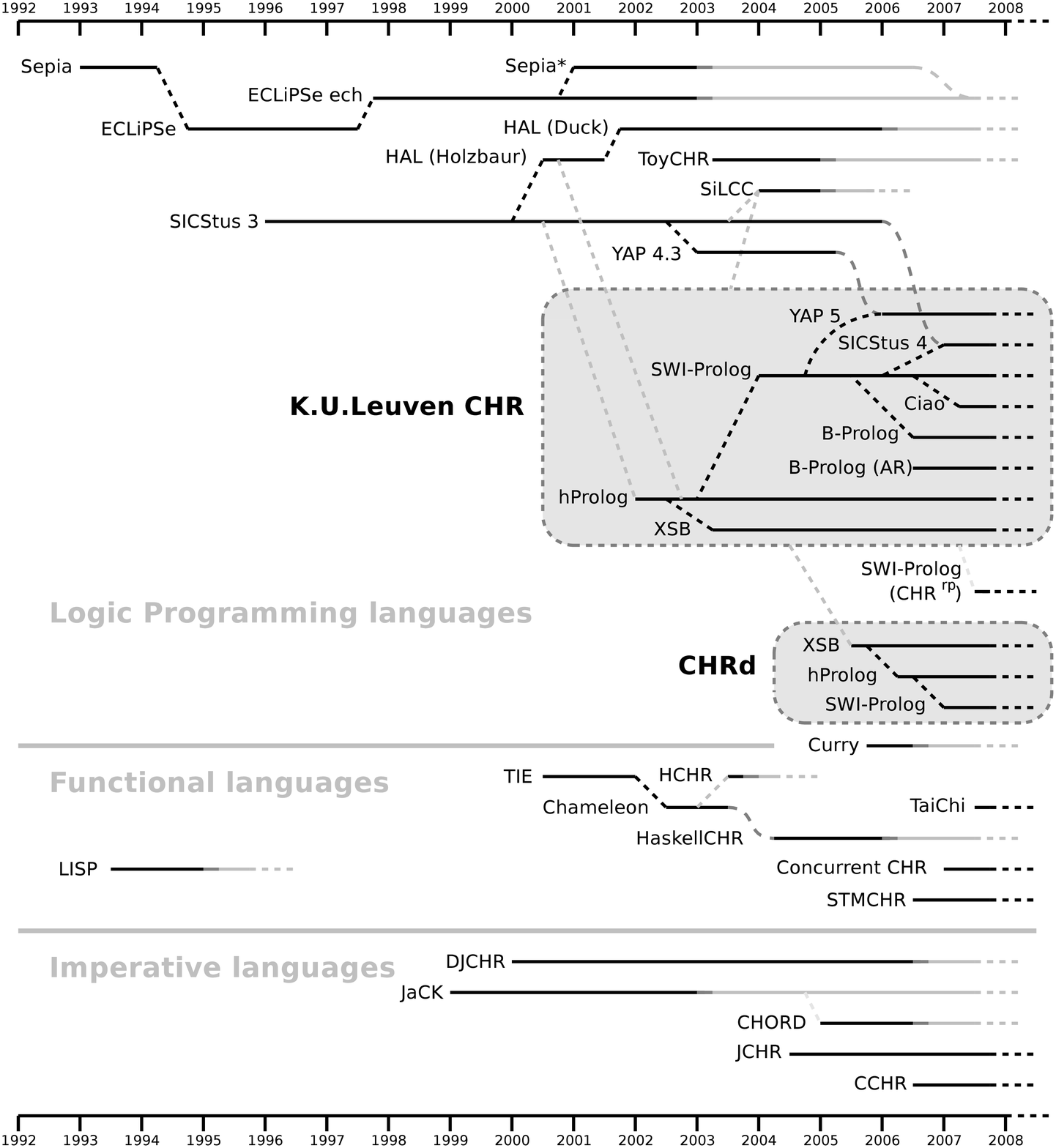}
        \caption{A timeline of CHR implementations.}
        \label{fig:timeline}
\end{figure}

Since the conception of CHR a large number of CHR systems
(compilers, interpreters and ports) have been developed. In particular, in the
last ten years the number of systems has exploded. Figure \ref{fig:timeline}
presents a timeline of system development, branches and influences. We discuss
these systems, grouped by host language or host paradigm, in more detail.

\subsubsection{CHR(LP)}
\label{sec:implementation:systems:lp}

Logic Programming is the natural host language paradigm for CHR. 
Hence, it is not surprising that the CHR(Prolog) implementations are the most established ones. 
\citeN{holz_fru_prolog_chr_compiler_aai00} have laid the groundwork with their general
compilation scheme for Prolog. 
This compilation scheme was first implemented in SICStus Prolog by Holzbaur, 
and later further refined in HAL by \citeN{holz_garc_stuck_duck_opt_comp_chr_hal_tplp05} 
and in hProlog by \citeN{schr_demoen_kulchr_chr04}. 
The latter system, called the K.U.Leuven CHR system, was subsequently
ported to many other Prolog systems and is currently available in 
XSB \cite{schr_warren_demoen_chr_xsb_ciclops03,schr_warren_chr_xsb_iclp04},
SWI-Prolog \cite{schr_wielemaker_demoen_chr_swi_wclp05}, 
YAP, B-Prolog (using Action Rules; \citeNP{schr_zhou_demoen_action_rules_chr06}), 
SICStus 4 and Ciao Prolog.
Another system directly based on the work of Holzbaur and Schrijvers
is the CHR library for SiLCC by \citeN{bouissou_chr_for_silcc_2004}.
	SiLCC is a programming language based on 
	linear logic and concurrent constraint programming.
All of these systems compile CHR 
programs to host language programs.
The only available {\em interpreter} for CHR(Prolog) is TOYCHR\footnote{by Gregory J. Duck, 2003.
Download: {\href{http://www.cs.mu.oz.au/~gjd/toychr/}{http://www.cs.mu.oz.au/$\sim$gjd/toychr/}}}.

Recently, systems with deviating operational semantics have been developed.
The CHRd system by \citeN{sarnastarosta_ramakrishnan_chrd_padl07} 
runs in XSB, SWI-Prolog and hProlog. It features a constraint
store with set semantics and is particularly suitable for tabled execution.
The \chrrp\ system by \citeN{dekoninck_stuck_duck_compiling-chrrp_08}
for SWI-Prolog provides rule priorities. 
\index{chrrp@{\chrrp}}

\subsubsection{CHR(FP)} \label{sec:implementation:systems:fp}

As type checking is one of the most successful applications of CHR in the
context of Functional Programming (see Section \ref{sec:applications:type}),
several CHR implementations were developed specifically for this purpose. Most
notable is the Chameleon system
\cite{stuck_sulz_theory_of_overloading_toplas05} which features CHR as the
programming language for its extensible type system. Internally, Chameleon uses
the HaskellCHR implementation\footnote{by Gregory J. Duck, 2004. Download: \href{http://www.cs.mu.oz.au/~gjd/haskellchr/}{http://www.cs.mu.oz.au/$\sim$gjd/haskellchr/}}. The earlier HCHR prototype
\cite{chin_sulzmann_wang_haskell_chr_03} had a rather heavy-weight and impractical
approach to logical variables.

The aim of a 2007 Google Summer of Code project was to transfer this CHR based type
checking approach to two Haskell compilers (YHC and nhc98).  
The project 
led to a new CHR interpreter for Haskell, 
called TaiChi \cite{boespflug_taichi_monadreader07}.

With the advent of software transactional memories (STM) in Haskell, 
two prototype systems with parallel execution strategies have been developed: 
STMCHR\footnote{by Michael Stahl, 2007. Download:
\href{http://www.cs.kuleuven.be/~dtai/projects/CHR/}{http://www.cs.kuleuven.be/$\sim$dtai/projects/CHR/}}
and Concurrent CHR \cite{lam_sulz_concurrent_chr_damp07}.
These systems are currently the only known CHR implementations
that exploit the inherent parallelism in CHR programs.
Concurrent CHR also serves as the basis for Haskell-Join-Rules
\cite{sulz_lam_haskelljoinrules_ifl07} (cf.\ Section~\ref{ssec:joincalculus}).

We also mention the Haskell library for the PAKCS implementation 
of the functional logic language Curry \cite{hanus_chr_curry_wlp06}. 
The PAKCS system actually compiles Curry code to SICStus Prolog, 
and its CHR library is essentially a front-end for the SICStus Prolog CHR library. 
The notable added value of the Curry front-end is the (semi-)typing of the CHR code.

\subsubsection{CHR(Java) and CHR(C)} \label{sec:systems:java}
\label{sec:implementation:systems:java}

Finally, CHR systems are available for both Java and C.
These multiparadigmatic integrations of CHR and mainstream programming languages
offer powerful synergetic advantages to the software developer:
they facilitate the development of application-tailored constraint systems
that cooperate efficiently with existing host language components.
For a detailed discussion on the different conceptual and technical challenges 
encountered when embedding CHR into an imperative host language,
we refer to \citeN{vanweert_wuille_et_al_chr_imperative_lnai08}.

\paragraph{CHR(Java).}
There are at least four implementations of CHR in Java.
The earliest is the Java Constraint Kit (JaCK)
by \citeN{abd_habilitation_2001} and others \cite{abd_kr_saft_schm_jack_wflp01:entcs02}.
It consists of three major components:
\begin{enumerate}
\item JCHR \cite{schmauss_jack:jchr_1999}
--- a CHR dialect intended to resemble Java,
	in order to provide an intuitive programming experience.
	No operational semantics is specified for this system,  
	and its behavior deviates from other CHR implementations.
\item VisualCHR \cite{abd_saft_jack:visualchr_wlpe01}
--- an interactive tool visualizing the execution of JCHR 
	(cf.\ Section~\ref{sec:implementation:environments}).
\item JASE \cite{kr_jack:jase_2001}
--- a ``Java Abstract Search Engine'' in which tree-based search strategies can be specified.
	The JASE library is added to the JaCK framework as an orthogonal component.
	It provides a number of utility classes that aid the user to implement search algorithms in 
	the Java host language.
	A typical algorithm consists of the following two operations, executed in a loop:
	a JCHR handler is run until it reaches a fix-point,
	after which a new choice is made.
	If an inconsistency is found, 
	backtracking is used to return to the previous choice point.
	JASE aids in maintaining the search tree,
	and can be configured to use either trailing or copying.
\end{enumerate}

DJCHR (Dynamic JCHR; \citeNP{wolf_adaptive_chr_java_cp01}) 
is an implementation of adaptive CHR (see Section~\ref{sec:extensions:adaptive}).
The incremental adaptation algorithm underlying DJCHR maintains 
\emph{justifications} for rule applications and constraint additions. 
\citeN{wolf_intelligent_search_tplp05} shows that these justifications,
and in particular those of any derived \texttt{false} constraint,
also serve as a basis for intelligent search strategies.
As in JaCK, 
the different search algorithms are implemented orthogonally to the CHR program.
Wolf's approach 
confirms 
that advanced search strategies are often more efficient than a low-level, 
built-in implementation of chronological backtracking (as in Prolog).

The K.U.Leuven JCHR system \cite{vanweert_schr_demoen_jchr_chr05} addresses
the main issue of JaCK, its poor performance.
The focus of K.U.Leuven JCHR is on both performance and 
integration with the host language.
K.U.Leuven JCHR handlers integrate neatly with existing Java code,
and it is currently one of the most efficient CHR systems available.
The current implementation does not feature search capabilities.

Finally, the CHORD system (Constraint Handling Object-oriented Rules with Disjunctive bodies)\footnote{%
	by Jairson Vitorino and Marcos Aurelio, 2005, \url{http://chord.sourceforge.net/}
}, developed as part of the ORCAS project \cite{robin_vitorino_orcas_wlp06},
is a Java implementation of \CHRv\ \cite{menezes_vitorino_aurelio_high_performance_chr_or_chr05}.

\paragraph{CHR(C).}

CCHR \cite{wuille_schr_demoen_cchr_chr07} implements CHR for C. 
It is an extremely efficient CHR system conforming 
to the \omegar~refined operational semantics.
It uses a syntax that is 
intuitive to both CHR adepts and imperative programmers.


\subsection{Compilation}
\label{sec:implementation:compilation}

Considerable research has been conducted on the efficient compilation of CHR.
Section~\ref{sec:implementation:compilation:schemes} provides 
an overview of the compilation schemes used by the different CHR systems;
Sections~\ref{sec:implementation:optimization} and \ref{sec:implementation:specialization_transformation}
survey existing analyses and optimizations.

\subsubsection{Compilation Schemes}
\label{sec:implementation:compilation:schemes}

The first CHR compilation scheme, for \eclipse{} Prolog, 
was described by \citeN{fru_brisset_highlevel_implementation_techrep95}. 
\citeANP{holz_fru_compiling_chr_attr_vars_ppdp99}
\citeyear{holz_fru_compiling_chr_attr_vars_ppdp99,holz_fru_prolog_chr_compiler_aai00}
have adapted this scheme from \eclipse{}'s fairly specific suspension mechanism to
the more primitive and flexible attributed variables feature found in SICStus Prolog. 
The latter form has been adopted by HALCHR, K.U.Leuven CHR, 
and was formalized in the refined operational semantics
\cite{duck_stuck_garc_holz_refined_op_sem_iclp04}. 
A good overview of the compilation scheme can be found in the Ph.D.\ theses of 
\citeN{duck_phdthesis05} and \citeN{schr_phdthesis05}, 
and in \cite{vanweert_wuille_et_al_chr_imperative_lnai08}.

In its essence, the scheme maps each constraint to a procedure. 
Imposing the constraint then corresponds to calling the procedure. 
This procedure puts the new constraint in the constraint store datastructure, 
and attempts to fire rules involving the new constraint. 
For the latter purpose, the scheme contains an
occurrence procedure for each occurrence of the constraint symbol in a rule.
The main constraint procedure calls these occurrence procedures 
in the textual order of the rules. 
The reactivation of a constraint is realized through calling the occurrence procedures anew. 
Each procedure looks up the required additional constraints in
the constraint store datastructure and checks both the guard and propagation history.
If all tests succeed, the rule is committed to: an entry is added to the
propagation history, the appropriate matching constraints are removed from the
constraint store and the body of the rule is executed.

The Prolog compilation scheme is specifically designed for the built-in
constraint theory of Herbrand equations. 
\citeN{duck_stuck_garcia_holz_extending_arbitrary_solvers_with_chr_ppdp03} show
how it can be extended to cover arbitrary constraint theories and solvers.

\citeN{schr_zhou_demoen_action_rules_chr06} experimented with
an action rules compilation scheme in BProlog. However, capturing 
the intricate reactivation behavior of CHR's refined
operational semantics turned out to be hard because the action rules' behavior
differs considerably on that account. 

\citeN{lam_sulz_concurrent_chr_damp07} showed that software transactional memories (STM),
as supported by the Glasgow Haskell Compiler, are a good match  
for the concurrent implementation of CHR.
\citeN{sulz_lam_lazy_concurr_search_chr07} also explored the use of 
Haskell's laziness and concurrency abstractions
for implementing the search of partner constraints.

\paragraph{CHR(Java).}
Both JaCK \cite{schmauss_jack:jchr_1999} and CHORD take a different approach compared to most other CHR compilers.
Their front-end transforms the CHR source files 
to Java code that initializes the data structures of a generic runtime.
CHR programs are then essentially interpreted.
No major optimizations are performed.

DJCHR uses a compilation scheme similar to the basic CHR(Prolog) scheme,
but extended with truth maintenance facilities required for adaptive constraint handling \cite{wolf_adaptive_chr_java_cp01}.
Justifications for constraints and rule applications are implemented efficiently using bit vectors.
The runtime also implements adaptive unification and entailment algorithms.
Following the approach of 
\citeANP{holz_fru_compiling_chr_attr_vars_ppdp99}
\citeyear{holz_fru_compiling_chr_attr_vars_ppdp99,holz_fru_prolog_chr_compiler_aai00},
fast partner constraint retrieval is achieved using 
a form of attributed variables \cite{wolf_attr_vars_inap01}.

\paragraph{K.U.Leuven JCHR and CCHR.}
The compilation schemes used by the K.U.Leuven JCHR and CCHR systems \cite{vanweert_schr_demoen_jchr_chr05,wuille_schr_demoen_cchr_chr07}
are based on the basic compilation scheme for CHR(Prolog), 
modified to fit an imperative language \cite{vanweert_wuille_et_al_chr_imperative_lnai08}.
%
%
%
Searching for partner constraints is done through explicit iteration.
Also, the data structures required for the implementation of the constraint
store are implemented more naturally and efficiently in an imperative
host language than in an LP language.
An important issue
is that the host language typically cannot handle recursive calls efficiently.
The common CHR(Prolog) scheme was therefore adjusted 
significantly to avoid frequent call stack overflows.
The compilation scheme used by the K.U.Leuven JCHR system is described 
in detail in \cite{vanweert_jchr_compilation_techrep08}.


\comment{
The constraint stack defined by the \omegar~refined operational semantics is,
in contrast with CHR(Prolog), mapped to an explicit stack in heap memory.
Although mapping it to the implicit call stack seems natural, 
practice has shown this quickly leads to stack overflows 
(cf.\ also \cite{vanweert_jchr_compilation_techrep08}).
Many CHR programs give rise to recursive constraint calls, 
necessitating very large stacks. 
Tail-call optimization would partially solve this problem.
Java Virtual Machines, however, do not implement these optimizations,
and common C compilers only implement severely restricted versions.
Substantially more stack frames fit in heap memory than in the implicit call stack.
}

The CCHR compiler performs some limited optimizations, 
like memory reuse, basic join ordering (see Section \ref{sec:implementation:optimization}), 
as well as imperative-language specific optimizations.
The K.U.Leuven JCHR compiler implements 
most of the optimizations mentioned in Section~\ref{sec:implementation:optimization}.
For more details, see \cite{vanweert_wuille_et_al_chr_imperative_lnai08}.

\paragraph{\chrrp{}.}
A compilation scheme for \chrrp{} that is strongly based on the one for regular
CHR in Prolog, is presented in \cite{dekoninck_stuck_duck_compiling-chrrp_08}.
It is formalized in the refined priority semantics 
\omegarp{}, which combines the refined operational semantics \omegar{} of 
regular CHR, with the priority semantics \omegap{} of \chrrp{}. 

The main differences are the following. The initial goal, as well as rule
bodies, are executed in {\em batch mode}, i.e., no rule can fire as long as there are
unprocessed goal or body constraints (i.e., the {\bf Apply} transition
is not allowed if the {\bf Introduce} transition is applicable, cf.~Fig.~\ref{fig:omegat}).
New constraints are scheduled for
activation at all priorities at which they have occurrences, instead of being
activated as soon as they are processed. Constraints are activated at a given
priority and as such only consider those rules that share this priority.
Finally, after each rule firing, it is checked whether 
a scheduled constraint needs to be activated.
\comment{
constraints are
scheduled at a higher priority than the rule priority, in which case the
highest priority one is activated. 
}
To deal with so-called \emph{dynamic}
priority rules, which are rules for which the actual priority is only
determined at runtime, a source-to-source transformation is given in
\cite{dekoninck_stuck_duck_compiling-chrrp_08}, that transforms these rules to
the desired form for the \omegarp{} semantics.
 
\subsubsection{Analysis and Optimizing Compilation}
\label{sec:implementation:optimization}

A number of (mostly static) analyses and optimizations have been proposed
to increase the performance of CHR systems
\cite{holz_garc_stuck_duck_opt_comp_chr_hal_tplp05,schr_phdthesis05,duck_phdthesis05,vanweert_wuille_et_al_chr_imperative_lnai08}.
Without going into the technical details, we very briefly discuss a 
list of recent optimizations:

\begin{description}
\item[Indexing.]
The efficient, selective lookup of candidate partner constraints
is indispensable for the efficient determination of matching rules.
The traditional CHR(Prolog) compilation scheme (see Section~\ref{sec:implementation:compilation:schemes}) 
uses attributed variables for a constant time lookup of the constraints containing a known, unbound variable.
\citeN{holz_garc_stuck_duck_opt_comp_chr_hal_tplp05} propose the use
of a balanced tree for the lookup of constraints via known ground arguments,
which allows for logarithmic worst-case time lookup.
\citeN{schr_phdthesis05} further improves this using 
hash tables to get amortized constant time constraint store operations.
\citeN{sney_schr_demoen_dijkstra_chr_wlp06} finally introduce array-based indexes
to obtain correct space and time complexity guarantees (see also \cite{sney_schr_demoen_chr_complexity_08}).
\citeN{sarnastarosta_schr_indexing_techrep07} show how indexing on compound term patterns
is reduced to the above indexing techniques via program transformation.

\item[Abstract interpretation.] \citeN{schr_stuck_duck_ai_chr_ppdp05} present a general
and systematic framework for program analysis of CHR for optimized compilation 
based on abstract interpretation. 
Two instances are given:
late storage analysis (for reducing constraint store updates) and groundness analysis.

\item[Functional dependencies.] Functional dependency analysis
\cite{duck_schr_accurate_funcdep_chr05} is a third instance of the abstract
interpretation framework. It aims at tracking the cardinality of constraints
(zero, one or more) for specializing constraint store indexes and related operations.

\item[Guard optimization.] 
The {\em guard optimization} \cite{sney_schr_demoen_guard_and_continuation_opt_iclp05}
removes redundant conjuncts in rule guards by reasoning
on the ramifications of the refined operational semantics.
More precisely, 
non-applicability of rules containing
earlier removed occurrences of 
a constraint, can be used to infer 
redundant guard conditions.

\item[Continuation optimization.] 
The {\em continuation optimization} \cite{sney_schr_demoen_guard_and_continuation_opt_iclp05}
uses a similar reasoning to skip occurrences that 
can never lead to rule firings.

\item[Delay avoidance.] 
\citeN{schr_demoen_delay_avoid_techrep04} and \citeN{holz_garc_stuck_duck_opt_comp_chr_hal_tplp05}
describe techniques for avoiding the unnecessary delay and reactivation of constraints.

\item[Memory reuse.] 
Two memory optimizations ({\em in-place updates} and {\em suspension reuse})
were introduced by \citeN{sney_schr_demoen_memory_reuse_iclp06}.
They significantly reduce the memory footprint and the time spent in garbage collection.

\item[Join ordering.]
The time complexity of executing a CHR program 
is often determined by the join ordering --- the order
in which partner constraints are looked up in order to find matching rules. 
\citeN{holz_garc_stuck_duck_opt_comp_chr_hal_tplp05} and \citeN{duck_phdthesis05}
discussed ad-hoc heuristics for join ordering.
\citeN{dekoninck_sney_join_ordering_chr07} proposed a more rigorous investigation.
\end{description}

\subsubsection{Code Specialization and Transformation}
\label{sec:implementation:specialization_transformation}

Most of the optimizations in the previous section are not expressible as
source-to-source transformations. Like compiler optimization, 
program transformation can also be used to improve performance.
The first proposal for CHR source-to-source transformation, by 
\citeN{fru_specialization_lopstr04}, adds redundant, specialized rules to a CHR
program; these rules capture the effect of the original program for a
particular goal. In more recent work, 
\citeN{tacchella_gabbrielli_meo_unfolding_ppdp07} adapt the conventional notion
of unfolding to CHR.

While the above study program transformation from a more theoretical point of view,
\citeN{sarnastarosta_schr_indexing_techrep07} show that various program transformation
techniques improve indexing performance.

\citeN{fru_holz_source2source_agp03} propose to express
CHR source-to-source transformations in CHR itself, and they show how
to implement various language extensions 
(such as probabilistic CHR) by transformation to plain CHR.
\citeN{vanweert_sney_demoen_aggregates_lopstr07} implemented an extension of CHR with aggregates
(see Section~\ref{ssec:aggregates}) using a similar approach, 
with a more expressive transformation language.

\subsection{Programming Environments}
\label{sec:implementation:environments}

Over the past decade there has been an exponential increase in the number of CHR systems
(Section~\ref{sec:implementation:systems}), 
and CHR compilation techniques have matured considerably 
(Section~\ref{sec:implementation:compilation}).
The support for advanced software development tools,
such as debuggers, refactoring tools, and automated analysis tools,
lags somewhat behind, 
and remains an important challenge for the CHR community.

VisualCHR \cite{abd_saft_jack:visualchr_wlpe01}, 
part of JaCK (see Section~\ref{sec:systems:java}), 
is an interactive tool visualizing the execution of CHR rules.
It can be used to debug and to improve the efficiency of constraint solvers.

Both Holzbaur's CHR implementation and the K.U.Leuven CHR system feature a
trace-based debugger that is integrated in the Prolog four port tracer.  
A generic trace analysis tool, with an instantiation for CHR, 
is presented in \cite{ducasse_opium_jlp99}.

Checked type annotations are useful both for documentation and debugging purposes.
CHR systems with a typed host language commonly perform type checking, or even type inference.
The K.U.Leuven CHR system also allows optional type declarations,
with both dynamic and static type checking.
\citeN{coquery_fages_type_system_chr05} present a generic type system for CHR(\HL)
(cf.\ also Section~\ref{sec:applications:type}).

\citeN{ring_schlenk_inference_wlp00} propose the automatic inference of
imported and exported symbols of CHR solvers, and the composition of solvers by
matching up their interfaces \cite{ring_schlenk_interfaces_rcorp00bis}. 

\citeN{torres_literate_programming_wflp02} presents a literate programming system for CHR. 
The system allows for generating from the same literate program source both 
an algorithm specification typeset in \LaTeX\ using mathematical notation, 
and the corresponding executable CHR source code.

Based on the theoretical results of \citeN{abd_fru_meuss_confluence_semantics_csr_constr99}
(see Section~\ref{sec:analysis:confluence}), 
\citeN{bouissou_chr_for_silcc_2004} implemented a confluence analyzer in CHR.
\citeN{duck_phdthesis05} presents and evaluates a confluence checker
based on the refined operational semantics.
We do not know of any practical implementations of the other analyses 
of Section~\ref{sec:analysis:confluence}.

%% file: 05-extensions.tex
\section{Extensions and Variants}
\label{sec:extensions}

Over the years, weaknesses and limitations of CHR have been identified, 
for instance regarding execution control, expressivity, modularity, incrementality, and search.
In this section we consider extensions and variants of CHR that were
proposed to tackle these issues.

\subsection{Deviating Operational Semantics}

We first discuss variants of CHR with an operational semantics that deviates from the
commonly used refined operational semantics discussed in Section~\ref{sec:semantics:operational:omegar}.

\subsubsection{Probabilistic CHR}
Probabilistic CHR (PCHR; \citeNP{fru_dipierro_wiklicky_probabilistic_chr_wflp02})
extends CHR with probabilistic choice between the applicable rules in
any state (though only for a given active constraint). It supports the 
implementation of algorithms like simulated annealing, which is often used for
constrained optimization. It also gives rise to new concepts like probabilistic
confluence and probabilistic termination. In PCHR, the probabilities are either
a fixed number or an arithmetic expression involving variables that appear in
the head. The latter are called \emph{parametrised probabilities}. The actual
probability that a rule is executed in a given state is found by dividing the
rule probability by the sum of the probabilities of all fireable rule 
instances. 
\citeN{fru_dipierro_wiklicky_probabilistic_chr_wflp02} give an implementation
of PCHR by means of a 
source-to-source transformation using the framework proposed by 
\citeN{fru_holz_source2source_agp03}.

As a simple example, the following PCHR program simulates tossing a coin:
\begin{verbatim}
   toss(Coin) <=>0.5: Coin=head.
   toss(Coin) <=>0.5: Coin=tail.
\end{verbatim}
Given a {\tt toss/1} constraint, one of the rules will be applied,
with equal probability.

\subsubsection{CHRd}
The CHRd system (CHR with distributed constraint store) of 
\citeN{sarnastarosta_ramakrishnan_chrd_padl07} alleviates the
limitations of conventional CHR systems for efficient tabled evaluation
encountered by \citeN{schr_warren_chr_xsb_iclp04}. 
For this purpose it implements a set-based operational semantics, i.e. the
constraint store is a set rather than a multiset.  Moreover, CHRd's constraint
store has no global access point; constraints can only be retrieved through 
their logical variables. This rules out (the efficient execution of)
of ground CHR programs. 

\subsubsection{\chrrp} \index{chrrp@{\chrrp}}
\label{sec:extensions:chrrp}
\chrrp{} \cite{dekoninck_schr_demoen_chrrp_ppdp07} is CHR extended with 
user-definable rule priorities. A rule's priority is either a number or an 
arithmetic expression involving variables that appear in the rule heads. The
latter allows different instances of a rule to be executed at different
priorities. The following \chrrp{}-related topics are dealt with in other sections: its 
operational semantics in 
Section~\ref{sec:semantics:operational:omegap}; its compilation schema in Section~\ref{sec:implementation:compilation:schemes}; 
and its implementation for SWI-Prolog in Section~\ref{sec:implementation:systems:lp}.

An example that illustrates the power of dynamic priorities
is the following \chrrp{} implementation of Dijkstra's algorithm:
\begin{verbatim}
   1 :: source(V) ==> dist(V,0).
   1 :: dist(V,D1) \ dist(V,D2) <=> D1 =< D2 | true.
   D+2 :: dist(V,D), edge(V,C,W) ==> dist(W,D+C).
\end{verbatim}
The priority of the last rule makes sure that
new distance labels are propagated in the right order:
first the nodes closest to the source node.

\subsubsection{Adaptive CHR}\label{sec:extensions:adaptive}
Constraint solving in a continuously changing, dynamic environment 
often requires immediate adaptation of the solutions, i.e. when constraints are added or removed.
%
By nature, CHR solvers already support efficient adaptation when constraints are added. 
Wolf \citeNN{wolf_phdthesis99,wolf_ea_incremental_adaptation_aai00}
introduces an extended incremental adaptation algorithm
which is capable of adapting CHR derivations after constraint deletions as well.
This algorithm is further improved by \citeN{wolf_projection_compulog00} with
the elimination of local variables using \emph{early projection}.
An efficient implementation exists in Java 
(Wolf \citeyearNP{wolf_adaptive_chr_java_cp01,wolf_attr_vars_inap01}; cf.\ Section~\ref{sec:systems:java}).


Interesting applications of adaptive CHR include
adaptive solving of soft constraints, discussed in Section~\ref{sec:applications:solvers},
and the realization of intelligent search strategies, discussed in 
Sections~\ref{sec:systems:java}~and~\ref{sec:extensions:disj}.

\subsection{Language Extensions}
We now discuss some additional language features that have been added to the CHR language.

\subsubsection{Disjunction and Search}
\label{sec:extensions:disj}
Most constraint solvers require search next to constraint simplification and propagation. 
However pure CHR does not offer any support for search. 
\citeN{abd_query_lang_fqas98} propose a solution to this problem:
an extension of CHR with disjunctions in rule bodies
(see also Abdennadher \citeyearNP{abd_chr_disjunction_rcorp00,abd_habilitation_2001}).
The resulting language is denoted \CHRv{} (pronounced ``CHR-or''),
and is capable of expressing several declarative evaluation strategies, 
including bottom-up evaluation, top-down evaluation,
model generation and abduction (see Section~\ref{sec:applications:abduction} for abduction).
Any (pure) Prolog program can be rephrased as an equivalent \CHRv{} program 
(Abdennadher \citeyearNP{abd_chr_disjunction_rcorp00,abd_habilitation_2001}).
An interesting aspect of \CHRv{} is that the extension comes for free
in CHR(Prolog) implementations by means of the built-in Prolog disjunction and
search mechanism.

As a typical example of programming in \CHRv{},
consider the following rule:
\begin{verbatim}
   labeling, X::Domain <=> member(X,Domain), labeling.
\end{verbatim}
Note the implicit disjunction in the call to the Prolog
predicate {\tt member/2}.

Various ways have been proposed to make the search in \CHRv{} programs more
flexible and efficient. 
\citeN{menezes_vitorino_aurelio_high_performance_chr_or_chr05} present a \CHRv{}
implementation for Java in which the search tree is made explicit 
and manipulated at runtime to improve efficiency. 
The nodes in the search tree can be reordered to avoid redundant work. 
\citeN{dekoninck_schr_demoen_search_chr06} extend both the theoretical and refined
operational semantics of CHR towards \CHRv{}.
The theoretical version leaves the search strategy undetermined, 
whereas the refined version allows the specification of various search 
strategies. In the same work, an implementation for different strategies in
CHR(Prolog) is realized by means of a source-to-source transformation. 

For CHR(Java) systems, unlike for CHR(LP) systems, 
the host language does not provide search capabilities.
The flexible specification of intelligent search strategies 
has therefore received considerable attention in several CHR(Java) systems 
\cite{kr_jack:jase_2001,wolf_intelligent_search_tplp05}.
As described in Section~\ref{sec:systems:java},
in these systems,
the search strategies are implemented and specified in the host language itself,
orthogonally to the actual CHR program.
\citeN{wolf_robin_vitorino_adaptive_chr_or_chr07} propose an implementation of 
\CHRv\ using the ideas of \cite{wolf_intelligent_search_tplp05},
in order to allow a more declarative formulation of search 
in the bodies of CHR rules,
while preserving efficiency and flexibility.
A refined operational semantics of the proposed execution strategy is presented as well.
Along these lines is the approach proposed by \citeN{robin_vitorino_wolf_CPA_proposal_jucs07}, 
where disjunctions in \CHRv{} are transformed 
into special purpose constraints that can be handled by an external search component
such as JASE \cite{kr_jack:jase_2001}.

\subsubsection{Negation and Aggregates} \label{ssec:aggregates}
CHR programmers often want to test for the absence of constraints.
CHR was therefore extended with \emph{negation as absence} 
by \citeN{vanweert_sney_schr_demoen_negation_chr06}.
Negation as absence was later generalized to a much more powerful language feature, 
called \emph{aggregates} \cite{sney_vanweert_demoen_aggregates_chr07}. 
Aggregates accumulate information over unbounded portions of the constraint store.
Predefined aggregates include \texttt{sum}, \texttt{count}, \texttt{findall}, and \texttt{min}.
The proposed extension also features nested aggregate expressions 
over guarded conjunctions of constraints,
and application-tailored user-defined aggregates.
Aggregates lead to increased expressivity and more concise programs. 
An implementation based on source-to-source transformations 
\cite{vanweert_sney_demoen_aggregates_lopstr07} is available.
The implementation uses efficient incremental aggregate computation,
and empirical results show that the desired runtime complexity is attainable 
with an acceptable constant time overhead.

As an example of nested aggregate expressions, consider
the following rule:
\begin{verbatim}
   eulerian, forall(node(N),(
                count(edge(N,_),X), count(edge(_,N),X)
             )) <=> writeln(graph_is_eulerian).
\end{verbatim}
The above rule is applicable if for every constraint
{\tt node(N)}, the number of out-going edges
in {\tt N} equals the number of incoming edges
(i.e. if the first number is {\tt X}, the other number
must also be {\tt X}).

\subsection{Solver Hierarchies}
\label{ssec:hierarchies}

While the theory of the CHR language generally considers arbitrary built-in
solvers, traditional CHR implementations restrict themselves to the Herbrand
equality constraint solver, with very little, if any, support for other constraint
solvers.

\citeN{duck_stuck_garcia_holz_extending_arbitrary_solvers_with_chr_ppdp03}
show how to build CHR solvers on top of arbitrary built-in
constraint solvers by means of \emph{ask} constraints. The ask constraints signal
the CHR solver when something has changed in the built-in store with respect to
variables of interest. Then the relevant CHR constraints may be reactivated.

\citeN{schr_demoen_duck_stuck_fru_implication_checking_entcs06} 
provide an automated means for deriving ask versions of CHR constraints. 
In this way full hierarchies of constraint solvers can be written in CHR, where
one CHR solver serves as the built-in solver for another CHR solver.

%% file: 06-other_formalisms.tex
\section{Relation to Other Formalisms}
\label{sec:other_formalisms}


The relation of CHR to other formalisms has recently received quite a lot of attention.
In this section we give a brief overview.

As a general remark, we should mention that most of the formalisms related to CHR 
are limited to ground terms and lack the equivalent of propagation rules.
In that sense, they are subsumed by CHR.
Also, CHR can be seen as an instance or a generalization of concurrent constraint programming,
constraint logic programming, constraint databases, and deductive databases.

\paragraph{Logical formalisms.}
In Section~\ref{sec:semantics:logical}, we discussed the logical semantics of CHR.
CHR 
can be given
a classical logic semantics \cite{fru_chr_overview_jlp98},
a linear logic semantics (Betz and Fr\"uhwirth \citeyearNP{betz_fru_linear_logic_semantics_cp05,betz_fru_linear_logic_chr_disj_chr07}),
and a transaction logic semantics \cite{meister_djelloul_robin_transaction_logic_semantics_lpnmr07}.
Also, it should be noted that \CHRv{} subsumes Prolog. 
Frame-Logic (F-Logic) is an object-oriented extension of classical predicate logic.
\citeN{kaeser_meister_flogic_chr06} explore the relation
between CHR and F-Logic by implementing (a fragment of) F-Logic in CHR.

\paragraph{Term rewriting.}
CHR can be considered as associative and commutative (AC) term rewriting of flat conjunctions.
The term rewriting literature inspired many results for CHR, for example
on confluence (see Section~\ref{sec:analysis:confluence})
and termination (see Section~\ref{sec:analysis:termination}).
Recently, \citeN{duck_stuck_brand_acd_term_rewriting_iclp06} proposed the formalism of ACD term rewriting,
which subsumes both AC term rewriting and CHR.
\label{ACD}

\subsection{Set-Based Formalisms}

Numerous formalisms have been proposed that are based on (multi-)set rewriting.

\subsubsection{Production Rules / Business Rules}

Production rules, or business rules as they are now often called,
is a rule-based paradigm closely related to CHR.
Classic matching algorithms, such as RETE and LEAPS,
have influenced early work on CHR compilation.
Production rules have also inspired the research towards 
extending CHR with aggregates (see Section~\ref{ssec:aggregates}).

\subsubsection{Join-Calculus} \label{ssec:joincalculus}

The join-calculus is a calculus for concurrent programming, 
with both stand-alone implementations 
and extensions of general purpose languages, 
such as JoCaml (OCaml), Join Java 
and Polyphonic C\#.

\citeN{sulz_lam_haskelljoinrules_ifl07} propose 
a Haskell language extension for supporting join-calculus-style concurrent programming,
based on CHR.
Join-calculus rules, called chords, are essentially guardless simplification
rules with linear match patterns. 
In a linear pattern,
different head conjuncts are not allowed to share variables.
Hence, CHR offers considerably increased expressivity over the join-calculus: 
propagation rules, general guards and non-linear patterns.

\subsubsection{Logical Algorithms}\label{ssec:LA}

As already mentioned in Section \ref{sec:analysis:complexity:meta}, CHR is strongly related to the
Logical Algorithms (LA) formalism by \citeN{ganzinger_mcallester_la_iclp02}.  
\citeN{dekoninck_schr_demoen_la-chr_iclp07} have showed how LA programs can
easily be translated into \chrrp{} programs. The opposite only holds for a
subset of \chrrp{} since the LA language lacks the ability to plug in any
built-in constraint theory, and also only supports ground constraints (called
\emph{assertions} in LA terminology). The correspondence between both languages
makes it possible to apply the meta-complexity result for LA to a subset of
\chrrp{} as explained in Section \ref{sec:analysis:complexity:meta}. It is also
interesting that the first actual implementation of LA is that of
\citeN{dekoninck_schr_demoen_la-chr_iclp07}, which compiles the language into 
(regular) CHR rules.

\subsubsection{Equivalent Transformation Rules}
The Equivalent Transformation (ET) computation model is a rewriting system which 
consists of the application of conditional multi-headed multi-body ET rules (ETR). 
Although CHR and ETR are similar in syntax, they have 
different theoretical bases: 
CHR is based on logical equivalence of logical formulas, 
whereas ETR is based on the set equivalence of descriptions. 
\citeN{shigeta_akama_mabuchi_koike_chr_to_etr_jaciii06}
investigate the relation between CHR and ETR.
\comment{In particular, they demonstrate that CHR is a proper subclass of ETR
by proposing a method for correctly converting CHR to ETR and giving
an example of an ETR program that supposedly does not correspond to any CHR program
--- a 
claim that seems to be based on restricting the
host language of CHR more than the ``ETR host language''.
}

\subsection{Graph-Based Formalisms}

Recently, CHR has also been related to a number of graph-based formalisms:

\subsubsection{Graph Transformation Systems}
\citeN{raiser_graph_transformation_systems_iclp07} describes an elegant 
embedding of Graph Transformation Systems (GTS) in CHR.
The confluence properties (see Section~\ref{sec:analysis:confluence})
for CHR and GTS are similar; in particular,
a sufficient criterion for confluence of a GTS is the
confluence of the corresponding CHR program.
Using a slightly weaker notion of confluence of CHR
(in the spirit of observable confluence; \citeNP{duck_stuck_sulz_observable_confluence_iclp07}),
standard CHR confluence checkers can be reused to decide GTS confluence.

\subsubsection{Petri Nets}
Petri nets are a well-known formalism for the modeling and analysis 
of concurrent processes.
\citeN{betz_petri_nets_chr07} provides a first study of the relation between CHR and Petri nets.
He provides a sound and complete translation of place/transition nets (P/T nets)
--- a standard variant of Petri nets ---
into a small segment of CHR.
P/T nets are, unlike CHR (cf.\ Section~\ref{sec:analysis:complexity}), 
not Turing complete.
A translation of a significant subsegment of CHR 
into \emph{colored} Petri nets is presented as well by \citeN{betz_petri_nets_chr07}.
This work is a promising first step towards cross-fertilization between both formalisms.
Results from Petri nets could for instance be applied 
to analyze concurrency properties of CHR programs.

\subsubsection{LMNtal}
LMNtal \cite{ueda_LMNtal_chr06} is a language based on hierarchical graph rewriting
which intends to unify constraint-based concurrency and CHR.
It uses logical variables to represent connectivity and so-called \emph{membranes}
to represent hierarchy.
Flat LMNtal rules (rules without membranes) can be seen as simplification rules in CHR.


%% file: 07-applications.tex
\section{Applications}
\label{sec:applications}

The main application domain considered in the previous CHR survey \cite{fru_chr_overview_jlp98} 
is the development of constraint solvers.
It also discusses two other applications of CHR in some depth.
The first one is related to the problem of finding an optimal
placement of wireless transmitters 
(Fr\"uhwirth and Brisset \citeyearNP{fru_brisset_wireless_cp98,fru_bri_base_wireless_ieee00});
the second one is an expert system for estimating the maximum fair rent
in the city of Munich, called the
\emph{Munich Rent Advisor} \cite{fru_abd_munich_rent_advisor_tplp01}.
In this section, we give an overview of more recent applications of CHR.

\subsection{Constraint Solvers}
\label{sec:applications:solvers}

CHR was originally designed specifically for writing constraint solvers.
We discuss some recent examples of constraint solvers written in CHR.
%
%
%
The following examples illustrate how CHR can be used to build effective prototypes of 
non-trivial constraint solvers:

\begin{description}
\item[Lexicographic order.]
\citeN{fru_lexico_csclp05:06} presented a constraint solver for
a lexicographic order constraint in terms of inequality constraints offered by
the underlying solver. The approach is general in that it can be used for any
constraint domain offering inequality (\emph{less-than}) constraints between
problem variables. Still, the program is very concise and elegant; it consists of just six rules.

\item[Rational trees.]
\citeN{meister_djelloul_fru_compl_tree_equations_csclp06} presented
a solver for existentially quantified conjunctions of non-flat equations over rational 
trees. The solver consists of a transformation to flat equations, after which
a classic CHR solver for rational trees can be used. This results in a 
complexity improvement with respect to previous work. 
In \cite{djelloul_dao_fru_1st_order_extension_prolog_unification_sac07}, 
the solver for rational trees is used as part of a more general solver for
(quantified) first-order constraints over finite and infinite trees.

\item[Sequences.]
Kosmatov~\citeyear{kosmatov_sequences_06,kosmatov_sequences_inap05} has constructed
a constraint solver for sequences, 
inspired by an earlier solver for lists by Fr\"uhwirth. 
The solver expresses many sequence constraints in terms
of two basic constraints, for sequence concatenation and size.
%

\item[Non-linear constraints.]
A general purpose CHR-based CLP system for non-linear (polynomial) 
constraints over the real numbers was presented by \citeN{dekoninck_schr_demoen_inclpr_wlp06}. 
The system, called INCLP($\mathbb{R}$), is based on interval arithmetic and uses an interval 
Newton method as well as constraint inversion to achieve respectively box and
hull consistency.

\item[Interactive constraint satisfaction.]
\citeN{alberti_et_al_chr_implementation_of_arc_consistency_tplp05} describe the implementation 
of a CLP language for expressing Interactive Constraint Satisfaction Problems (ICSP).
In the ICSP model incremental constraint propagation is possible 
even when variable domains are not fully known, 
performing acquisition of domain elements only when necessary.

\item[Solvers derived from union-find.]
\citeN{fru_deriving_linear_algorithms_from_uf_chr06}
proposes linear-time algorithms for solving certain boolean equations
and linear polynomial equations in two variables.
These solvers are derived from the classic union-find algorithm
(see Section~\ref{sec:applications:uf}).
\end{description}

In the rest of this subsection we discuss, in a bit more detail,
some typical application domains in which CHR has been used to
implement constraint solvers.

\subsubsection{Soft Constraints and Scheduling}
An important class of constraints are the so-called \emph{soft constraints}
which are used to represent preferences amongst solutions to a problem.
Unlike hard (required) constraints which must hold in any solution, 
soft (preferential) constraints must only be satisfied as far as possible.
\citeN{bistarelli_fru_marte_rossi_soft_constraint_propagation_CI04} present
a series of constraint solvers for (mostly) extensionally defined finite domain soft
constraints, based on the framework of \emph{c-semirings}.
In this framework, soft constraints can be combined and projected onto a
subset of their variables, by using the two operators of a c-semiring. 
A node and arc consistency solver is presented, 
as well as complete solvers based on variable elimination or branch and bound
optimization. 

Another well-known formalism for describing over-constrained systems 
is that of \emph{constraint hierarchies},
where constraints with hierarchical strengths or preferences can be specified,
and non-trivial error functions can be used to determine the solutions.
\citeN{wolf_rule_based_hierarchies_rcorp00} 
proposes an approach for solving dynamically changing constraint hierarchies.
Constraint hierarchies over finite domains are transformed into equivalent constraint systems, 
which are then solved using an adaptive CHR solver
(\citeNP{wolf_ea_incremental_adaptation_aai00,wolf_adaptive_chr_java_cp01};
see Section~\ref{sec:extensions:adaptive}).

\paragraph{Scheduling.}
\citeN{abd_marte_timetabling_chr_aai00} have successfully used CHR 
for scheduling courses at the university of Munich. 
Their approach is based on a form of soft constraints, implemented in CHR, to deal with teacher's preferences.
A related problem, namely that of assigning classrooms to courses given a timetable, 
is dealt with by \citeN{abd_saft_will_classroom_assignment_paclp00}.
An overview of both applications is found in \cite{abd_habilitation_2001}.

\subsubsection{Spatio-Temporal Reasoning}

In the context of autonomous mobile robot navigation,
a crucial research topic is automated qualitative reasoning about 
spatio-temporal information, 
including orientation, named or compared distances, cardinal directions, topology and time.
The use of CHR for spatio-temporal reasoning has received considerable research attention.
We mention in particular the contributions of Escrig et al.\
(Escrig and Toledo \citeyearNP{escrig_toledo_framework_qual_orientation_vlc98,escrig_toledo_qual_spatial_reasoning_book98};
\citeNP{museros_escrig_modeling_motion_jucs03}).

\citeN{meyer_diagrammatic_reasoning_aai00} has applied CHR for 
the constraint-based specification and implementation of diagrammatic environments.
Grammar-based specifications of diagrammatic objects are translated to directly executable CHR rules.
This approach is very powerful and flexible.
The use of CHR allows the integration with other constraint domains,
and additional CHR rules can easily be added to model more complex diagrammatic systems.
Similar results are obtained with CHRG in the context of natural language processing 
(see Section~\ref{sec:applications:parsing_and_nlp}).

\subsubsection{Multi-Agent Systems}
FLUX (Thielscher \citeyearNP{thielscher_actions_iclp02,thielscher_flux_tplp05}) 
is a high-level programming system,
implemented in CHR and based on fluent calculus, 
for cognitive agents that reason logically about actions in the context of incomplete information.
An interesting application of this system is FLUXPLAYER \cite{schiffel_thielscher_fluxplayer_aaai07},
which won the 2006 General Game Playing (GGP) competition at AAAI'06.
\citeN{seitz_bauer_berger_MAS_icai02} and Alberti et al.\ 
\shortcite{alberti_et_al_social_integrity_constraints_lcmas03,alberti_et_al_agent_interaction_protocols_sac04,alberti_et_al_compliance_agents_tool_aai06} 
also applied CHR in the context of multi-agent systems.

\citeN{lam_sulz_linear_logic_agents_chr06} explore the use of CHR as an agent specification language,
founded on CHR's linear logic semantics (see Section~\ref{sec:semantics:logical:linear}).
They introduce a monadic operational semantics for CHR,
where special \emph{action constraints} have to be processed in sequence.
They reason about the termination and confluence properties of the resulting language.
They were also the first to propose the use of invariants for CHR confluence testing,
an approach that was later formalized by \citeN{duck_stuck_sulz_observable_confluence_iclp07} 
--- cf.\ Section~\ref{sec:analysis:confluence}.

\subsubsection{Semantic Web and Web 3.0}
One of the core problems related to the so-called {\em Semantic Web}
is the integration and combination of data from diverse information sources.
\citeN{bressan_goh_coin_fqas98} describe an implementation 
of the {\sc coin} ({\sc co}ntext {\sc in}terchange) mediator that uses CHR
for solving integrity constraints.
In more recent work, CHR is used for implementing 
an extension of the {\sc coin} framework,
capable of handling more data source 
heterogeneity
\cite{firat_ecoin_phdthesis03}.
\citeN{badea_et_al_semantic_web_reasoning_ppswr04} present 
an improved mediator-based integration system.
It allows forward propagation rules involving model predicates, 
whereas {\sc coin} only allows integrity constraints on source predicates.

The Web Ontology Language (OWL) is based on Description Logic (DL). Various rule-based formalisms have been
considered for combination and integration with OWL or other description logics. 
\citeN{fru_description_logic_chr07} proposes a CHR-based approach to DL and DL rules.
Simply encoding the first-order logic theory of the DL in CHR results in a concise, correct,
confluent and concurrent CHR program with performance guarantees.

The Cuypers Multimedia Transformation Engine \cite{geurts_vanOss_hardman_multimedia_presentation_cuypers_mm01}
is a prototype system for automatic generation of Web-based presentations
adapted to device-specific capabilities and user preferences.
It uses CHR and traditional CLP to solve qualitative and quantitative spatio-temporal constraints.

\subsubsection{Automatic Generation of Solvers}

\label{sec:extensions:codegen}
Many authors have investigated the automatic generation of CHR rules,
constituting a constraint solver, from a formal specification. 
Most authors consider extensionally defined constraints over (small) finite
domains as the specification.

A first line of work is that of \citeN{apt_monfroy_cp_viewed_as_rulebased_tplp01}.  
From an extensional definition of a finite domain constraint, a set of
propagation rules is derived. These rules reduce the domain of one variable 
based on the domains of the other variables involved in the same constraint.
As an extension, \citeN{brand_monfroy_generation_propagation_rules_entcs03}
propose to transform the derived rules to obtain stronger propagation rules.
This technique is useful to obtain derived versions of constraints, such as the
conjunction of two constraints.

A second line of work is that of Abdennadher and Rigotti.  In
\citeNN{abd_rigo_automatic_gen_of_solvers_tocl04} they derive propagation rules
from extensionally defined constraints. There are two main differences with
the previous line of work. Firstly, the rules are assembled from given parts,
and, secondly, propagation rules are transformed into simplification rules when
valid. 
This algorithm is implemented by the Automatic Rule Miner tool
\cite{abd_et_al_arm_lopstr06}. 
They extend their approach to intensional constraint definitions, where constraints
are defined by logic programs \citeNN{abd_rigotti_automatic_generation_tplp05},
and further to symbolically derive rules from the logic programs, rather than from
given parts \citeNN{abd_sobhi_generation_combined_lopstr07}.


\citeN{brand_redundant_rules_csclp02} has proposed a method to
eliminate redundant propagation rules and applies it to rules generated
by RuleMiner: one of the algorithms that forms the basis of the
Automatic Rule Miner tool by Abdennadher et al.~\citeNN{abd_et_al_arm_lopstr06}.

\subsection{Union-Find and Other Classic Algorithms}
\label{sec:applications:uf}

CHR is used increasingly as a general-purpose programming language.
Starting a trend of investigating this side of CHR,
Schrijvers and Fr\"uhwirth \citeNN{schr_fru_analysing_union_find_wclp05,schr_fru_opt_union_find_tplp06}
implemented and analyzed the classic union-find algorithm in CHR.
In particular, they showed how the optimal complexity of this algorithm can be achieved in CHR
--- a non-trivial achievement since this is believed to be impossible in pure Prolog.
This work lead to parallel versions of the union-find algorithm \cite{fru_parallel_union_find_iclp05}
and several derived algorithms \cite{fru_deriving_linear_algorithms_from_uf_chr06}.
Inspired by the specific optimal complexity result for the union-find algorithm,
\citeN{sney_schr_demoen_chr_complexity_08}
have generalized this to arbitrary (RAM-machine) algorithms
(see Section~\ref{sec:analysis:complexity:meta}).

The question of finding elegant and natural implementations of classic
algorithms in CHR remains 
nevertheless
an interesting research topic.
Examples of recent work in this area are implementations of 
Dijkstra's shortest path algorithm using Fibonacci heaps
\cite{sney_schr_demoen_dijkstra_chr_wlp06}
and the preflow-push maximal flow algorithm \cite{meister_preflow_push_wlp06}.

\subsection{Programming Language Development}

Another application area in which CHR has proved to be useful 
is the development of programming languages.
CHR has been applied to implement additional programming 
language features such as type systems (Section~\ref{sec:applications:type}),
meta-programming (Section~\ref{sec:applications:meta}), and abduction (Section~\ref{sec:applications:abduction}).
Also, CHR has been used to implement new programming languages, 
especially in the context of computational linguistics 
(Section~\ref{sec:applications:parsing_and_nlp}).
Finally, CHR has been used for testing and verification
(Section~\ref{sec:applications:testing}).

\subsubsection{Type Systems}
\label{sec:applications:type}

CHR's aptness for symbolic constraint solving has led to many applications in
the context of type system design, type checking and type inference. 
While 
the basic Hindley-Milner type system requires no more than a
simple Herbrand equality constraint, more advanced type systems require custom
constraint solvers.

\citeN{alves_florido_type_inference_chr_wflp01:entcs02} 
presented the first work on using Prolog and CHR 
for implementing the type inference framework HM(X), 
i.e.\ type inference for extensions of the Hindley-Milner type system.
This work was followed up by TypeTool \cite{simoes_florido_typetool_wflp04}, 
a tool for visualizing the type inference process.

The most successful use of CHR in this area is for Haskell type classes.
Type classes are a principled approach to ad hoc function
overloading based on type-level constraints. By defining these type class
constraints in terms of a CHR program
\cite{stuck_sulz_theory_of_overloading_toplas05} the essential properties of
the type checker (soundness, completeness and termination) can easily be
established. 
Moreover, various extensions, such as multi-parameter type classes
\cite{sulz_schr_stuck_aplas06} and functional dependencies
\cite{sulz_duck_peyton_stuck_func_dep_via_chr_fp07} are easily expressed. 
At several occasions Sulzmann argues for HM(CHR), 
where the programmer can directly implement custom type system extensions in CHR.
\citeN{wazny_phdthesis06} has made considerable contributions in this
setting, in particular to type error diagnosis based on justifications.

Coquery and Fages \citeNN{coquery_fages_type_system_wlpe03,coquery_fages_type_system_chr05} 
presented TCLP, a CHR-based type checker for Prolog and CHR(Prolog)  
that deals with parametric polymorphism, subtyping and overloading. 
\citeN{schr_bruynooghe_polymorphic_type_reconstruction_ppdp06} reconstruct type
definitions for untyped functional and logic programs.

Finally, \citeN{chin_et_al_florin_sigplan06} presented
a control-flow-based approach for variant parametric polymorphism in Java.
 

\subsubsection{Abduction}	\label{sec:applications:abduction}

Abduction is the inference of a cause to explain a consequence: given $B$
determine $A$ such that $A \to B$. It has applications in many areas:
diagnosis, recognition, natural language processing, type inference, \ldots

The earliest paper connecting CHR with abduction is that of
\citeN{abd_christ_abduction_fqas00}. It shows how to model logic programs with
abducibles and integrity constraints in \CHRv. The disjunction is used for
(naively) enumerating the alternatives of the abducibles, while integrity
constraints are implemented as propagation rules. 
The HYPROLOG system of \citeN{christiansen_dahl_hyprolog_iclp05} combines the
above approach to abductive reasoning with abductive-based logic programming in
one system. Both the abducibles and the assumptions are implemented as CHR
constraints.
\citeN{christiansen_clima06} also proposes the use of CHR for the implementation of
{\em global abduction}, an extended form of logical abduction for reasoning about a dynamic world.

\citeN{gavanelli_et_al_interpreting_abduction_agp03} propose two variant
approaches to implementing abductive logic programming. The first is similar to
the above approach, but tries to leverage as much as possible from a more
efficient boolean constraint solver, rather than CHR. The second approach
propagates abducibles based on integrity constraints.

The system of \citeN{alberti_et_al_sys_gen_conf_hypoth_wclp05} extends the
abductive reasoning procedure with the dynamic acquisition of new facts. These
new facts serve to confirm or disconfirm earlier hypotheses.

\citeN{sulz_wazny_stuck_abduction_chr05} show that advanced type system
features give rise to implications of constraints, or, in other words,
constraint abduction. An extension of their CHR-based type
checking algorithm is required to deal with these implications.

%

\subsubsection{Computational Linguistics} 
\label{sec:applications:parsing_and_nlp}

CHR allows flexible combinations of top-down and bottom-up computation \cite{abd_query_lang_fqas98},
and abduction fits naturally in CHR as well (see Section~\ref{sec:applications:abduction}).
It is therefore not surprising that CHR 
has proven a powerful implementation and specification tool for language processors.

\citeN{penn_hpsg_rcorp00} focuses on another benefit CHR provides
to computational linguists, 
namely the possibility of delaying constraints until their arguments
are sufficiently instantiated.
As a comprehensive case study he considers a grammar development system 
for HPSG, a popular constraint-based linguistic theory.

\citeN{morawietz_blache_unpublished02} show that CHR allows a flexible and perspicuous implementation 
of a series of standard chart parsing algorithms (cf.\ also \citeN{morawietz_chart_parsing_coling00}),
as well as more advanced grammar formalisms such as minimalist grammars and property grammars.
Items of a conventional chart parser are modeled as CHR constraints,
and new constraints are deduced using constraint propagation. 
The constraint store represents the chart,
from which the parse tree can be determined.
Along the same lines is the CHR implementation of a context-sensitive,
rule-based grammar formalism by \citeN{garat_wonsever_parser_contextual_rules_sccc02}.

A more recent application of CHR in the context of natural language processing
is \cite{christ_have_use_cases_to_uml_ranlp07}, where
a combination of Definite Clause Grammars (DCG) and CHR is used
to automatically derive UML class diagrams from use cases written in a restricted natural language.

\paragraph{CHR Grammars.}
The most successful approach to CHR-based language processing
is given by CHR grammars (CHRG), a highly expressive, bottom-up grammar 
specification language proposed by \citeN{christ_chr_grammars_tplp05}.
Contrary to the aforementioned approaches, 
which mostly use CHR as a general-purpose implementation language,
Christiansen recognizes that the CHR language itself 
can be used as a powerful grammar formalism.
CHRG's,
built as a relatively transparent layer of syntactic sugar over CHR,
are to CHR what DCG's are to Prolog.

CHRG's inherent support for context-sensitive rules readily allows 
linguistic phenomena such as long-distance reference and coordination to be modeled naturally
\cite{christ_chr_grammars_tplp05,as_dahl_coordination_iberamia04,dahl_abductive_dependencies_cslp04}.
%
CHRG grammar rules can also use extra-grammatical hypotheses, modeled as regular CHR constraints.
This caters, e.g.,\ for straightforward implementations of assumption grammars 
and abductive language interpretation with integrity constraints. 

\paragraph{Applications of CHRG.}
Using CHRG, \citeN{dahl_blache_extracting_phrases_cslp05}
develop directly executable specifications of property grammars.
They show this combination of grammar formalisms to be robust, 
and able to handle various levels of granularity, 
as well as incomplete and incorrect input.
In \cite{dahl_gu_property_gram_biomed_iclp06}, 
an extension of this approach is used to extract concepts and relations from biomedical texts.
 
\citeN{dahl_voll_concept_formation_rules_nlucs04} generalize the property grammar parsing methodology 
into a general concept formation system, providing a cognitive sciences view of problem solving.
Applications of this formalism include 
early lung cancer diagnosis \cite[Chapter~4]{alma_thesis05},
error detection and correction of radiology reports obtained from speech recognition \cite[Section~5.2.8]{voll_thesis06},
and the analysis of biological sequences \cite{bavarian_dahl_bio_seq_analysis_jucs06}.

\citeN{christ_dahl_diagnosis_and_repair_ijait03} use
an abductive model based on CHRG to diagnose and correct grammatical errors.
Other applications of CHRG include 
the characterization of the grammar of ancient Egyptian 
hieroglyphs \cite{roskilde_students_ancient_egyptian_grammar_02},
linguistic discourse analysis \cite{christ_dahl_meaning_in_context_context05}, 
and the disambiguation of biological text \cite{dahl_gu_chrg_amb_bio_texts_cslp07}.
An approach similar to CHRG is taken by \citeN{bes_dahl_balanced_2003} 
for the parsing of balanced parentheses in natural language.

%

\subsubsection{Meta-Programming}\label{sec:applications:meta}

\citeN{christiansen_meta_logic_aai00} develops a meta-programming environment,
\textsc{DemoII}, that relies on CHR for its powerful features. Firstly, the
meta-interpreter is made reversible in order to both evaluate queries and
generate programs. Secondly, soundness of negation-as-failure is achieved
through incremental evaluation.

\subsubsection{Testing and Verification}\label{sec:applications:testing}
Another application domain for which CHR has proved useful is
software testing and verification.
\citeN{ribeiro_et_al_security_policy_consistency_rcorp00} present
a CHR-based tool for detecting security policy inconsistencies.
%
\citeN{lotzbeyer_pretschner_autofocus_lpse00} and
\citeN{pretschner_et_al_model-based_testing_sttt04}
propose a model-based testing methodology,
in which test cases are automatically generated from abstract models using CLP and CHR.
They consider the ability to formulate arbitrary test case specifications
by means of CHR to be one of the strengths of their approach.
\citeN{gouraud_gotlieb_javacard_padl06} use a similar approach
for the automatic generation of test cases for the Java Card Virtual Machine (JCVM).
A formal model of the JCVM is automatically translated into CHR,
and the generated CHR program is used to generate test cases.

More of an exploration than testing application is the \textsc{JmmSolve}
framework \cite{schr_JmmSolve_iclp04}. Its purpose is to explore and test the behavior of
declarative memory models for Java, based on the Concurrent Constraint-based
Memory Machines proposal of V.\ Saraswat.



\subsection{Industrial CHR Users} \label{sec:applications:industrial}

Although most CHR systems are essentially still research prototypes,
there are a few systems that can be considered to be robust enough
for industrial application. We give a few examples of companies
that are currently using CHR.

The New-Zealand-based company 
\citeN{sssltd_2008} is one of the main industrial users of CHR.  The company
uses CHR throughout its flagship product the SecuritEase stock broking
system\footnote{\url{http://www.securitease.com/}}. SecuritEase provides
front office (order entry) and back-office (settlement and delivery) functions
for stock brokers in Australia and New Zealand.
Inside SecuritEase CHR is used for: 
\begin{enumerate}
\item implementing
the logic to recognize advantageous market conditions to
automatically place orders in equity markets,
\item translating high-level queries to SQL,
\item
describing complex relationships
between mutually dependent fields on user input screens, and calculating the
consequences of user input actions, and
\item 
realizing a Financial Information eXchange (FIX) server.
\end{enumerate}

The Canadian company Cornerstone Technology
Inc.\footnote{\url{http://www.cornerstonemold.com/}}\ has created an inference engine
for solving and optimizing collections of design constraints, using Prolog and
CHR. The design constraints work together to determine what design
configuration to use, select components from catalogs, compute dimensions for
custom components, and arrange the components into assemblies. The engine
allows for generating, interactive editing, and validating of injection
mould designs. Part of the system is covered by US Patent 7,117,055.

BSSE System and Software Engineering\footnote{\url{http://www.bsse.biz/}},
a German company specializing in the discipline of full automation of software development,
uses K.U.Leuven JCHR for the generation of test data for unit tests.

At the MITRE Corporation\footnote{\url{http://www.mitre.org/}},
CHR is used in the context of optical network design.
It is used to implement constraint-based optimization, network
configuration analysis, and as a tool coordination framework.
\comment{
\jon{
``I use CHR and SWI-prolog for my own work on scheduling tools (at
 www.starlikedesign.com) and in my company work (at MITRE) dealing with
 optical network design. In the latter, I use it to implement
 constraint-based optimization, network configuration analysis, and as a
 rudimentary tool coordination framework.''
Meer uitleg komt (hopelijk) nog. 
}
}

%% file: 08-conclusions.tex
\section{Conclusions}
\label{sec:conclusions}

\comment{
In this survey we gave an overview of recent CHR-related research.
We discussed the logical and operational semantics of CHR and program analysis properties
like confluence, termination, and complexity.  We described the different CHR systems 
and their compilation techniques and optimizations.
We saw some extensions and variants of CHR, and we related CHR to other formalisms.
Finally, we enumerated the main trends in applications of CHR.
}

In this section, we first try to assess to what extent
we have covered the CHR literature in this survey (Section~\ref{ssec:coverage}).
Next, in Section~\ref{ssec:retrospection}, we look back
at the research topics that were mentioned in the previous CHR survey 
\cite{fru_chr_overview_jlp98} as being open issues.
Finally, to conclude this survey, we propose four remaining ``grand challenges''
for CHR researchers.

\input{09-biblio-meta}

\subsection{Retrospection}
\label{ssec:retrospection}

The first CHR survey \cite{fru_chr_overview_jlp98} ended with a list of 
research
topics from the first draft paper on CHR in 1991, noting that most of those
topics were still open in 1998. We re-examine that list:

\begin{description}


\item[$\bullet$  Termination and confluence.]
As seen in Sections~\ref{sec:analysis:confluence}~and~\ref{sec:analysis:termination},
both confluence \cite{abd_fru_meuss_confluence_semantics_csr_constr99,duck_phdthesis05,duck_stuck_sulz_observable_confluence_iclp07,raiser_tacchella_confluence_non_terminating_chr07}
and termination \cite{fru_termination_compulog00,pilozzi_schr_deschreye_termination_wst07,voets_pilozzi_deschreye_termination_chr07}
have received a great deal of attention in the past ten years. 
While substantial results have been obtained for confluence, 
this notion turned out to be rather impractical.
The recent proposal of observable confluence seems to be a first step towards a more practical
notion of confluence. 
In the area of termination analysis there appears to be a much wider scope for improvement. 
Existing work, 
mostly carrying over results of other programming languages,
only works for a limited fraction of programs. 
Propagation rules and logical variables,
part of typical CHR programs, cannot be dealt with. 
An important breakthrough is still ahead of us.

\item[$\bullet$  Negation and entailment of constraints.]
Negation as absence \cite{vanweert_sney_schr_demoen_negation_chr06}
was recently explored (see Section~\ref{ssec:aggregates}), but this has little
relation to the logical negation of constraints.
The topic of entailment is closely related to building solver hierarchies
\cite{duck_stuck_garcia_holz_extending_arbitrary_solvers_with_chr_ppdp03,schr_demoen_duck_stuck_fru_implication_checking_entcs06} 
(see Section~\ref{ssec:hierarchies}).
Both topics are still an important issue today.

\item[$\bullet$  Combination and communication of solvers.]
This topic is again related to solver hierarchies (see the item above),
but also to integration of solvers \cite{abd_fru_integration_lopstr03}
(see Section~\ref{sec:analysis:confluence}).
Another potential approach could be based on a compositional semantics for CHR
\cite{delz_gab_meo_comp_sem_chr_ppdp05} (see Section~\ref{sec:semantics:operational}).
In any case, this is still an important open research topic.

\item[$\bullet$  Correctness w.r.t. specifications, debugging.] 
Although some progress has been made (Section~\ref{sec:implementation:environments}),
these topics are still mostly open.

\item[$\bullet$  Soft constraints with priorities.]
As we discussed in Section~\ref{sec:applications:solvers}, two distinct approaches
were proposed for dealing with (prioritized) soft constraints in CHR.

\item[$\bullet$  Dynamic constraints, removable constraints.]
\citeN{wolf_ea_incremental_adaptation_aai00} designed an incremental adaptation
algorithm that supports dynamic and removable constraints.
An efficient implementation of adaptive CHR exists for Java \cite{wolf_adaptive_chr_java_cp01}.
We discussed this in Sections~\ref{sec:extensions:adaptive}~and~\ref{sec:systems:java} respectively.

\item[$\bullet$  Automatic labeling, variable projection.]
While progress has been made on several accounts of constraint solver support,
these topics have not been addressed yet. The topic of projection,
the elimination of existentially quantified variables,
is particularly challenging to generalize to arbitrary CHR solvers.
Any progress would be highly significant.

\item[$\bullet$  Partial evaluation.]
Section~\ref{sec:implementation:specialization_transformation}
mentions recent work in this area
\cite{fru_specialization_lopstr04,tacchella_gabbrielli_meo_unfolding_ppdp07,sarnastarosta_schr_indexing_techrep07}.
For now, it is clear that the multi-headedness of CHR makes 
a straightforward application of partial evaluation
techniques for conventional languages to CHR programs nearly impossible.
Further investigation of techniques tailored towards CHR are necessary to make
substantial improvements.

\item[$\bullet$  Abstract interpretation.]
A general framework for abstract interpretation in the context of CHR was
proposed \cite{schr_stuck_duck_ai_chr_ppdp05} (see
Section~\ref{sec:implementation:optimization}). However, only a limited number
of analyses have been formulated in the framework so far. Moreover, little
information is present in a CHR program on its own. In order to make analysis
results more accurate, we require an analysis framework that encompasses both
CHR and its host language. Such a framework should prove to be beneficial to
the accuracy of the analyses for both languages.
\end{description}

\subsection{Grand Challenges}

Much progress has been made in the last ten years and many of the open problems
have been resolved. However, a few difficult questions are still unresolved,
and in the meantime many more problems have become apparent. 

In our view the following four topics are {\em grand challenges} that must be
addressed by the CHR research community in the next decade. These four grand
challenges are not only of technical interest, they are also vital for the
further adoption of the CHR community and user-base.

\begin{enumerate}

\item
{\bf  Programming environments and tools.}
If measured by current standards, which dictate that a language is only as good
as its tools, CHR is a poor language indeed.  While several strong theoretical
results have been obtained in the field of program analysis for CHR, little (if
any) effort has been made to embody these results into a practical tool for
day-to-day programming. For example, programmers have to manually check for
confluence, and, in the case of non-confluence, complete their solvers by hand.


\item
{\bf  Execution control.}
Compared to the refined operational semantics,
the rule priorities of the \chrrp{} semantics are a step in the realization
of Kowalski's slogan ``Algorithm = Logic + Control'' in the context of CHR.
However, there are still many challenges in finding satisfactory ways to
allow programmers to fine-tune the execution strategy of their programs.

\item
{\bf Parallelism, concurrency.}
Recent theoretical work \cite{fru_parallel_union_find_iclp05,meister_preflow_push_wlp06} 
confirms CHR's inherent aptness for parallel programming.
Truly leveraging the full power of current and future multi-core processors through CHR,
however, requires practical, efficient, concurrent implementations.
Currently, these implementations are still in early stages 
(cf.\ Section~\ref{sec:implementation:systems:fp} for a discussion 
of some early Haskell-based prototypes).
Many important problems are still to be researched in this domain, 
from language features and semantics, 
to analysis, implementation, and optimization.

\item
{\bf  Scaling to industrial applications.}
Strong theoretical results have been obtained concerning the performance of CHR (cf.\ Section~\ref{sec:analysis:complexity:meta}), 
and these have also been reflected in the actual runtimes of CHR programs.
However, CHR is still at least one or two orders of magnitude slower than most
conventional programming languages and constraint solvers. 
This becomes particularly apparent 
for CHR applications that surpass the toy research programs of 10 lines:
industrial applications for instance, such as those mentioned in Section~\ref{sec:applications:industrial},
easily count 100 to 1000 lines. 
The refined semantics compilation scheme (see Section~\ref{sec:implementation}) 
was not designed or benchmarked with such program sizes in mind.
Some potential scalability aspects are:
\begin{itemize}
\item huge constraint stores that have to be persistent and/or distributed;
\item (dynamic) optimizations, also for variants and extensions of CHR;
\item incremental compilation, run-time rule assertion, reflection;
\item higher-order / meta-programming.
\end{itemize}        
These and other aspects must be investigated to achieve further industrial adoption.
\end{enumerate}


%% file: 09-biblio-meta.tex
\subsection{Survey Coverage and Bibliographic Meta-Information}
\label{ssec:coverage}

\begin{figure}
        \centering
	\includegraphics[width=\textwidth]{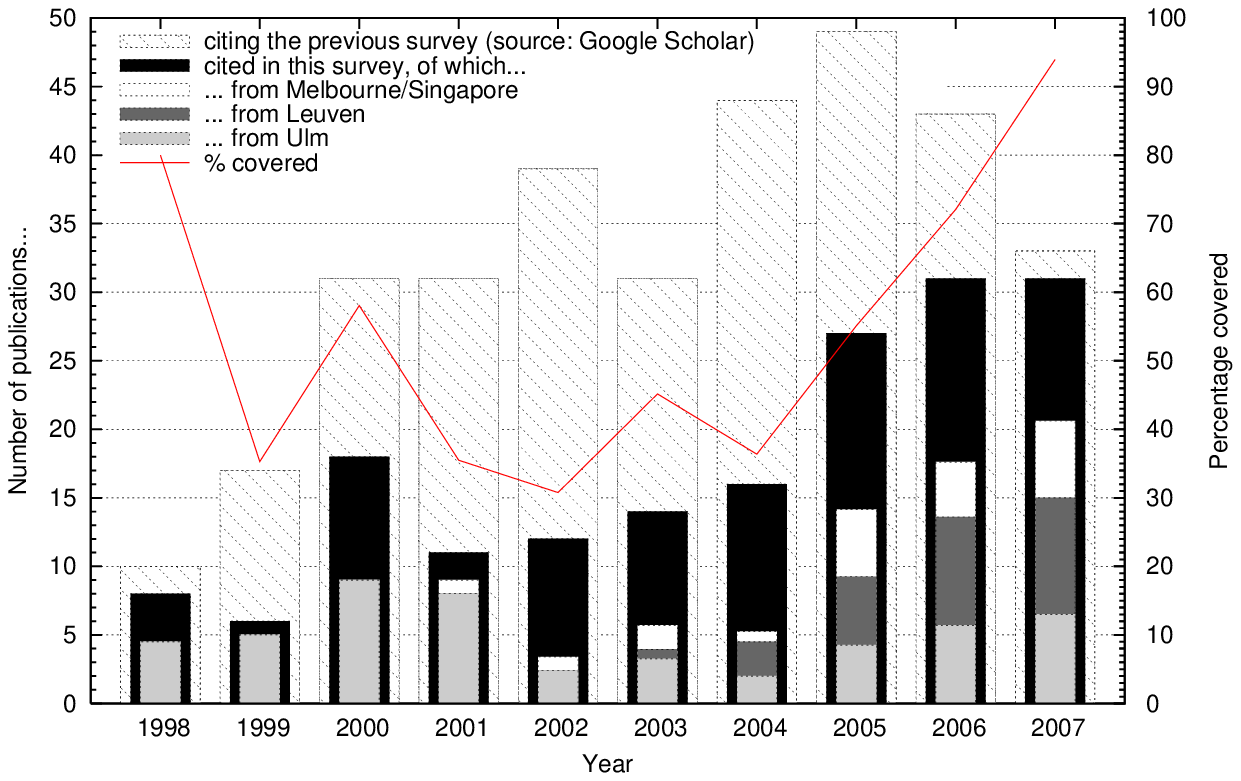}
\caption{Estimation of the number of CHR-related publications 
and the number of CHR-related publications cited in this survey (and their origin).
}
\label{fig:nbpapers}
\end{figure}

This survey cites 182 publications 
related to CHR.
For convenience, we define the total number of CHR-related publications 
as the number of publications that cite \cite{fru_chr_overview_jlp98},
according to Google Scholar (with some manual corrections for errors in the Google Scholar result list).
Figure~\ref{fig:nbpapers} compares the number of publications we cite in this survey
with the total number of CHR-related publications since 1998. 
Globally, we cite roughly half of the CHR-related publications; the remaining half
\comment{ -- was:
are mostly either preliminary versions of cited publications, or about an application
of CHR which 
we did not include in Section~\ref{sec:applications}.
}%
are preliminary versions of cited publications
and papers that use CHR or refer to CHR in only a relatively minor way.

Figure~\ref{fig:authors} shows how CHR authors have collaborated. This graph is derived
from the joint authorships of papers cited in this survey.



\begin{figure}%
\hspace*{-0.05\textwidth}%
\includegraphics[width=1.1\textwidth]{chr_authors}
        \caption{CHR author collaboration graph.
        Two authors are connected if they have co-authored a CHR-related paper cited in this survey.
        The number of co-authored papers is reflected in the edge thickness.
	}\label{fig:authors}
\end{figure}

%% file: chr_survey.bbl
\begin{thebibliography}{}

\bibitem[\protect\citeauthoryear{Abdennadher}{Abdennadher}{2000}]{abd_chr_disj%
unction_rcorp00}
{\sc Abdennadher, S.} 2000.
\newblock A language for experimenting with declarative paradigms.
\newblock In {\em RCoRP '00(bis): Proc.\ 2nd Workshop on Rule-Based Constraint
  Reasoning and Programming}, {T.~Fr{\"u}hwirth} {et~al\mbox{.}}, Eds.

\bibitem[\protect\citeauthoryear{Abdennadher}{Abdennadher}{2001}]{abd_habilita%
tion_2001}
{\sc Abdennadher, S.} 2001.
\newblock Rule-based constraint programming: Theory and practice.
\newblock Habilitationsschrift.
\newblock Institute of Computer Science, LMU, Munich, Germany.

\bibitem[\protect\citeauthoryear{Abdennadher and Christiansen}{Abdennadher and
  Christiansen}{2000}]{abd_christ_abduction_fqas00}
{\sc Abdennadher, S.} {\sc and} {\sc Christiansen, H.} 2000.
\newblock An experimental {CLP} platform for integrity constraints and
  abduction.
\newblock In {\em FQAS '00: Proc.\ 4th Intl.\ Conf.\ Flexible Query Answering
  Systems}. Springer, 141--152.

\bibitem[\protect\citeauthoryear{Abdennadher and Fr\"{u}hwirth}{Abdennadher and
  Fr\"{u}hwirth}{1998}]{abd_fru_completion_cp98}
{\sc Abdennadher, S.} {\sc and} {\sc Fr\"{u}hwirth, T.} 1998.
\newblock On completion of {C}onstraint {H}andling {R}ules.
\newblock In {\em CP '98}, {M.~J. Maher} {and} {J.-F. Puget}, Eds. LNCS, vol.
  1520. Springer, 25--39.

\bibitem[\protect\citeauthoryear{Abdennadher and Fr{\"u}hwirth}{Abdennadher and
  Fr{\"u}hwirth}{1999}]{abd_fru_equivalence_cp99}
{\sc Abdennadher, S.} {\sc and} {\sc Fr{\"u}hwirth, T.} 1999.
\newblock Operational equivalence of {CHR} programs and constraints.
\newblock In {\em CP '99}, {J.~Jaffar}, Ed. LNCS, vol. 1713. Springer, 43--57.

\bibitem[\protect\citeauthoryear{Abdennadher and Fr{\"u}hwirth}{Abdennadher and
  Fr{\"u}hwirth}{2004}]{abd_fru_integration_lopstr03}
{\sc Abdennadher, S.} {\sc and} {\sc Fr{\"u}hwirth, T.} 2004.
\newblock Integration and optimization of rule-based constraint solvers.
\newblock In {\em LOPSTR '03}, {M.~Bruynooghe}, Ed. LNCS, vol. 3018. Springer,
  198--213.

\bibitem[\protect\citeauthoryear{Abdennadher, Fr\"uhwirth, and
  Holzbaur}{Abdennadher, Fr\"uhwirth et~al\mbox{.}}{2005}]{tplp05}
{\sc Abdennadher, S.}, {\sc Fr\"uhwirth, T.}, {\sc and} {\sc Holzbaur, C.},
  Eds. 2005.
\newblock {\em Special Issue on Constraint Handling Rules}. Theory and Practice
  of Logic Programming, vol. 5(4--5).

\bibitem[\protect\citeauthoryear{Abdennadher, Fr{\"u}hwirth, and
  Meuss}{Abdennadher, Fr{\"u}hwirth
  et~al\mbox{.}}{1999}]{abd_fru_meuss_confluence_semantics_csr_constr99}
{\sc Abdennadher, S.}, {\sc Fr{\"u}hwirth, T.}, {\sc and} {\sc Meuss, H.} 1999.
\newblock Confluence and semantics of constraint simplification rules.
\newblock {\em Constraints\/}~{\em 4,\/}~2, 133--165.

\bibitem[\protect\citeauthoryear{Abdennadher, Kr{\"a}mer, Saft, and
  Schmau\ss{}}{Abdennadher, Kr{\"a}mer
  et~al\mbox{.}}{2002}]{abd_kr_saft_schm_jack_wflp01:entcs02}
{\sc Abdennadher, S.}, {\sc Kr{\"a}mer, E.}, {\sc Saft, M.}, {\sc and} {\sc
  Schmau\ss{}, M.} 2002.
\newblock {JACK}: A {J}ava {C}onstraint {K}it.
\newblock In {\em WFLP '01: Proc.\ 10th Intl.\ Workshop on Functional and
  (Constraint) Logic Programming, Selected Papers}, {M.~Hanus}, Ed. ENTCS,
  vol.~64. Elsevier, 1--17.
\newblock See also {\tt http://pms.ifi.lmu.de/software/jack/}.

\bibitem[\protect\citeauthoryear{Abdennadher and Marte}{Abdennadher and
  Marte}{2000}]{abd_marte_timetabling_chr_aai00}
{\sc Abdennadher, S.} {\sc and} {\sc Marte, M.} 2000.
\newblock University course timetabling using {C}onstraint {H}andling {R}ules.
\newblock In \citeN{aai00}, 311--325.

\bibitem[\protect\citeauthoryear{Abdennadher, Olama, Salem, and
  Thabet}{Abdennadher, Olama et~al\mbox{.}}{2006}]{abd_et_al_arm_lopstr06}
{\sc Abdennadher, S.}, {\sc Olama, A.}, {\sc Salem, N.}, {\sc and} {\sc Thabet,
  A.} 2006.
\newblock {ARM}: {A}utomatic {R}ule {M}iner.
\newblock In {\em LOPSTR '06, Revised Selected Papers}. LNCS, vol. 4407.
  Springer.

\bibitem[\protect\citeauthoryear{Abdennadher and Rigotti}{Abdennadher and
  Rigotti}{2004}]{abd_rigo_automatic_gen_of_solvers_tocl04}
{\sc Abdennadher, S.} {\sc and} {\sc Rigotti, C.} 2004.
\newblock Automatic generation of rule-based constraint solvers over finite
  domains.
\newblock {\em ACM TOCL\/}~{\em 5,\/}~2, 177--205.

\bibitem[\protect\citeauthoryear{Abdennadher and Rigotti}{Abdennadher and
  Rigotti}{2005}]{abd_rigotti_automatic_generation_tplp05}
{\sc Abdennadher, S.} {\sc and} {\sc Rigotti, C.} 2005.
\newblock Automatic generation of {CHR} constraint solvers.
\newblock In \citeN{tplp05}, 403--418.

\bibitem[\protect\citeauthoryear{Abdennadher and Saft}{Abdennadher and
  Saft}{2001}]{abd_saft_jack:visualchr_wlpe01}
{\sc Abdennadher, S.} {\sc and} {\sc Saft, M.} 2001.
\newblock A visualization tool for {C}onstraint {H}andling {R}ules.
\newblock In {\em WLPE '01}, {A.~Kusalik}, Ed.

\bibitem[\protect\citeauthoryear{Abdennadher, Saft, and Will}{Abdennadher, Saft
  et~al\mbox{.}}{2000}]{abd_saft_will_classroom_assignment_paclp00}
{\sc Abdennadher, S.}, {\sc Saft, M.}, {\sc and} {\sc Will, S.} 2000.
\newblock Classroom assignment using constraint logic programming.
\newblock In {\em PACLP '00: Proc.\ 2nd Intl.\ Conf.\ and Exhibition on
  Practical Application of Constraint Technologies and Logic Programming}.

\bibitem[\protect\citeauthoryear{Abdennadher and Sch{\"u}tz}{Abdennadher and
  Sch{\"u}tz}{1998}]{abd_query_lang_fqas98}
{\sc Abdennadher, S.} {\sc and} {\sc Sch{\"u}tz, H.} 1998.
\newblock {CHR$^\vee$}, a flexible query language.
\newblock In {\em FQAS '98: Proc.\ 3rd Intl.\ Conf.\ on Flexible Query
  Answering Systems}, {T.~Andreasen}, {H.~Christiansen}, {and} {H.~Larsen},
  Eds. LNAI, vol. 1495. Springer, 1--14.

\bibitem[\protect\citeauthoryear{Abdennadher and Sobhi}{Abdennadher and
  Sobhi}{2008}]{abd_sobhi_generation_combined_lopstr07}
{\sc Abdennadher, S.} {\sc and} {\sc Sobhi, I.} 2008.
\newblock Generation of rule-based constraint solvers: Combined approach.
\newblock In {\em LOPSTR '07, Revised Selected Papers}, {A.~King}, Ed. LNCS,
  vol. 4915.

\bibitem[\protect\citeauthoryear{Aguilar-Solis and Dahl}{Aguilar-Solis and
  Dahl}{2004}]{as_dahl_coordination_iberamia04}
{\sc Aguilar-Solis, D.} {\sc and} {\sc Dahl, V.} 2004.
\newblock Coordination revisited -- a {C}onstraint {H}andling {R}ule approach.
\newblock In {\em IBERAMIA '04: Proc.\ 9th Ibero-American Conf.\ on AI}. LNCS,
  vol. 3315. 315--324.

\bibitem[\protect\citeauthoryear{Alberti, Chesani, Gavanelli, and
  Lamma}{Alberti, Chesani
  et~al\mbox{.}}{2005}]{alberti_et_al_sys_gen_conf_hypoth_wclp05}
{\sc Alberti, M.}, {\sc Chesani, F.}, {\sc Gavanelli, M.}, {\sc and} {\sc
  Lamma, E.} 2005.
\newblock The {CHR}-based implementation of a system for generation and
  confirmation of hypotheses.
\newblock In \citeN{pwclp05}, 111--122.

\bibitem[\protect\citeauthoryear{Alberti, Daolio, Torroni, Gavanelli, Lamma,
  and Mello}{Alberti, Daolio
  et~al\mbox{.}}{2004}]{alberti_et_al_agent_interaction_protocols_sac04}
{\sc Alberti, M.}, {\sc Daolio, D.}, {\sc Torroni, P.}, {\sc Gavanelli, M.},
  {\sc Lamma, E.}, {\sc and} {\sc Mello, P.} 2004.
\newblock Specification and verification of agent interaction protocols in a
  logic-based system.
\newblock In {\em SAC '04: Proc. 19th ACM Symp. Applied Computing}, {H.~Haddad}
  {et~al\mbox{.}}, Eds. ACM Press, 72--78.

\bibitem[\protect\citeauthoryear{Alberti, Gavanelli, Lamma, Chesani, Mello, and
  Torroni}{Alberti, Gavanelli
  et~al\mbox{.}}{2006}]{alberti_et_al_compliance_agents_tool_aai06}
{\sc Alberti, M.}, {\sc Gavanelli, M.}, {\sc Lamma, E.}, {\sc Chesani, F.},
  {\sc Mello, P.}, {\sc and} {\sc Torroni, P.} 2006.
\newblock Compliance verification of agent interaction: a logic-based software
  tool.
\newblock {\em Applied Artificial Intelligence\/}~{\em 20,\/}~2--4, 133--157.

\bibitem[\protect\citeauthoryear{Alberti, Gavanelli, Lamma, Mello, and
  Milano}{Alberti, Gavanelli
  et~al\mbox{.}}{2005}]{alberti_et_al_chr_implementation_of_arc_consistency_tp%
lp05}
{\sc Alberti, M.}, {\sc Gavanelli, M.}, {\sc Lamma, E.}, {\sc Mello, P.}, {\sc
  and} {\sc Milano, M.} 2005.
\newblock A {CHR}-based implementation of known arc-consistency.
\newblock In \citeN{tplp05}, 419--440.

\bibitem[\protect\citeauthoryear{Alberti, Gavanelli, Lamma, Mello, and
  Torroni}{Alberti, Gavanelli
  et~al\mbox{.}}{2004}]{alberti_et_al_social_integrity_constraints_lcmas03}
{\sc Alberti, M.}, {\sc Gavanelli, M.}, {\sc Lamma, E.}, {\sc Mello, P.}, {\sc
  and} {\sc Torroni, P.} 2004.
\newblock Specification and verification of agent interaction using social
  integrity constraints.
\newblock In {\em LCMAS'03: Logic and Communication in Multi-Agent Systems}.
  ENTCS, vol. 85(2). Elsevier, 94--116.

\bibitem[\protect\citeauthoryear{Alves and Florido}{Alves and
  Florido}{2002}]{alves_florido_type_inference_chr_wflp01:entcs02}
{\sc Alves, S.} {\sc and} {\sc Florido, M.} 2002.
\newblock Type inference using {C}onstraint {H}andling {R}ules.
\newblock In {\em WFLP '01: Proc.\ 10th Intl.\ Workshop on Functional and
  (Constraint) Logic Programming, Selected Papers}, {M.~Hanus}, Ed. ENTCS,
  vol.~64. Elsevier, 56--72.

\bibitem[\protect\citeauthoryear{Apt and Monfroy}{Apt and
  Monfroy}{2001}]{apt_monfroy_cp_viewed_as_rulebased_tplp01}
{\sc Apt, K.~R.} {\sc and} {\sc Monfroy, E.} 2001.
\newblock Constraint programming viewed as rule-based programming.
\newblock {\em TPLP\/}~{\em 1,\/}~6, 713--750.

\bibitem[\protect\citeauthoryear{Badea, Tilivea, and Hotaran}{Badea, Tilivea
  et~al\mbox{.}}{2004}]{badea_et_al_semantic_web_reasoning_ppswr04}
{\sc Badea, L.}, {\sc Tilivea, D.}, {\sc and} {\sc Hotaran, A.} 2004.
\newblock {Semantic Web Reasoning for Ontology-Based Integration of Resources}.
\newblock {\em PPSWR '04: Proc.\ 2nd Intl.\ Workshop on Principles And Practice
  Of Semantic Web Reasoning\/}~{\em 3208}, 61--75.

\bibitem[\protect\citeauthoryear{Barranco-Mendoza}{Barranco-Mendoza}{2005}]{al%
ma_thesis05}
{\sc Barranco-Mendoza, A.} 2005.
\newblock Stochastic and heuristic modelling for analysis of the growth of
  pre-invasive lesions and for a multidisciplinary approach to early cancer
  diagnosis.
\newblock Ph.D. thesis, Simon Fraser University, Canada.

\bibitem[\protect\citeauthoryear{Bavarian and Dahl}{Bavarian and
  Dahl}{2006}]{bavarian_dahl_bio_seq_analysis_jucs06}
{\sc Bavarian, M.} {\sc and} {\sc Dahl, V.} 2006.
\newblock Constraint based methods for biological sequence analysis.
\newblock {\em J. UCS\/}~{\em 12,\/}~11, 1500--1520.

\bibitem[\protect\citeauthoryear{B\`es and Dahl}{B\`es and
  Dahl}{2003}]{bes_dahl_balanced_2003}
{\sc B\`es, G.~G.} {\sc and} {\sc Dahl, V.} 2003.
\newblock Balanced parentheses in {NL} texts: a useful cue in the
  syntax/semantics interface.
\newblock In {\em Proc.\ Lorraine-Saarland Workshop on Prospects and Advances
  in the Syntax/Semantics Interface}.
\newblock Poster Paper.

\bibitem[\protect\citeauthoryear{Betz}{Betz}{2007}]{betz_petri_nets_chr07}
{\sc Betz, H.} 2007.
\newblock Relating coloured {Petri} nets to {C}onstraint {H}andling {R}ules.
\newblock In \citeN{pchr07}, 33--47.

\bibitem[\protect\citeauthoryear{Betz and Fr{\"u}hwirth}{Betz and
  Fr{\"u}hwirth}{2005}]{betz_fru_linear_logic_semantics_cp05}
{\sc Betz, H.} {\sc and} {\sc Fr{\"u}hwirth, T.} 2005.
\newblock A linear-logic semantics for {C}onstraint {H}andling {R}ules.
\newblock In {\em CP '05}. LNCS, vol. 3709. Springer, 137--151.

\bibitem[\protect\citeauthoryear{Betz and Fr{\"u}hwirth}{Betz and
  Fr{\"u}hwirth}{2007}]{betz_fru_linear_logic_chr_disj_chr07}
{\sc Betz, H.} {\sc and} {\sc Fr{\"u}hwirth, T.} 2007.
\newblock A linear-logic semantics for {C}onstraint {H}andling {R}ules with
  disjunction.
\newblock In \citeN{pchr07}, 17--31.

\bibitem[\protect\citeauthoryear{Bistarelli, Fr{\"u}hwirth, Marte, and
  Rossi}{Bistarelli, Fr{\"u}hwirth
  et~al\mbox{.}}{2004}]{bistarelli_fru_marte_rossi_soft_constraint_propagation%
_CI04}
{\sc Bistarelli, S.}, {\sc Fr{\"u}hwirth, T.}, {\sc Marte, M.}, {\sc and} {\sc
  Rossi, F.} 2004.
\newblock Soft constraint propagation and solving in {C}onstraint {H}andling
  {R}ules.
\newblock {\em Computational Intelligence: Special Issue on Preferences in AI
  and CP\/}~{\em 20,\/}~2 (May), 287--307.

\bibitem[\protect\citeauthoryear{Boespflug}{Boespflug}{2007}]{boespflug_taichi%
_monadreader07}
{\sc Boespflug, M.} 2007.
\newblock {TaiChi:how to check your types with serenity}.
\newblock {\em The Monad.Reader\/}~{\em 9}, 17--31.

\bibitem[\protect\citeauthoryear{Bouissou}{Bouissou}{2004}]{bouissou_chr_for_s%
ilcc_2004}
{\sc Bouissou, O.} 2004.
\newblock A {CHR} library for {SiLCC}.
\newblock Diplomathesis.
\newblock Technical University of Berlin, Germany.

\bibitem[\protect\citeauthoryear{Brand}{Brand}{2002}]{brand_redundant_rules_cs%
clp02}
{\sc Brand, S.} 2002.
\newblock A note on redundant rules in rule-based constraint programming.
\newblock In {\em Joint ERCIM/CologNet Intl.\ Workshop on Constraint Solving
  and Constraint Logic Programming, Selected papers}. LNCS, vol. 2627.
  Springer, 279--336.

\bibitem[\protect\citeauthoryear{Brand and Monfroy}{Brand and
  Monfroy}{2003}]{brand_monfroy_generation_propagation_rules_entcs03}
{\sc Brand, S.} {\sc and} {\sc Monfroy, E.} 2003.
\newblock Deductive generation of constraint propagation rules.
\newblock In {\em RULE '03: 4th Intl. Workshop on Rule-Based Programming},
  {G.~Vidal}, Ed. ENTCS, vol. 86(2). Elsevier, 45--60.

\bibitem[\protect\citeauthoryear{Bressan and Goh}{Bressan and
  Goh}{1998}]{bressan_goh_coin_fqas98}
{\sc Bressan, S.} {\sc and} {\sc Goh, C.~H.} 1998.
\newblock Answering queries in context.
\newblock In {\em FQAS '98: Proc.\ 3rd Intl.\ Conf.\ on Flexible Query
  Answering Systems}, {T.~Andreasen}, {H.~Christiansen}, {and} {H.~Larsen},
  Eds. LNAI, vol. 1495. Springer, 68--82.

\bibitem[\protect\citeauthoryear{Cabedo and Escrig}{Cabedo and
  Escrig}{2003}]{museros_escrig_modeling_motion_jucs03}
{\sc Cabedo, L.~M.} {\sc and} {\sc Escrig, M.~T.} 2003.
\newblock Modeling motion by the integration of topology and time.
\newblock {\em J. UCS\/}~{\em 9,\/}~9, 1096--1122.

\bibitem[\protect\citeauthoryear{Chin, Craciun, Khoo, and Popeea}{Chin, Craciun
  et~al\mbox{.}}{2006}]{chin_et_al_florin_sigplan06}
{\sc Chin, W.-N.}, {\sc Craciun, F.}, {\sc Khoo, S.-C.}, {\sc and} {\sc Popeea,
  C.} 2006.
\newblock A flow-based approach for variant parametric types.
\newblock {\em SIGPLAN Not.\/}~{\em 41,\/}~10, 273--290.

\bibitem[\protect\citeauthoryear{Chin, Sulzmann, and Wang}{Chin, Sulzmann
  et~al\mbox{.}}{2003}]{chin_sulzmann_wang_haskell_chr_03}
{\sc Chin, W.-N.}, {\sc Sulzmann, M.}, {\sc and} {\sc Wang, M.} 2003.
\newblock A type-safe embedding of {C}onstraint {H}andling {R}ules into
  {H}askell.
\newblock Honors Thesis.
\newblock School of Computing, National University of Singapore.

\bibitem[\protect\citeauthoryear{Christiansen}{Christiansen}{2005}]{christ_chr%
_grammars_tplp05}
{\sc Christiansen, H.} 2005.
\newblock {CHR} grammars.
\newblock In \citeN{tplp05}, 467--501.

\bibitem[\protect\citeauthoryear{Christiansen}{Christiansen}{2006}]{christians%
en_clima06}
{\sc Christiansen, H.} 2006.
\newblock On the implementation of global abduction.
\newblock In {\em CLIMA '06: 7th Intl.\ Workshop on Computational Logic in
  Multi-Agent Systems -- Revised, Selected and Invited Papers}, {K.~Inoue},
  {K.~Satoh}, {and} {F.~Toni}, Eds. LNCS, vol. 4371. Springer, 226--245.

\bibitem[\protect\citeauthoryear{Christiansen and Dahl}{Christiansen and
  Dahl}{2003}]{christ_dahl_diagnosis_and_repair_ijait03}
{\sc Christiansen, H.} {\sc and} {\sc Dahl, V.} 2003.
\newblock Logic grammars for diagnosis and repair.
\newblock {\em Intl.\ J. Artificial Intelligence Tools\/}~{\em 12,\/}~3,
  227--248.

\bibitem[\protect\citeauthoryear{Christiansen and Dahl}{Christiansen and
  Dahl}{2005a}]{christiansen_dahl_hyprolog_iclp05}
{\sc Christiansen, H.} {\sc and} {\sc Dahl, V.} 2005a.
\newblock {HYPROLOG}: A new logic programming language with assumptions and
  abduction.
\newblock In \citeN{piclp05}, 159--173.

\bibitem[\protect\citeauthoryear{Christiansen and Dahl}{Christiansen and
  Dahl}{2005b}]{christ_dahl_meaning_in_context_context05}
{\sc Christiansen, H.} {\sc and} {\sc Dahl, V.} 2005b.
\newblock Meaning in context.
\newblock In {\em CONTEXT '05: Proc.\ 4th Intl.\ and Interdisciplinary Conf.\
  Modeling and Using Context}. LNAI, vol. 3554. Springer, 97--111.

\bibitem[\protect\citeauthoryear{Christiansen and Have}{Christiansen and
  Have}{2007}]{christ_have_use_cases_to_uml_ranlp07}
{\sc Christiansen, H.} {\sc and} {\sc Have, C.~T.} 2007.
\newblock From use cases to {UML} class diagrams using logic grammars and
  constraints.
\newblock In {\em RANLP '07: Proc.\ Intl.\ Conf.\ Recent Adv.\ Nat.\ Lang.\
  Processing}. 128--132.

\bibitem[\protect\citeauthoryear{Christiansen and Martinenghi}{Christiansen and
  Martinenghi}{2000}]{christiansen_meta_logic_aai00}
{\sc Christiansen, H.} {\sc and} {\sc Martinenghi, D.} 2000.
\newblock Symbolic constraints for meta-logic programming.
\newblock In \citeN{aai00}, 345--367.

\bibitem[\protect\citeauthoryear{Coquery and Fages}{Coquery and
  Fages}{2003}]{coquery_fages_type_system_wlpe03}
{\sc Coquery, E.} {\sc and} {\sc Fages, F.} 2003.
\newblock {TCLP}: A type checker for {CLP}($\mathcal{X}$).
\newblock In {\em WLPE '03}, {F.~Mesnard} {and} {A.~Serebrenik}, Eds.
  K.U.Leuven, Dept.\ Comp.\ Sc., Technical report CW 371. 17--30.

\bibitem[\protect\citeauthoryear{Coquery and Fages}{Coquery and
  Fages}{2005}]{coquery_fages_type_system_chr05}
{\sc Coquery, E.} {\sc and} {\sc Fages, F.} 2005.
\newblock A type system for {CHR}.
\newblock In \citeN{pchr05}, 19--33.

\bibitem[\protect\citeauthoryear{Dahl}{Dahl}{2004}]{dahl_abductive_dependencie%
s_cslp04}
{\sc Dahl, V.} 2004.
\newblock An abductive treatment of long distance dependencies in {CHR}.
\newblock In {\em CSLP '04: Proc.\ First Intl.\ Workshop on Constraint Solving
  and Language Processing}. LNCS, vol. 3438. Springer, 17--31.
\newblock Invited Paper.

\bibitem[\protect\citeauthoryear{Dahl and Blache}{Dahl and
  Blache}{2005}]{dahl_blache_extracting_phrases_cslp05}
{\sc Dahl, V.} {\sc and} {\sc Blache, P.} 2005.
\newblock Extracting selected phrases through constraint satisfaction.
\newblock In {\em Proc.\ 2nd Intl.\ Workshop on Constraint Solving and Language
  Processing}.

\bibitem[\protect\citeauthoryear{Dahl and Gu}{Dahl and
  Gu}{2006}]{dahl_gu_property_gram_biomed_iclp06}
{\sc Dahl, V.} {\sc and} {\sc Gu, B.} 2006.
\newblock Semantic property grammars for knowledge extraction from biomedical
  text.
\newblock In \citeN{piclp06}, 442--443.
\newblock Poster Paper.

\bibitem[\protect\citeauthoryear{Dahl and Gu}{Dahl and
  Gu}{2007}]{dahl_gu_chrg_amb_bio_texts_cslp07}
{\sc Dahl, V.} {\sc and} {\sc Gu, B.} 2007.
\newblock A {CHRG} analysis of ambiguity in biological texts.
\newblock In {\em CSLP '07: Proc.\ 4th Intl.\ Workshop on Constraints and
  Language Processing}.
\newblock Extended Abstract.

\bibitem[\protect\citeauthoryear{Dahl and Niemel\"a}{Dahl and
  Niemel\"a}{2007}]{piclp07}
{\sc Dahl, V.} {\sc and} {\sc Niemel\"a, I.}, Eds. 2007.
\newblock {\em ICLP '07: Proc.\ 23rd Intl.\ Conf.\ Logic Programming}. LNCS,
  vol. 4670. Springer.

\bibitem[\protect\citeauthoryear{Dahl and Voll}{Dahl and
  Voll}{2004}]{dahl_voll_concept_formation_rules_nlucs04}
{\sc Dahl, V.} {\sc and} {\sc Voll, K.} 2004.
\newblock Concept formation rules: An executable cognitive model of knowledge
  construction.
\newblock In {\em NLUCS '04: Proc.\ First Intl.\ Workshop on Natural Language
  Understanding and Cognitive Sciences}.

\bibitem[\protect\citeauthoryear{{De Koninck}, Schrijvers, and Demoen}{{De
  Koninck}, Schrijvers
  et~al\mbox{.}}{2006a}]{dekoninck_schr_demoen_inclpr_wlp06}
{\sc {De Koninck}, L.}, {\sc Schrijvers, T.}, {\sc and} {\sc Demoen, B.} 2006a.
\newblock {INCLP($\mathbb{R}$)} - {I}nterval-based nonlinear constraint logic
  programming over the reals.
\newblock In \citeN{pwlp06}, 91--100.

\bibitem[\protect\citeauthoryear{{De Koninck}, Schrijvers, and Demoen}{{De
  Koninck}, Schrijvers
  et~al\mbox{.}}{2006b}]{dekoninck_schr_demoen_search_chr06}
{\sc {De Koninck}, L.}, {\sc Schrijvers, T.}, {\sc and} {\sc Demoen, B.} 2006b.
\newblock Search strategies in {CHR(Prolog)}.
\newblock In \citeN{pchr06}, 109--123.

\bibitem[\protect\citeauthoryear{{De Koninck}, Schrijvers, and Demoen}{{De
  Koninck}, Schrijvers
  et~al\mbox{.}}{2007a}]{dekoninck_schr_demoen_la-chr_iclp07}
{\sc {De Koninck}, L.}, {\sc Schrijvers, T.}, {\sc and} {\sc Demoen, B.} 2007a.
\newblock The correspondence between the {L}ogical {A}lgorithms language and
  {CHR}.
\newblock In \citeN{piclp07}, 209--223.

\bibitem[\protect\citeauthoryear{{De Koninck}, Schrijvers, and Demoen}{{De
  Koninck}, Schrijvers
  et~al\mbox{.}}{2007b}]{dekoninck_schr_demoen_chrrp_ppdp07}
{\sc {De Koninck}, L.}, {\sc Schrijvers, T.}, {\sc and} {\sc Demoen, B.} 2007b.
\newblock User-definable rule priorities for {CHR}.
\newblock In {\em PPDP '07}, {M.~Leuschel} {and} {A.~Podelski}, Eds. ACM Press,
  25--36.

\bibitem[\protect\citeauthoryear{{De Koninck} and Sneyers}{{De Koninck} and
  Sneyers}{2007}]{dekoninck_sney_join_ordering_chr07}
{\sc {De Koninck}, L.} {\sc and} {\sc Sneyers, J.} 2007.
\newblock Join ordering for {C}onstraint {H}andling {R}ules.
\newblock In \citeN{pchr07}, 107--121.

\bibitem[\protect\citeauthoryear{{De Koninck}, Stuckey, and Duck}{{De Koninck},
  Stuckey et~al\mbox{.}}{2008}]{dekoninck_stuck_duck_compiling-chrrp_08}
{\sc {De Koninck}, L.}, {\sc Stuckey, P.~J.}, {\sc and} {\sc Duck, G.~J.} 2008.
\newblock Optimizing compilation of {CHR} with rule priorities.
\newblock In {\em Proc.\ 9th Intl.\ Symp.\ Functional and Logic Programming},
  {J.~Garrigue} {and} {M.~Hermenegildo}, Eds. LNCS, vol. 4989. Springer,
  32--47.

\bibitem[\protect\citeauthoryear{Delzanno, Gabbrielli, and Meo}{Delzanno,
  Gabbrielli et~al\mbox{.}}{2005}]{delz_gab_meo_comp_sem_chr_ppdp05}
{\sc Delzanno, G.}, {\sc Gabbrielli, M.}, {\sc and} {\sc Meo, M.~C.} 2005.
\newblock A compositional semantics for {CHR}.
\newblock In {\em PPDP '05}, {P.~Barahona} {and} {A.~Felty}, Eds. ACM Press,
  209--217.

\bibitem[\protect\citeauthoryear{Demoen and Lifschitz}{Demoen and
  Lifschitz}{2004}]{piclp04}
{\sc Demoen, B.} {\sc and} {\sc Lifschitz, V.}, Eds. 2004.
\newblock {\em ICLP '04: Proc.\ 20th Intl.\ Conf.\ Logic Programming}. LNCS,
  vol. 3132. Springer.

\bibitem[\protect\citeauthoryear{Djelloul, Dao, and Fr{\"u}hwirth}{Djelloul,
  Dao
  et~al\mbox{.}}{2007}]{djelloul_dao_fru_1st_order_extension_prolog_unificatio%
n_sac07}
{\sc Djelloul, K.}, {\sc Dao, T.-B.-H.}, {\sc and} {\sc Fr{\"u}hwirth, T.}
  2007.
\newblock Toward a first-order extension of {P}rolog's unification using {CHR}:
  a {CHR} first-order constraint solver over finite or infinite trees.
\newblock In {\em SAC '07: Proc.\ 2007 ACM Symp.\ Applied computing}. ACM
  Press, 58--64.

\bibitem[\protect\citeauthoryear{Djelloul, Duck, and Sulzmann}{Djelloul, Duck
  et~al\mbox{.}}{2007}]{pchr07}
{\sc Djelloul, K.}, {\sc Duck, G.~J.}, {\sc and} {\sc Sulzmann, M.}, Eds. 2007.
\newblock {\em CHR '07: Proc.\ 4th Workshop on Constraint Handling Rules}.

\bibitem[\protect\citeauthoryear{Ducass{\'e}}{Ducass{\'e}}{1999}]{ducasse_opiu%
m_jlp99}
{\sc Ducass{\'e}, M.} 1999.
\newblock Opium: an extendable trace analyzer for {P}rolog.
\newblock {\em J. Logic Programming\/}~{\em 39,\/}~1--3, 177--223.

\bibitem[\protect\citeauthoryear{Duck}{Duck}{2005}]{duck_phdthesis05}
{\sc Duck, G.~J.} 2005.
\newblock Compilation of {C}onstraint {H}andling {R}ules.
\newblock Ph.D. thesis, University of Melbourne, Victoria, Australia.

\bibitem[\protect\citeauthoryear{Duck and Schrijvers}{Duck and
  Schrijvers}{2005}]{duck_schr_accurate_funcdep_chr05}
{\sc Duck, G.~J.} {\sc and} {\sc Schrijvers, T.} 2005.
\newblock Accurate functional dependency analysis for {C}onstraint {H}andling
  {R}ules.
\newblock In \citeN{pchr05}, 109--124.

\bibitem[\protect\citeauthoryear{Duck, Stuckey, and Brand}{Duck, Stuckey
  et~al\mbox{.}}{2006}]{duck_stuck_brand_acd_term_rewriting_iclp06}
{\sc Duck, G.~J.}, {\sc Stuckey, P.~J.}, {\sc and} {\sc Brand, S.} 2006.
\newblock {ACD} term rewriting.
\newblock In \citeN{piclp06}, 117--131.

\bibitem[\protect\citeauthoryear{Duck, Stuckey, {Garc{\'i}a de la Banda}, and
  Holzbaur}{Duck, Stuckey
  et~al\mbox{.}}{2003}]{duck_stuck_garcia_holz_extending_arbitrary_solvers_wit%
h_chr_ppdp03}
{\sc Duck, G.~J.}, {\sc Stuckey, P.~J.}, {\sc {Garc{\'i}a de la Banda}, M.},
  {\sc and} {\sc Holzbaur, C.} 2003.
\newblock Extending arbitrary solvers with {C}onstraint {H}andling {R}ules.
\newblock In {\em PPDP '03}. ACM Press, 79--90.

\bibitem[\protect\citeauthoryear{Duck, Stuckey, {Garc{\'i}a de la Banda}, and
  Holzbaur}{Duck, Stuckey
  et~al\mbox{.}}{2004}]{duck_stuck_garc_holz_refined_op_sem_iclp04}
{\sc Duck, G.~J.}, {\sc Stuckey, P.~J.}, {\sc {Garc{\'i}a de la Banda}, M.},
  {\sc and} {\sc Holzbaur, C.} 2004.
\newblock The refined operational semantics of {C}onstraint {H}andling {R}ules.
\newblock In \citeN{piclp04}, 90--104.

\bibitem[\protect\citeauthoryear{Duck, Stuckey, and Sulzmann}{Duck, Stuckey
  et~al\mbox{.}}{2007}]{duck_stuck_sulz_observable_confluence_iclp07}
{\sc Duck, G.~J.}, {\sc Stuckey, P.~J.}, {\sc and} {\sc Sulzmann, M.} 2007.
\newblock Observable confluence for {C}onstraint {H}andling {R}ules.
\newblock In \citeN{piclp07}, 224--239.

\bibitem[\protect\citeauthoryear{Escrig and Toledo}{Escrig and
  Toledo}{1998a}]{escrig_toledo_framework_qual_orientation_vlc98}
{\sc Escrig, M.~T.} {\sc and} {\sc Toledo, F.} 1998a.
\newblock A framework based on {CLP} extended with {CHR}s for reasoning with
  qualitative orientation and positional information.
\newblock {\em J. Visual Languages and Computing\/}~{\em 9,\/}~1, 81--101.

\bibitem[\protect\citeauthoryear{Escrig and Toledo}{Escrig and
  Toledo}{1998b}]{escrig_toledo_qual_spatial_reasoning_book98}
{\sc Escrig, M.~T.} {\sc and} {\sc Toledo, F.} 1998b.
\newblock {\em Qualitative Spatial Reasoning: Theory and Practice ---
  Application to Robot Navigation}.
\newblock IOS Press.

\bibitem[\protect\citeauthoryear{Etalle and Truszczynski}{Etalle and
  Truszczynski}{2006}]{piclp06}
{\sc Etalle, S.} {\sc and} {\sc Truszczynski, M.}, Eds. 2006.
\newblock {\em ICLP '06: Proc.\ 22nd Intl.\ Conf.\ Logic Programming}. LNCS,
  vol. 4079. Springer.

\bibitem[\protect\citeauthoryear{Fink, Tompits, and Woltran}{Fink, Tompits
  et~al\mbox{.}}{2006}]{pwlp06}
{\sc Fink, M.}, {\sc Tompits, H.}, {\sc and} {\sc Woltran, S.}, Eds. 2006.
\newblock {\em WLP '06: Proc.\ 20th Workshop on Logic Programming}. T.U.Wien,
  Austria, INFSYS Research report 1843-06-02.

\bibitem[\protect\citeauthoryear{Firat}{Firat}{2003}]{firat_ecoin_phdthesis03}
{\sc Firat, A.} 2003.
\newblock Information integration using contextual knowledge and ontology
  merging.
\newblock Ph.D. thesis, MIT Sloan School of Management, Cambridge,
  Massachusetts, USA.

\bibitem[\protect\citeauthoryear{Fr{\"u}hwirth}{Fr{\"u}hwirth}{1992}]{fru_cons%
traint_simplification_rules_techrep92}
{\sc Fr{\"u}hwirth, T.} 1992.
\newblock Constraint simplification rules.
\newblock Tech. Rep. ECRC-92-18, European Computer-Industry Research Centre,
  Munich, Germany. July.

\bibitem[\protect\citeauthoryear{Fr{\"u}hwirth}{Fr{\"u}hwirth}{1995}]{fru_chr_%
cp95}
{\sc Fr{\"u}hwirth, T.} 1995.
\newblock {C}onstraint {H}andling {R}ules.
\newblock In {\em Constraint Programming: Basic and Trends --- Selected Papers
  of the 22nd Spring School in Theoretical Computer Sciences, May 16--20,
  1994}, {A.~Podelski}, Ed. LNCS, vol. 910. Springer, 90--107.

\bibitem[\protect\citeauthoryear{Fr{\"u}hwirth}{Fr{\"u}hwirth}{1998}]{fru_chr_%
overview_jlp98}
{\sc Fr{\"u}hwirth, T.} 1998.
\newblock Theory and practice of {C}onstraint {H}andling {R}ules.
\newblock {\em J. Logic Programming, Special Issue on Constraint Logic
  Programming\/}~{\em 37,\/}~1--3, 95--138.

\bibitem[\protect\citeauthoryear{Fr{\"u}hwirth}{Fr{\"u}hwirth}{2000}]{fru_term%
ination_compulog00}
{\sc Fr{\"u}hwirth, T.} 2000.
\newblock Proving termination of constraint solver programs.
\newblock In {\em New Trends in Constraints, Joint ERCIM/Compulog Net Workshop,
  October 1999, Selected papers}, {K.~Apt}, {A.~Kakas}, {E.~Monfroy}, {and}
  {F.~Rossi}, Eds. LNCS, vol. 1865. Springer, 298--317.

\bibitem[\protect\citeauthoryear{Fr{\"u}hwirth}{Fr{\"u}hwirth}{2001}]{fru_numb%
er_entcs01}
{\sc Fr{\"u}hwirth, T.} 2001.
\newblock On the number of rule applications in constraint programs.
\newblock In {\em Declarative Programming - Selected Papers from AGP 2000},
  {A.~Dovier}, {M.~C. Meo}, {and} {A.~Omicini}, Eds. ENTCS, vol.~48. Elsevier,
  147--166.

\bibitem[\protect\citeauthoryear{Fr{\"u}hwirth}{Fr{\"u}hwirth}{2002a}]{fru_com%
plexity_kr02}
{\sc Fr{\"u}hwirth, T.} 2002a.
\newblock As time goes by: Automatic complexity analysis of simplification
  rules.
\newblock In {\em KR '02: Proc.\ 8th Intl.\ Conf. Princ.\ Knowledge
  Representation and Reasoning}, {D.~Fensel}, {F.~Giunchiglia},
  {D.~McGuinness}, {and} {M.-A. Williams}, Eds. Morgan Kaufmann, 547--557.

\bibitem[\protect\citeauthoryear{Fr{\"u}hwirth}{Fr{\"u}hwirth}{2002b}]{fru_com%
plexity2_entcs02}
{\sc Fr{\"u}hwirth, T.} 2002b.
\newblock As time goes by {II}: More automatic complexity analysis of
  concurrent rule programs.
\newblock In {\em QAPL '01: Proc.\ First Intl.\ Workshop on Quantitative
  Aspects of Programming Languages}, {A.~D. Pierro} {and} {H.~Wiklicky}, Eds.
  ENTCS, vol. 59(3). Elsevier.

\bibitem[\protect\citeauthoryear{Fr{\"u}hwirth}{Fr{\"u}hwirth}{2005a}]{fru_lex%
icographic_chr05}
{\sc Fr{\"u}hwirth, T.} 2005a.
\newblock Logical rules for a lexicographic order constraint solver.
\newblock In \citeN{pchr05}, 79--91.

\bibitem[\protect\citeauthoryear{Fr{\"u}hwirth}{Fr{\"u}hwirth}{2005b}]{fru_par%
allel_union_find_iclp05}
{\sc Fr{\"u}hwirth, T.} 2005b.
\newblock Parallelizing union-find in {C}onstraint {H}andling {R}ules using
  confluence.
\newblock In \citeN{piclp05}, 113--127.

\bibitem[\protect\citeauthoryear{Fr{\"u}hwirth}{Fr{\"u}hwirth}{2005c}]{fru_spe%
cialization_lopstr04}
{\sc Fr{\"u}hwirth, T.} 2005c.
\newblock Specialization of concurrent guarded multi-set transformation rules.
\newblock In {\em LOPSTR '04}, {S.~Etalle}, Ed. LNCS, vol. 3573. Springer,
  133--148.

\bibitem[\protect\citeauthoryear{Fr{\"u}hwirth}{Fr{\"u}hwirth}{2006a}]{fru_lex%
ico_csclp05:06}
{\sc Fr{\"u}hwirth, T.} 2006a.
\newblock Complete propagation rules for lexicographic order constraints over
  arbitrary domains.
\newblock In {\em Recent Advances in Constraints, CSCLP '05: Joint
  ERCIM/CoLogNET Intl.\ Workshop on Constraint Solving and CLP, Revised
  Selected and Invited Papers}. LNAI, vol. 3978. Springer.

\bibitem[\protect\citeauthoryear{Fr{\"u}hwirth}{Fr{\"u}hwirth}{2006b}]{fru_der%
iving_linear_algorithms_from_uf_chr06}
{\sc Fr{\"u}hwirth, T.} 2006b.
\newblock Deriving linear-time algorithms from union-find in {CHR}.
\newblock In \citeN{pchr06}, 49--60.

\bibitem[\protect\citeauthoryear{Fr{\"u}hwirth}{Fr{\"u}hwirth}{2007}]{fru_desc%
ription_logic_chr07}
{\sc Fr{\"u}hwirth, T.} 2007.
\newblock Description logic and rules the {CHR} way.
\newblock In \citeN{pchr07}, 49--61.

\bibitem[\protect\citeauthoryear{Fr{\"uh}wirth}{Fr{\"uh}wirth}{2009}]{fru_chr_%
2008}
{\sc Fr{\"uh}wirth, T.} 2009.
\newblock {\em Constraint Handling Rules}.
\newblock Cambridge University Press.
\newblock To appear.

\bibitem[\protect\citeauthoryear{Fr{\"u}hwirth et~al\mbox{.}}{Fr{\"u}hwirth
  et~al\mbox{.}}{2000}]{rcorp00}
{\sc Fr{\"u}hwirth, T.} {\sc et~al\mbox{.}}, Eds. 2000.
\newblock {\em RCoRP '00: Proc.\ 1st Workshop on Rule-Based Constraint
  Reasoning and Programming}.

\bibitem[\protect\citeauthoryear{Fr\"{u}hwirth and Abdennadher}{Fr\"{u}hwirth
  and Abdennadher}{2001}]{fru_abd_munich_rent_advisor_tplp01}
{\sc Fr\"{u}hwirth, T.} {\sc and} {\sc Abdennadher, S.} 2001.
\newblock The {Munich} rent advisor: A success for logic programming on the
  internet.
\newblock {\em TPLP\/}~{\em 1,\/}~3, 303--319.

\bibitem[\protect\citeauthoryear{Fr{\"u}hwirth and Abdennadher}{Fr{\"u}hwirth
  and Abdennadher}{2003}]{fru_abd_essentials_of_cp_book03}
{\sc Fr{\"u}hwirth, T.} {\sc and} {\sc Abdennadher, S.} 2003.
\newblock {\em Essentials of Constraint Programming}.
\newblock Springer.

\bibitem[\protect\citeauthoryear{Fr{\"u}hwirth and Brisset}{Fr{\"u}hwirth and
  Brisset}{1995}]{fru_brisset_highlevel_implementation_techrep95}
{\sc Fr{\"u}hwirth, T.} {\sc and} {\sc Brisset, P.} 1995.
\newblock High-level implementations of {{C}onstraint {H}andling {R}ules}.
\newblock Tech. Rep. ECRC-95-20, European Computer-Industry Research Centre.

\bibitem[\protect\citeauthoryear{Fr{\"u}hwirth and Brisset}{Fr{\"u}hwirth and
  Brisset}{1998}]{fru_brisset_wireless_cp98}
{\sc Fr{\"u}hwirth, T.} {\sc and} {\sc Brisset, P.} 1998.
\newblock Optimal placement of base stations in wireless indoor
  telecommunication.
\newblock In {\em CP '98}, {M.~J. Maher} {and} {J.-F. Puget}, Eds. LNCS, vol.
  1520. Springer, 476--480.

\bibitem[\protect\citeauthoryear{Fr{\"u}hwirth and Brisset}{Fr{\"u}hwirth and
  Brisset}{2000}]{fru_bri_base_wireless_ieee00}
{\sc Fr{\"u}hwirth, T.} {\sc and} {\sc Brisset, P.} 2000.
\newblock Placing base stations in wireless indoor communication networks.
\newblock {\em IEEE Intelligent Systems and Their Applications\/}~{\em
  15,\/}~1, 49--53.

\bibitem[\protect\citeauthoryear{Fr{\"u}hwirth, {Di Pierro}, and
  Wiklicky}{Fr{\"u}hwirth, {Di Pierro}
  et~al\mbox{.}}{2002}]{fru_dipierro_wiklicky_probabilistic_chr_wflp02}
{\sc Fr{\"u}hwirth, T.}, {\sc {Di Pierro}, A.}, {\sc and} {\sc Wiklicky, H.}
  2002.
\newblock Probabilistic {C}onstraint {H}andling {R}ules.
\newblock In {\em WFLP '02: Proc.\ 11th Intl.\ Workshop on Functional and
  (Constraint) Logic Programming, Selected Papers}, {M.~Comini} {and}
  {M.~Falaschi}, Eds. ENTCS, vol.~76. Elsevier.

\bibitem[\protect\citeauthoryear{Fr{\"u}hwirth and Holzbaur}{Fr{\"u}hwirth and
  Holzbaur}{2003}]{fru_holz_source2source_agp03}
{\sc Fr{\"u}hwirth, T.} {\sc and} {\sc Holzbaur, C.} 2003.
\newblock Source-to-source transformation for a class of expressive rules.
\newblock In {\em AGP '03: Joint Conf.\ Declarative Programming
  APPIA-GULP-PRODE}, {F.~Buccafurri}, Ed. 386--397.

\bibitem[\protect\citeauthoryear{Fr{\"u}hwirth and Meister}{Fr{\"u}hwirth and
  Meister}{2004}]{pchr04}
{\sc Fr{\"u}hwirth, T.} {\sc and} {\sc Meister, M.}, Eds. 2004.
\newblock {\em CHR '04: 1st Workshop on Constraint Handling Rules: Selected
  Contributions}.

\bibitem[\protect\citeauthoryear{Gabbrielli and Gupta}{Gabbrielli and
  Gupta}{2005}]{piclp05}
{\sc Gabbrielli, M.} {\sc and} {\sc Gupta, G.}, Eds. 2005.
\newblock {\em ICLP '05: Proc.\ 21st Intl.\ Conf.\ Logic Programming}. LNCS,
  vol. 3668. Springer.

\bibitem[\protect\citeauthoryear{Gabbrielli and Meo}{Gabbrielli and
  Meo}{2009}]{gabbr_meo_compos_semantics_tocl08}
{\sc Gabbrielli, M.} {\sc and} {\sc Meo, M.~C.} 2009.
\newblock A compositional semantics for {CHR}.
\newblock {\em ACM TOCL\/}~{\em 10,\/}~2.

\bibitem[\protect\citeauthoryear{Ganzinger and McAllester}{Ganzinger and
  McAllester}{2002}]{ganzinger_mcallester_la_iclp02}
{\sc Ganzinger, H.} {\sc and} {\sc McAllester, D.~A.} 2002.
\newblock Logical algorithms.
\newblock In \citeN{piclp02}, 209--223.

\bibitem[\protect\citeauthoryear{Garat and Wonsever}{Garat and
  Wonsever}{2002}]{garat_wonsever_parser_contextual_rules_sccc02}
{\sc Garat, D.} {\sc and} {\sc Wonsever, D.} 2002.
\newblock A constraint parser for contextual rules.
\newblock In {\em Proc.\ 22nd Intl.\ Conf.\ of the Chilean Computer Science
  Society}. IEEE Computer Society, 234--242.

\bibitem[\protect\citeauthoryear{Gavanelli, Lamma, Mello,
  et~al\mbox{.}}{Gavanelli, Lamma
  et~al\mbox{.}}{2003}]{gavanelli_et_al_interpreting_abduction_agp03}
{\sc Gavanelli, M.}, {\sc Lamma, E.}, {\sc Mello, P.}, {\sc et~al\mbox{.}}
  2003.
\newblock Interpreting abduction in {CLP}.
\newblock In {\em AGP '03: Joint Conf.\ Declarative Programming
  APPIA-GULP-PRODE}, {F.~Buccafurri}, Ed. 25--35.

\bibitem[\protect\citeauthoryear{Geurts, {van Ossenbruggen}, and
  Hardman}{Geurts, {van Ossenbruggen}
  et~al\mbox{.}}{2001}]{geurts_vanOss_hardman_multimedia_presentation_cuypers_%
mm01}
{\sc Geurts, J.}, {\sc {van Ossenbruggen}, J.}, {\sc and} {\sc Hardman, L.}
  2001.
\newblock Application-specific constraints for multimedia presentation
  generation.
\newblock In {\em MMM '01: Proc.\ 8th Intl.\ Conf.\ on Multimedia Modeling}.
  247--266.

\bibitem[\protect\citeauthoryear{Gouraud and Gotlieb}{Gouraud and
  Gotlieb}{2006}]{gouraud_gotlieb_javacard_padl06}
{\sc Gouraud, S.-D.} {\sc and} {\sc Gotlieb, A.} 2006.
\newblock Using {CHR}s to generate functional test cases for the {J}ava card
  virtual machine.
\newblock In {\em PADL '06: Proc.\ 8th Intl.\ Symp.\ Practical Aspects of
  Declarative Languages}, {P.~{Van Hentenryck}}, Ed. LNCS, vol. 3819. Springer,
  1--15.

\bibitem[\protect\citeauthoryear{Haemmerl{\'e} and Fages}{Haemmerl{\'e} and
  Fages}{2007}]{haemm_fages_abstract_critical_pairs_rta07}
{\sc Haemmerl{\'e}, R.} {\sc and} {\sc Fages, F.} 2007.
\newblock Abstract critical pairs and confluence of arbitrary binary relations.
\newblock In {\em RTA '07: Proc.\ 18th Intl.\ Conf.\ Term Rewriting and
  Applications}. LNCS, vol. 4533. Springer.

\bibitem[\protect\citeauthoryear{Hanus}{Hanus}{2006}]{hanus_chr_curry_wlp06}
{\sc Hanus, M.} 2006.
\newblock Adding {C}onstraint {H}andling {R}ules to {Curry}.
\newblock In \citeN{pwlp06}, 81--90.

\bibitem[\protect\citeauthoryear{Hecksher, Nielsen, and Pigeon}{Hecksher,
  Nielsen et~al\mbox{.}}{2002}]{roskilde_students_ancient_egyptian_grammar_02}
{\sc Hecksher, T.}, {\sc Nielsen, S.~T.}, {\sc and} {\sc Pigeon, A.} 2002.
\newblock A {CHRG} model of the ancient {E}gyptian grammar.
\newblock Unpublished student project report, Roskilde University, Denmark.

\bibitem[\protect\citeauthoryear{Holzbaur and Fr\"uhwirth}{Holzbaur and
  Fr\"uhwirth}{1998}]{holz_fru_CHR_manual_techrep98}
{\sc Holzbaur, C.} {\sc and} {\sc Fr\"uhwirth, T.} 1998.
\newblock {C}onstraint {H}andling {R}ules reference manual, release 2.2.
\newblock Tech. Rep. TR-98-01, \"Osterreichisches Forschungsinstitut f\"ur
  Artificial Intelligence, Wien.

\bibitem[\protect\citeauthoryear{Holzbaur and Fr{\"u}hwirth}{Holzbaur and
  Fr{\"u}hwirth}{1999}]{holz_fru_compiling_chr_attr_vars_ppdp99}
{\sc Holzbaur, C.} {\sc and} {\sc Fr{\"u}hwirth, T.} 1999.
\newblock Compiling {C}onstraint {H}andling {R}ules into {Prolog} with
  attributed variables.
\newblock In {\em PPDP '99}, {G.~Nadathur}, Ed. LNCS, vol. 1702. Springer,
  117--133.

\bibitem[\protect\citeauthoryear{Holzbaur and Fr{\"u}hwirth}{Holzbaur and
  Fr{\"u}hwirth}{2000a}]{holz_fru_prolog_chr_compiler_aai00}
{\sc Holzbaur, C.} {\sc and} {\sc Fr{\"u}hwirth, T.} 2000a.
\newblock A {P}rolog {C}onstraint {H}andling {R}ules compiler and runtime
  system.
\newblock In \citeN{aai00}, 369--388.

\bibitem[\protect\citeauthoryear{Holzbaur and Fr{\"u}hwirth}{Holzbaur and
  Fr{\"u}hwirth}{2000b}]{aai00}
{\sc Holzbaur, C.} {\sc and} {\sc Fr{\"u}hwirth, T.}, Eds. 2000b.
\newblock {\em Special Issue on Constraint Handling Rules}. Journal of Applied
  Artificial Intelligence, vol. 14(4).

\bibitem[\protect\citeauthoryear{Holzbaur, {Garc{\'i}a de la Banda}, Stuckey,
  and Duck}{Holzbaur, {Garc{\'i}a de la Banda}
  et~al\mbox{.}}{2005}]{holz_garc_stuck_duck_opt_comp_chr_hal_tplp05}
{\sc Holzbaur, C.}, {\sc {Garc{\'i}a de la Banda}, M.}, {\sc Stuckey, P.~J.},
  {\sc and} {\sc Duck, G.~J.} 2005.
\newblock Optimizing compilation of {C}onstraint {H}andling {R}ules in {HAL}.
\newblock In \citeN{tplp05}, 503--531.

\bibitem[\protect\citeauthoryear{K\"aser and Meister}{K\"aser and
  Meister}{2006}]{kaeser_meister_flogic_chr06}
{\sc K\"aser, M.} {\sc and} {\sc Meister, M.} 2006.
\newblock Implementation of an {F-Logic} kernel in {CHR}.
\newblock In \citeN{pchr06}, 33--47.

\bibitem[\protect\citeauthoryear{Kosmatov}{Kosmatov}{2006a}]{kosmatov_sequence%
s_06}
{\sc Kosmatov, N.} 2006a.
\newblock A constraint solver for sequences and its applications.
\newblock In {\em Proc.\ 2006 ACM Symp.\ on Applied Computing}. ACM Press,
  404--408.

\bibitem[\protect\citeauthoryear{Kosmatov}{Kosmatov}{2006b}]{kosmatov_sequence%
s_inap05}
{\sc Kosmatov, N.} 2006b.
\newblock Constraint solving for sequences in software validation and
  verification.
\newblock In {\em INAP '05: Proc.\ 16th Intl.\ Conf.\ Applications of
  Declarative Programming and Knowledge Management}. LNCS, vol. 4369. Springer,
  25--37.

\bibitem[\protect\citeauthoryear{Kr{\"a}mer}{Kr{\"a}mer}{2001}]{kr_jack:jase_2%
001}
{\sc Kr{\"a}mer, E.} 2001.
\newblock A generic search engine for a {J}ava {C}onstraint {K}it.
\newblock Diplomarbeit.
\newblock Institute of Computer Science, LMU, Munich, Germany.

\bibitem[\protect\citeauthoryear{Lam and Sulzmann}{Lam and
  Sulzmann}{2006}]{lam_sulz_linear_logic_agents_chr06}
{\sc Lam, E.~S.} {\sc and} {\sc Sulzmann, M.} 2006.
\newblock Towards agent programming in {CHR}.
\newblock In \citeN{pchr06}, 17--31.

\bibitem[\protect\citeauthoryear{Lam and Sulzmann}{Lam and
  Sulzmann}{2007}]{lam_sulz_concurrent_chr_damp07}
{\sc Lam, E.~S.} {\sc and} {\sc Sulzmann, M.} 2007.
\newblock A concurrent {C}onstraint {H}andling {R}ules semantics and its
  implementation with software transactional memory.
\newblock In {\em DAMP '07: Proc.\ ACM SIGPLAN Workshop on Declarative Aspects
  of Multicore Programming}. ACM Press.
\newblock System's homepage at {\tt
  http://taichi.ddns.comp.nus.edu.sg/taichiwiki/CCHR/}.

\bibitem[\protect\citeauthoryear{L{\"o}tzbeyer and Pretschner}{L{\"o}tzbeyer
  and Pretschner}{2000}]{lotzbeyer_pretschner_autofocus_lpse00}
{\sc L{\"o}tzbeyer, H.} {\sc and} {\sc Pretschner, A.} 2000.
\newblock {\sc {A}uto{F}ocus} on constraint logic programming.
\newblock In {\em LPSE '00: Proc.\ Intl.\ Workshop on (Constraint) Logic
  Programming and Software Engineering}.

\bibitem[\protect\citeauthoryear{Maher}{Maher}{2002}]{propagation_completeness%
_maher_iclp02}
{\sc Maher, M.~J.} 2002.
\newblock Propagation completeness of reactive constraints.
\newblock In \citeN{piclp02}, 148--162.

\bibitem[\protect\citeauthoryear{Meister}{Meister}{2006}]{meister_preflow_push%
_wlp06}
{\sc Meister, M.} 2006.
\newblock Fine-grained parallel implementation of the preflow-push algorithm in
  {CHR}.
\newblock In \citeN{pwlp06}, 172--181.

\bibitem[\protect\citeauthoryear{Meister, Djelloul, and Fr{\"u}hwirth}{Meister,
  Djelloul
  et~al\mbox{.}}{2006}]{meister_djelloul_fru_compl_tree_equations_csclp06}
{\sc Meister, M.}, {\sc Djelloul, K.}, {\sc and} {\sc Fr{\"u}hwirth, T.} 2006.
\newblock Complexity of a {CHR} solver for existentially quantified
  conjunctions of equations over trees.
\newblock In {\em CSCLP '06: Proc.\ 11th Annual ERCIM Workshop on Constraint
  Solving and Constraint Programming}, {F.~Azevedo} {et~al\mbox{.}}, Eds. LNCS,
  vol. 4651. Springer, 139--153.

\bibitem[\protect\citeauthoryear{Meister, Djelloul, and Robin}{Meister,
  Djelloul
  et~al\mbox{.}}{2007}]{meister_djelloul_robin_transaction_logic_semantics_lpn%
mr07}
{\sc Meister, M.}, {\sc Djelloul, K.}, {\sc and} {\sc Robin, J.} 2007.
\newblock A unified semantics for {C}onstraint {H}andling {R}ules in
  transaction logic.
\newblock In {\em LPNMR '07: Proc.\ 9th Intl.\ Conf.\ Logic Programming and
  Nonmonotonic Reasoning}, {C.~Baral}, {G.~Brewka}, {and} {J.~S. Schlipf}, Eds.
  LNCS, vol. 4483. Springer, 201--213.

\bibitem[\protect\citeauthoryear{Menezes, Vitorino, and Aurelio}{Menezes,
  Vitorino
  et~al\mbox{.}}{2005}]{menezes_vitorino_aurelio_high_performance_chr_or_chr05}
{\sc Menezes, L.}, {\sc Vitorino, J.}, {\sc and} {\sc Aurelio, M.} 2005.
\newblock A high performance {CHR}${}^\lor$ execution engine.
\newblock In \citeN{pchr05}, 35--45.

\bibitem[\protect\citeauthoryear{Meyer}{Meyer}{2000}]{meyer_diagrammatic_reaso%
ning_aai00}
{\sc Meyer, B.} 2000.
\newblock A constraint-based framework for diagrammatic reasoning.
\newblock In \citeN{aai00}, 327--344.

\bibitem[\protect\citeauthoryear{Morawietz}{Morawietz}{2000}]{morawietz_chart_%
parsing_coling00}
{\sc Morawietz, F.} 2000.
\newblock Chart parsing and constraint programming.
\newblock In {\em COLING '00: Proc.\ 18th Intl.\ Conf.\ on Computational
  Linguistics}, {M.~Kay}, Ed. Morgan Kaufmann.

\bibitem[\protect\citeauthoryear{Morawietz and Blache}{Morawietz and
  Blache}{2002}]{morawietz_blache_unpublished02}
{\sc Morawietz, F.} {\sc and} {\sc Blache, P.} 2002.
\newblock Parsing natural languages with {CHR}.
\newblock Unpublished Draft.

\bibitem[\protect\citeauthoryear{Penn}{Penn}{2000}]{penn_hpsg_rcorp00}
{\sc Penn, G.} 2000.
\newblock Applying {C}onstraint {H}andling {R}ules to {HPSG}.
\newblock In \citeN{rcorp00}.

\bibitem[\protect\citeauthoryear{Pilozzi, Schrijvers, and {De
  Schreye}}{Pilozzi, Schrijvers
  et~al\mbox{.}}{2007}]{pilozzi_schr_deschreye_termination_wst07}
{\sc Pilozzi, P.}, {\sc Schrijvers, T.}, {\sc and} {\sc {De Schreye}, D.} 2007.
\newblock Proving termination of {CHR} in {Prolog}: A transformational
  approach.
\newblock In {\em WST '07: 9th Intl.\ Workshop on Termination}.

\bibitem[\protect\citeauthoryear{Pretschner, Slotosch, Aiglstorfer, and
  Kriebel}{Pretschner, Slotosch
  et~al\mbox{.}}{2004}]{pretschner_et_al_model-based_testing_sttt04}
{\sc Pretschner, A.}, {\sc Slotosch, O.}, {\sc Aiglstorfer, E.}, {\sc and} {\sc
  Kriebel, S.} 2004.
\newblock Model-based testing for real.
\newblock {\em J. Software Tools for Technology Transfer (STTT)\/}~{\em
  5,\/}~2--3, 140--157.

\bibitem[\protect\citeauthoryear{Raiser}{Raiser}{2007}]{raiser_graph_transform%
ation_systems_iclp07}
{\sc Raiser, F.} 2007.
\newblock Graph transformation systems in {CHR}.
\newblock In \citeN{piclp07}, 240--254.

\bibitem[\protect\citeauthoryear{Raiser and Tacchella}{Raiser and
  Tacchella}{2007}]{raiser_tacchella_confluence_non_terminating_chr07}
{\sc Raiser, F.} {\sc and} {\sc Tacchella, P.} 2007.
\newblock On confluence of non-terminating {CHR} programs.
\newblock In \citeN{pchr07}, 63--76.

\bibitem[\protect\citeauthoryear{Ribeiro, Z{\'u}quete, Ferreira, and
  Guedes}{Ribeiro, Z{\'u}quete
  et~al\mbox{.}}{2000}]{ribeiro_et_al_security_policy_consistency_rcorp00}
{\sc Ribeiro, C.}, {\sc Z{\'u}quete, A.}, {\sc Ferreira, P.}, {\sc and} {\sc
  Guedes, P.} 2000.
\newblock Security policy consistency.
\newblock In \citeN{rcorp00}.

\bibitem[\protect\citeauthoryear{Ringwelski and Schlenker}{Ringwelski and
  Schlenker}{2000a}]{ring_schlenk_inference_wlp00}
{\sc Ringwelski, G.} {\sc and} {\sc Schlenker, H.} 2000a.
\newblock Type inference in {CHR} programs for the composition of constraint
  systems.
\newblock In {\em WLP '00: Proc.\ 15th Workshop on Logic Programming},
  {S.~Abdennadher}, {U.~Geske}, {and} {D.~Seipel}, Eds. 137--146.

\bibitem[\protect\citeauthoryear{Ringwelski and Schlenker}{Ringwelski and
  Schlenker}{2000b}]{ring_schlenk_interfaces_rcorp00bis}
{\sc Ringwelski, G.} {\sc and} {\sc Schlenker, H.} 2000b.
\newblock Using typed interfaces to compose {CHR} programs.
\newblock In {\em RCoRP '00(bis): Proc.\ 2nd Workshop on Rule-Based Constraint
  Reasoning and Programming}, {T.~Fr{\"u}hwirth} {et~al\mbox{.}}, Eds.

\bibitem[\protect\citeauthoryear{Robin and Vitorino}{Robin and
  Vitorino}{2006}]{robin_vitorino_orcas_wlp06}
{\sc Robin, J.} {\sc and} {\sc Vitorino, J.} 2006.
\newblock {ORCAS}: Towards a {CHR}-based model-driven framework of reusable
  reasoning components.
\newblock In \citeN{pwlp06}, 192--199.

\bibitem[\protect\citeauthoryear{Robin, Vitorino, and Wolf}{Robin, Vitorino
  et~al\mbox{.}}{2007}]{robin_vitorino_wolf_CPA_proposal_jucs07}
{\sc Robin, J.}, {\sc Vitorino, J.}, {\sc and} {\sc Wolf, A.} 2007.
\newblock Constraint programming architectures: Review and a new proposal.
\newblock {\em J. UCS\/}~{\em 13,\/}~6, 701--720.

\bibitem[\protect\citeauthoryear{Sarna-Starosta and
  Ramakrishnan}{Sarna-Starosta and
  Ramakrishnan}{2007}]{sarnastarosta_ramakrishnan_chrd_padl07}
{\sc Sarna-Starosta, B.} {\sc and} {\sc Ramakrishnan, C.} 2007.
\newblock Compiling {C}onstraint {H}andling {R}ules for efficient tabled
  evaluation.
\newblock In {\em PADL '07: Proc.\ 9th Intl.\ Symp.\ Practical Aspects of
  Declarative Languages}, {M.~Hanus}, Ed. LNCS, vol. 4354. Springer, 170--184.
\newblock System's homepage at {\tt http://www.cse.msu.edu/$\sim$bss/chr\_d}.

\bibitem[\protect\citeauthoryear{Sarna-Starosta and Schrijvers}{Sarna-Starosta
  and Schrijvers}{2007}]{sarnastarosta_schr_indexing_techrep07}
{\sc Sarna-Starosta, B.} {\sc and} {\sc Schrijvers, T.} 2007.
\newblock Indexing techniques for {CHR} based on program transformation.
\newblock Tech. Rep. CW 500, K.U.Leuven, Dept.\ Comp.\ Sc. Aug.

\bibitem[\protect\citeauthoryear{Schiffel and Thielscher}{Schiffel and
  Thielscher}{2007}]{schiffel_thielscher_fluxplayer_aaai07}
{\sc Schiffel, S.} {\sc and} {\sc Thielscher, M.} 2007.
\newblock Fluxplayer: A successful general game player.
\newblock In {\em AAAI '07: Proc. 22nd AAAI Conf. Artificial Intelligence}.
  AAAI Press, 1191--1196.

\bibitem[\protect\citeauthoryear{Schmau\ss{}}{Schmau\ss{}}{1999}]{schmauss_jac%
k:jchr_1999}
{\sc Schmau\ss{}, M.} 1999.
\newblock An implementation of {CHR} in {J}ava.
\newblock Diplomarbeit.
\newblock Institute of Computer Science, LMU, Munich, Germany.

\bibitem[\protect\citeauthoryear{Schrijvers}{Schrijvers}{2004}]{schr_JmmSolve_%
iclp04}
{\sc Schrijvers, T.} 2004.
\newblock Jmmsolve: A generative java memory model implemented in {P}rolog and
  {CHR}.
\newblock In \citeN{piclp04}, 475--476.

\bibitem[\protect\citeauthoryear{Schrijvers}{Schrijvers}{2005}]{schr_phdthesis%
05}
{\sc Schrijvers, T.} 2005.
\newblock Analyses, optimizations and extensions of {C}onstraint {H}andling
  {R}ules.
\newblock Ph.D. thesis, K.U.Leuven, Leuven, Belgium.

\bibitem[\protect\citeauthoryear{Schrijvers and Bruynooghe}{Schrijvers and
  Bruynooghe}{2006}]{schr_bruynooghe_polymorphic_type_reconstruction_ppdp06}
{\sc Schrijvers, T.} {\sc and} {\sc Bruynooghe, M.} 2006.
\newblock Polymorphic algebraic data type reconstruction.
\newblock In {\em PPDP '06}, {A.~Bossi} {and} {M.~Maher}, Eds. ACM Press,
  85--96.

\bibitem[\protect\citeauthoryear{Schrijvers and Demoen}{Schrijvers and
  Demoen}{2004a}]{schr_demoen_delay_avoid_techrep04}
{\sc Schrijvers, T.} {\sc and} {\sc Demoen, B.} 2004a.
\newblock Antimonotony-based delay avoidance for {CHR}.
\newblock Tech. Rep. CW 385, K.U.Leuven, Dept.\ Comp.\ Sc. July.

\bibitem[\protect\citeauthoryear{Schrijvers and Demoen}{Schrijvers and
  Demoen}{2004b}]{schr_demoen_kulchr_chr04}
{\sc Schrijvers, T.} {\sc and} {\sc Demoen, B.} 2004b.
\newblock {T}he {K}.{U}.{L}euven {CHR} system: {I}mplementation and
  application.
\newblock In \citeN{pchr04}, 8--12.
\newblock System's homepage at {\tt http://www.cs.kuleuven.be/$\sim$toms/CHR/}.

\bibitem[\protect\citeauthoryear{Schrijvers, Demoen, Duck, Stuckey, and
  Fr{\"u}hwirth}{Schrijvers, Demoen
  et~al\mbox{.}}{2006}]{schr_demoen_duck_stuck_fru_implication_checking_entcs0%
6}
{\sc Schrijvers, T.}, {\sc Demoen, B.}, {\sc Duck, G.~J.}, {\sc Stuckey,
  P.~J.}, {\sc and} {\sc Fr{\"u}hwirth, T.} 2006.
\newblock Automatic implication checking for {CHR} constraints.
\newblock In {\em RULE '05: 6th Intl.\ Workshop on Rule-Based Programming}.
  ENTCS, vol. 147(1). Elsevier, 93--111.

\bibitem[\protect\citeauthoryear{Schrijvers and Fr{\"u}hwirth}{Schrijvers and
  Fr{\"u}hwirth}{2005a}]{schr_fru_analysing_union_find_wclp05}
{\sc Schrijvers, T.} {\sc and} {\sc Fr{\"u}hwirth, T.} 2005a.
\newblock Analysing the {CHR} implementation of union-find.
\newblock In \citeN{pwclp05}, 135--146.

\bibitem[\protect\citeauthoryear{Schrijvers and Fr{\"u}hwirth}{Schrijvers and
  Fr{\"u}hwirth}{2005b}]{pchr05}
{\sc Schrijvers, T.} {\sc and} {\sc Fr{\"u}hwirth, T.}, Eds. 2005b.
\newblock {\em CHR '05: Proc.\ 2nd Workshop on Constraint Handling Rules}.
  K.U.Leuven, Dept.\ Comp.\ Sc., Technical report CW 421.

\bibitem[\protect\citeauthoryear{Schrijvers and Fr{\"u}hwirth}{Schrijvers and
  Fr{\"u}hwirth}{2006}]{pchr06}
{\sc Schrijvers, T.} {\sc and} {\sc Fr{\"u}hwirth, T.}, Eds. 2006.
\newblock {\em CHR '06: Proc.\ 3rd Workshop on Constraint Handling Rules}.
  K.U.Leuven, Dept.\ Comp.\ Sc., Technical report CW 452.

\bibitem[\protect\citeauthoryear{Schrijvers and Fr\"{u}hwirth}{Schrijvers and
  Fr\"{u}hwirth}{2006}]{schr_fru_opt_union_find_tplp06}
{\sc Schrijvers, T.} {\sc and} {\sc Fr\"{u}hwirth, T.} 2006.
\newblock Optimal union-find in {C}onstraint {H}andling {R}ules.
\newblock {\em TPLP\/}~{\em 6,\/}~1--2, 213--224.

\bibitem[\protect\citeauthoryear{Schrijvers, Stuckey, and Duck}{Schrijvers,
  Stuckey et~al\mbox{.}}{2005}]{schr_stuck_duck_ai_chr_ppdp05}
{\sc Schrijvers, T.}, {\sc Stuckey, P.~J.}, {\sc and} {\sc Duck, G.~J.} 2005.
\newblock Abstract interpretation for {C}onstraint {H}andling {R}ules.
\newblock In {\em PPDP '05}, {P.~Barahona} {and} {A.~Felty}, Eds. ACM Press,
  218--229.

\bibitem[\protect\citeauthoryear{Schrijvers and Warren}{Schrijvers and
  Warren}{2004}]{schr_warren_chr_xsb_iclp04}
{\sc Schrijvers, T.} {\sc and} {\sc Warren, D.~S.} 2004.
\newblock {C}onstraint {H}andling {R}ules and tabled execution.
\newblock In \citeN{piclp04}, 120--136.

\bibitem[\protect\citeauthoryear{Schrijvers, Warren, and Demoen}{Schrijvers,
  Warren et~al\mbox{.}}{2003}]{schr_warren_demoen_chr_xsb_ciclops03}
{\sc Schrijvers, T.}, {\sc Warren, D.~S.}, {\sc and} {\sc Demoen, B.} 2003.
\newblock {CHR} for {XSB}.
\newblock In {\em CICLOPS '03: Proc. 3rd Intl. Colloq. on Implementation of
  Constraint and Logic Programming Systems}, {R.~Lopes} {and} {M.~Ferreira},
  Eds. University of Porto, Portugal, Dept.\ Comp.\ Sc., Technical report
  DCC-2003-05. 7--20.

\bibitem[\protect\citeauthoryear{Schrijvers, Wielemaker, and
  Demoen}{Schrijvers, Wielemaker
  et~al\mbox{.}}{2005}]{schr_wielemaker_demoen_chr_swi_wclp05}
{\sc Schrijvers, T.}, {\sc Wielemaker, J.}, {\sc and} {\sc Demoen, B.} 2005.
\newblock {P}oster: {C}onstraint {H}andling {R}ules for {SWI}-{P}rolog.
\newblock In \citeN{pwclp05}.

\bibitem[\protect\citeauthoryear{Schrijvers, Zhou, and Demoen}{Schrijvers, Zhou
  et~al\mbox{.}}{2006}]{schr_zhou_demoen_action_rules_chr06}
{\sc Schrijvers, T.}, {\sc Zhou, N.-F.}, {\sc and} {\sc Demoen, B.} 2006.
\newblock Translating {C}onstraint {H}andling {R}ules into {Action Rules}.
\newblock In \citeN{pchr06}, 141--155.

\bibitem[\protect\citeauthoryear{Schumann}{Schumann}{2002}]{torres_literate_pr%
ogramming_wflp02}
{\sc Schumann, E.~T.} 2002.
\newblock A literate programming system for logic programs with constraints.
\newblock In {\em WFLP '02: Proc.\ 11th Intl.\ Workshop on Functional and
  (Constraint) Logic Programming}, {M.~Comini} {and} {M.~Falaschi}, Eds.
  University of Udine, Research Report UDMI/18/2002/RR.

\bibitem[\protect\citeauthoryear{{Scientific Software \& Systems
  Ltd.}}{{Scientific Software \& Systems Ltd.}}{2008}]{sssltd_2008}
{\sc {Scientific Software \& Systems Ltd.}} 2008.
\newblock {Company Profile: Solving problems with proven solutions}.
\newblock Available at {\tt http://www.sss.co.nz/}.

\bibitem[\protect\citeauthoryear{Seitz, Bauer, and Berger}{Seitz, Bauer
  et~al\mbox{.}}{2002}]{seitz_bauer_berger_MAS_icai02}
{\sc Seitz, C.}, {\sc Bauer, B.}, {\sc and} {\sc Berger, M.} 2002.
\newblock Planning and scheduling in multi agent systems using {C}onstraint
  {H}andling {R}ules.
\newblock In {\em IC-AI '02: Proc.\ Intl.\ Conf.\ Artificial Intelligence}.
  CSREA Press.

\bibitem[\protect\citeauthoryear{Shigeta, Akama, Mabuchi, and Koike}{Shigeta,
  Akama et~al\mbox{.}}{2006}]{shigeta_akama_mabuchi_koike_chr_to_etr_jaciii06}
{\sc Shigeta, Y.}, {\sc Akama, K.}, {\sc Mabuchi, H.}, {\sc and} {\sc Koike,
  H.} 2006.
\newblock Converting {C}onstraint {H}andling {R}ules to {Equivalent
  Transformation Rules}.
\newblock {\em JACIII\/}~{\em 10,\/}~3, 339--348.

\bibitem[\protect\citeauthoryear{Sim{\~{o}}es and Florido}{Sim{\~{o}}es and
  Florido}{2004}]{simoes_florido_typetool_wflp04}
{\sc Sim{\~{o}}es, H.} {\sc and} {\sc Florido, M.} 2004.
\newblock {TypeTool}: A type inference visualization tool.
\newblock In {\em WFLP '04: Proc.\ 13th Intl.\ Workshop on Functional and
  (Constraint) Logic Programming}, {H.~Kuchen}, Ed. RWTH Aachen, Dept.\ Comp.\
  Sc., Technical report AIB-2004-05. 48--61.

\bibitem[\protect\citeauthoryear{Sneyers, Schrijvers, and Demoen}{Sneyers,
  Schrijvers
  et~al\mbox{.}}{2005}]{sney_schr_demoen_guard_and_continuation_opt_iclp05}
{\sc Sneyers, J.}, {\sc Schrijvers, T.}, {\sc and} {\sc Demoen, B.} 2005.
\newblock Guard and continuation optimization for occurrence representations of
  {CHR}.
\newblock In \citeN{piclp05}, 83--97.

\bibitem[\protect\citeauthoryear{Sneyers, Schrijvers, and Demoen}{Sneyers,
  Schrijvers et~al\mbox{.}}{2006a}]{sney_schr_demoen_dijkstra_chr_wlp06}
{\sc Sneyers, J.}, {\sc Schrijvers, T.}, {\sc and} {\sc Demoen, B.} 2006a.
\newblock Dijkstra's algorithm with {Fibonacci} heaps: An executable
  description in {CHR}.
\newblock In \citeN{pwlp06}, 182--191.

\bibitem[\protect\citeauthoryear{Sneyers, Schrijvers, and Demoen}{Sneyers,
  Schrijvers et~al\mbox{.}}{2006b}]{sney_schr_demoen_memory_reuse_iclp06}
{\sc Sneyers, J.}, {\sc Schrijvers, T.}, {\sc and} {\sc Demoen, B.} 2006b.
\newblock Memory reuse for {CHR}.
\newblock In \citeN{piclp06}, 72--86.

\bibitem[\protect\citeauthoryear{Sneyers, Schrijvers, and Demoen}{Sneyers,
  Schrijvers et~al\mbox{.}}{2009}]{sney_schr_demoen_chr_complexity_08}
{\sc Sneyers, J.}, {\sc Schrijvers, T.}, {\sc and} {\sc Demoen, B.} 2009.
\newblock The computational power and complexity of {C}onstraint {H}andling
  {R}ules.
\newblock {\em ACM TOPLAS\/}~{\em 31,\/}~2.

\bibitem[\protect\citeauthoryear{Sneyers, {Van Weert}, and Schrijvers}{Sneyers,
  {Van Weert} et~al\mbox{.}}{2007}]{sney_vanweert_demoen_aggregates_chr07}
{\sc Sneyers, J.}, {\sc {Van Weert}, P.}, {\sc and} {\sc Schrijvers, T.} 2007.
\newblock Aggregates for {C}onstraint {H}andling {R}ules.
\newblock In \citeN{pchr07}, 91--105.

\bibitem[\protect\citeauthoryear{Stuckey}{Stuckey}{2002}]{piclp02}
{\sc Stuckey, P.~J.}, Ed. 2002.
\newblock {\em ICLP '02: Proc.\ 18th Intl.\ Conf.\ Logic Programming}. LNCS,
  vol. 2401. Springer.

\bibitem[\protect\citeauthoryear{Stuckey and Sulzmann}{Stuckey and
  Sulzmann}{2005}]{stuck_sulz_theory_of_overloading_toplas05}
{\sc Stuckey, P.~J.} {\sc and} {\sc Sulzmann, M.} 2005.
\newblock A theory of overloading.
\newblock {\em ACM TOPLAS\/}~{\em 27,\/}~6, 1216--1269.

\bibitem[\protect\citeauthoryear{Sulzmann, Duck, Peyton-Jones, and
  Stuckey}{Sulzmann, Duck
  et~al\mbox{.}}{2007}]{sulz_duck_peyton_stuck_func_dep_via_chr_fp07}
{\sc Sulzmann, M.}, {\sc Duck, G.~J.}, {\sc Peyton-Jones, S.}, {\sc and} {\sc
  Stuckey, P.~J.} 2007.
\newblock Understanding functional dependencies via {C}onstraint {H}andling
  {R}ules.
\newblock {\em J. Functional Prog.\/}~{\em 17,\/}~1, 83--129.

\bibitem[\protect\citeauthoryear{Sulzmann and Lam}{Sulzmann and
  Lam}{2007a}]{sulz_lam_lazy_concurr_search_chr07}
{\sc Sulzmann, M.} {\sc and} {\sc Lam, E.~S.} 2007a.
\newblock Compiling {C}onstraint {H}andling {R}ules with lazy and concurrent
  search techniques.
\newblock In \citeN{pchr07}, 139--149.

\bibitem[\protect\citeauthoryear{Sulzmann and Lam}{Sulzmann and
  Lam}{2007b}]{sulz_lam_haskelljoinrules_ifl07}
{\sc Sulzmann, M.} {\sc and} {\sc Lam, E.~S.} 2007b.
\newblock {Haskell - Join - Rules}.
\newblock In {\em IFL '07: 19th Intl.\ Symp.\ Implementation and Application of
  Functional Languages}, {O.~Chitil}, Ed. 195--210.

\bibitem[\protect\citeauthoryear{Sulzmann, Schrijvers, and Stuckey}{Sulzmann,
  Schrijvers et~al\mbox{.}}{2006}]{sulz_schr_stuck_aplas06}
{\sc Sulzmann, M.}, {\sc Schrijvers, T.}, {\sc and} {\sc Stuckey, P.~J.} 2006.
\newblock Principal type inference for {GHC}-style multi-parameter type
  classes.
\newblock In {\em APLAS '06: Proc.\ 4th Asian Symp.\ on Programming Languages
  and Systems}, {N.~Kobayashi}, Ed. LNCS, vol. 4279. Springer, 26--43.

\bibitem[\protect\citeauthoryear{Sulzmann, Wazny, and Stuckey}{Sulzmann, Wazny
  et~al\mbox{.}}{2005}]{sulz_wazny_stuck_abduction_chr05}
{\sc Sulzmann, M.}, {\sc Wazny, J.}, {\sc and} {\sc Stuckey, P.~J.} 2005.
\newblock Constraint abduction and constraint handling rules.
\newblock In \citeN{pchr05}, 63--78.

\bibitem[\protect\citeauthoryear{Tacchella, Gabbrielli, and Meo}{Tacchella,
  Gabbrielli et~al\mbox{.}}{2007}]{tacchella_gabbrielli_meo_unfolding_ppdp07}
{\sc Tacchella, P.}, {\sc Gabbrielli, M.}, {\sc and} {\sc Meo, M.~C.} 2007.
\newblock Unfolding in {CHR}.
\newblock In {\em PPDP '07}, {M.~Leuschel} {and} {A.~Podelski}, Eds. ACM Press,
  179--186.

\bibitem[\protect\citeauthoryear{Thielscher}{Thielscher}{2002}]{thielscher_act%
ions_iclp02}
{\sc Thielscher, M.} 2002.
\newblock Reasoning about actions with {CHR}s and finite domain constraints.
\newblock In \citeN{piclp02}, 70--84.

\bibitem[\protect\citeauthoryear{Thielscher}{Thielscher}{2005}]{thielscher_flu%
x_tplp05}
{\sc Thielscher, M.} 2005.
\newblock {FLUX}: A logic programming method for reasoning agents.
\newblock In \citeN{tplp05}, 533--565.

\bibitem[\protect\citeauthoryear{Ueda et~al\mbox{.}}{Ueda
  et~al\mbox{.}}{2006}]{ueda_LMNtal_chr06}
{\sc Ueda, K.} {\sc et~al\mbox{.}} 2006.
\newblock {LMNtal} as a unifying declarative language.
\newblock In \citeN{pchr06}, 1--15.
\newblock Invited talk.

\bibitem[\protect\citeauthoryear{Van~Weert}{Van~Weert}{2008}]{vanweert_jchr_co%
mpilation_techrep08}
{\sc Van~Weert, P.} 2008.
\newblock Compiling {C}onstraint {H}andling {R}ules to {J}ava: A
  reconstruction.
\newblock Tech. Rep. CW 521, K.U.Leuven, Dept.\ Comp.\ Sc. Aug.

\bibitem[\protect\citeauthoryear{{Van Weert}, Schrijvers, and Demoen}{{Van
  Weert}, Schrijvers et~al\mbox{.}}{2005}]{vanweert_schr_demoen_jchr_chr05}
{\sc {Van Weert}, P.}, {\sc Schrijvers, T.}, {\sc and} {\sc Demoen, B.} 2005.
\newblock {K.U.Leuven JCHR}: a user-friendly, flexible and efficient {CHR}
  system for {Java}.
\newblock In \citeN{pchr05}, 47--62.
\newblock System's homepage at {\tt
  http://www.cs.kuleuven.be/$\sim$petervw/JCHR/}.

\bibitem[\protect\citeauthoryear{{Van Weert}, Sneyers, and Demoen}{{Van Weert},
  Sneyers et~al\mbox{.}}{2008}]{vanweert_sney_demoen_aggregates_lopstr07}
{\sc {Van Weert}, P.}, {\sc Sneyers, J.}, {\sc and} {\sc Demoen, B.} 2008.
\newblock Aggregates for {CHR} through program transformation.
\newblock In {\em LOPSTR '07, Revised Selected Papers}, {A.~King}, Ed. LNCS,
  vol. 4915.

\bibitem[\protect\citeauthoryear{{Van Weert}, Sneyers, Schrijvers, and
  Demoen}{{Van Weert}, Sneyers
  et~al\mbox{.}}{2006}]{vanweert_sney_schr_demoen_negation_chr06}
{\sc {Van Weert}, P.}, {\sc Sneyers, J.}, {\sc Schrijvers, T.}, {\sc and} {\sc
  Demoen, B.} 2006.
\newblock Extending {CHR} with negation as absence.
\newblock In \citeN{pchr06}, 125--140.

\bibitem[\protect\citeauthoryear{Van~Weert, Wuille, Schrijvers, and
  Demoen}{Van~Weert, Wuille
  et~al\mbox{.}}{2008}]{vanweert_wuille_et_al_chr_imperative_lnai08}
{\sc Van~Weert, P.}, {\sc Wuille, P.}, {\sc Schrijvers, T.}, {\sc and} {\sc
  Demoen, B.} 2008.
\newblock {CHR} for imperative host languages.
\newblock In {\em Special Issue on Constraint Handling Rules}. LNAI, vol. 5388.
  Springer.

\bibitem[\protect\citeauthoryear{Voets, Pilozzi, and {De Schreye}}{Voets,
  Pilozzi et~al\mbox{.}}{2007}]{voets_pilozzi_deschreye_termination_chr07}
{\sc Voets, D.}, {\sc Pilozzi, P.}, {\sc and} {\sc {De Schreye}, D.} 2007.
\newblock A new approach to termination analysis of {C}onstraint {H}andling
  {R}ules.
\newblock In \citeN{pchr07}, 77--89.

\bibitem[\protect\citeauthoryear{Voll}{Voll}{2006}]{voll_thesis06}
{\sc Voll, K.} 2006.
\newblock A methodology of error detection: Improving speech recognition in
  radiology.
\newblock Ph.D. thesis, Simon Fraser University, Canada.

\bibitem[\protect\citeauthoryear{Wazny}{Wazny}{2006}]{wazny_phdthesis06}
{\sc Wazny, J.} 2006.
\newblock Type inference and type error diagnosis for {H}indley/{M}ilner with
  extensions.
\newblock Ph.D. thesis, University of Melbourne, Australia.

\bibitem[\protect\citeauthoryear{Wolf}{Wolf}{1999}]{wolf_phdthesis99}
{\sc Wolf, A.} 1999.
\newblock Adaptive {C}onstraintverarbeitung mit {Constraint-Handling-Rules} --
  ein allgemeiner {A}nsatz zur {L}{\"o}sung dynamischer {C}onstraint-probleme.
\newblock Ph.D. thesis, Technical University Berlin, Berlin, Germany.

\bibitem[\protect\citeauthoryear{Wolf}{Wolf}{2000a}]{wolf_projection_compulog0%
0}
{\sc Wolf, A.} 2000a.
\newblock Projection in adaptive constraint handling.
\newblock In {\em New Trends in Constraints, Joint ERCIM/Compulog Net Workshop,
  October 1999, Selected papers}, {K.~Apt}, {A.~Kakas}, {E.~Monfroy}, {and}
  {F.~Rossi}, Eds. LNCS, vol. 1865. Springer, 318--338.

\bibitem[\protect\citeauthoryear{Wolf}{Wolf}{2000b}]{wolf_rule_based_hierarchi%
es_rcorp00}
{\sc Wolf, A.} 2000b.
\newblock Toward a rule-based solution of dynamic constraint hierarchies over
  finite domains.
\newblock In \citeN{rcorp00}.

\bibitem[\protect\citeauthoryear{Wolf}{Wolf}{2001a}]{wolf_adaptive_chr_java_cp%
01}
{\sc Wolf, A.} 2001a.
\newblock Adaptive constraint handling with {CHR} in {J}ava.
\newblock In {\em CP '01}, {T.~Walsh}, Ed. LNCS, vol. 2239. Springer, 256--270.

\bibitem[\protect\citeauthoryear{Wolf}{Wolf}{2001b}]{wolf_attr_vars_inap01}
{\sc Wolf, A.} 2001b.
\newblock Attributed variables for dynamic constraint solving.
\newblock In {\em Proc.\ 14th Intl.\ Conf.\ Applications of Prolog}. Prolog
  Association of Japan, 211--219.

\bibitem[\protect\citeauthoryear{Wolf}{Wolf}{2005}]{wolf_intelligent_search_tp%
lp05}
{\sc Wolf, A.} 2005.
\newblock Intelligent search strategies based on adaptive {C}onstraint
  {H}andling {R}ules.
\newblock In \citeN{tplp05}, 567--594.

\bibitem[\protect\citeauthoryear{Wolf, Fr{\"u}hwirth, and Meister}{Wolf,
  Fr{\"u}hwirth et~al\mbox{.}}{2005}]{pwclp05}
{\sc Wolf, A.}, {\sc Fr{\"u}hwirth, T.}, {\sc and} {\sc Meister, M.}, Eds.
  2005.
\newblock {\em W(C)LP '05: {P}roc. 19th {W}orkshop on ({C}onstraint) {L}ogic
  {P}rogramming}. Ulmer Informatik-Berichte, vol. 2005-01. Universit{\"a}t Ulm,
  Germany.

\bibitem[\protect\citeauthoryear{Wolf, Gruenhagen, and Geske}{Wolf, Gruenhagen
  et~al\mbox{.}}{2000}]{wolf_ea_incremental_adaptation_aai00}
{\sc Wolf, A.}, {\sc Gruenhagen, T.}, {\sc and} {\sc Geske, U.} 2000.
\newblock On incremental adaptation of {CHR} derivations.
\newblock In \citeN{aai00}, 389--416.

\bibitem[\protect\citeauthoryear{Wolf, Robin, and Vitorino}{Wolf, Robin
  et~al\mbox{.}}{2007}]{wolf_robin_vitorino_adaptive_chr_or_chr07}
{\sc Wolf, A.}, {\sc Robin, J.}, {\sc and} {\sc Vitorino, J.} 2007.
\newblock Adaptive {CHR} meets {CHR}$^{\lor}$: An extended refined operational
  semantics for {CHR}$^{\lor}$ based on justifications.
\newblock In \citeN{pchr07}, 1--15.

\bibitem[\protect\citeauthoryear{Wuille, Schrijvers, and Demoen}{Wuille,
  Schrijvers et~al\mbox{.}}{2007}]{wuille_schr_demoen_cchr_chr07}
{\sc Wuille, P.}, {\sc Schrijvers, T.}, {\sc and} {\sc Demoen, B.} 2007.
\newblock {CCHR}: the fastest {CHR} implementation, in {C}.
\newblock In \citeN{pchr07}, 123--137.
\newblock System's homepage at {\tt
  http://www.cs.kuleuven.be/$\sim$pieterw/CCHR/}.

\end{thebibliography}
